\documentclass[pra, a4paper, aps, twocolumn, notitlepage, superscriptaddress, floatfix, longbibliography]{revtex4-2}

\usepackage{chngcntr}
\usepackage[breaklinks,colorlinks,linkcolor=blue,urlcolor=blue, citecolor=blue]{hyperref}
\usepackage{etoolbox}
\appto\UrlBreaks{\do\-}
\usepackage{amsmath, amssymb, amsthm}
\usepackage{mathtools}
\usepackage{color}
\usepackage{braket}
\usepackage{bbm}
\usepackage{nicefrac}
\usepackage{booktabs}
\DeclareMathOperator{\poly}{poly}
\DeclareMathOperator{\polylog}{polylog}

\DeclareMathOperator{\spn}{span}
\DeclareMathOperator{\css}{{\mathcal{Q}}_{\mathrm{CSS}}}
\DeclareMathOperator{\Q}{\mathcal{Q}}
\DeclareMathOperator{\C}{\mathcal{C}}
\allowdisplaybreaks[4]

\theoremstyle{definition}
\newtheorem{theorem}{Theorem}

\newtheorem{proposition}[theorem]{Proposition}
\theoremstyle{definition}

\theoremstyle{remark}

\begin{document}

\title{Time-Efficient Constant-Space-Overhead Fault-Tolerant Quantum Computation}

\author{Hayata Yamasaki}
\email{hayata.yamasaki@gmail.com}
\affiliation{Department of Physics, Graduate School of Science, The University of Tokyo, Hongo 7--3--1, Bunkyo-ku, Tokyo 113--0033, Japan}
\affiliation{Institute for Quantum Optics and Quantum Information --- IQOQI Vienna, Austrian Academy of Sciences, Boltzmanngasse 3, 1090 Vienna, Austria}
\affiliation{Atominstitut,  Technische  Universit{\"a}t  Wien, Stadionallee 2, 1020  Vienna,  Austria}
\author{Masato Koashi}
\affiliation{Department of Applied Physics, Graduate School of Engineering, The University of Tokyo, 7--3--1  Hongo, Bunkyo-ku, Tokyo 113--8656, Japan}
\affiliation{Photon Science Center, Graduate School of Engineering, The University of Tokyo, 7--3--1 Hongo, Bunkyo-ku, Tokyo 113--8656, Japan}

\begin{abstract}
Scaling up quantum computers to attain substantial speedups over classical computing requires fault tolerance. Conventionally, protocols for fault-tolerant quantum computation demand excessive space overheads by using many physical qubits for each logical qubit. A more recent protocol using quantum analogues of low-density parity-check codes needs only a constant space overhead that does not grow with the number of logical qubits. However, the overhead in the processing time required to implement this protocol grows polynomially with the number of computational steps. To address these problems, here we introduce an alternative approach to constant-space-overhead fault-tolerant quantum computing using a concatenation of multiple small-size quantum codes rather than a single large-size quantum low-density parity-check code. We develop techniques for concatenating different quantum Hamming codes with growing sizes. As a result, we construct a low-overhead protocol to achieve constant space overhead and only quasi-polylogarithmic time overhead simultaneously. Our protocol is fault tolerant even if a decoder has a non-constant runtime, unlike the existing constant-space-overhead protocol. This code concatenation approach will make possible a large class of quantum speedups within feasibly bounded space overhead yet negligibly short time overhead.
\end{abstract}

\maketitle

Fault-tolerant quantum computation (FTQC) establishes a way to realize quantum computation achieving useful computational acceleration compared to conventional classical computation, even in the presence of intrinsic noise~\cite{G,N4}.
To solve computational problems for an $M$-bit input, quantum computation may exploit a quantum circuit of polynomial size $O(\poly(M))$ in width and depth.
If we run this original circuit directly on physical qubits, noise-induced errors may destroy the result of quantum computation.
A fault-tolerant protocol reduces the effect of errors by simulating the original circuit on logical qubits of a quantum error-correcting code using an adequate number of physical qubits.
Conventionally, using concatenated codes such as Steane's $7$-qubit code~\cite{PhysRevLett.77.793,S3}, or quantum low-density parity-check (LDPC) codes such as the surface code~\cite{Kitaev_1997,K2,Bravyi_1998}, fault-tolerant protocols can arbitrarily suppress the error rate on logical qubits if that on physical qubits is below a certain threshold~\cite{10.1145/258533.258579,10.1137/S0097539799359385,548464,Kitaev_1997,doi:10.1098/rspa.1998.0166,10.5555/2011665.2011666,10.1007/11786986_6,PhysRevA.71.012336,doi:10.1063/1.1499754,PhysRevLett.109.180502}.
However, error suppression in the conventional protocols requires a growing ratio of the number of physical qubits per logical qubit, a \textit{space overhead}~\cite{gottesman2014faulttolerant}, which scales polylogarithmically in $M$ and diverges to infinity.
In practice, the number of physical qubits available for a quantum device is severely limited, and the space overhead has been a major obstacle to realizing quantum computation~\cite{Preskill2018quantumcomputingin,PRXQuantum.1.020312,Gidney2021howtofactorbit}.

The fault-tolerant protocols also require extra runtime for implementing logical quantum gates in terms of the circuit depth, a \textit{time overhead}~\cite{gottesman2014faulttolerant}.
A class of conventional protocols can achieve a polylogarithmic time overhead with transversal implementation of Clifford gates and gate teleportation of non-Clifford gates~\cite{G,N4,PhysRevA.57.127,K5,K6,gottesmanchuang1999,PhysRevA.62.052316}.
The gate teleportation is assisted by auxiliary qubits that are to be prepared in fixed logical quantum states in a fault-tolerant way while executing the computation.
In such conventional protocols, the gates are applicable to all logical qubits at a time, i.e., \textit{parallelizable}.
Each logical gate is applied within a polylogarithmic time overhead, which can be considered to be negligibly small compared to a polynomial runtime of quantum computation.

The reduction of space and time overheads in FTQC is crucial for realizing wide-ranged applications of quantum information processing and hence has been of great interest from both practical and theoretical perspectives.
In contrast to the conventional protocols, theoretical progress in Refs.~\cite{PhysRevA.87.020304,gottesman2014faulttolerant,8555154} has shown that the space overhead can indeed be made constant, using quantum expander codes~\cite{6671468,6284206,10.1145/3188745.3188886,7354429}, a family of quantum LDPC codes with a non-vanishing rate of logical qubits per physical qubit.
However, unlike the conventional protocols, the protocol in Refs.~\cite{PhysRevA.87.020304,gottesman2014faulttolerant,8555154} has a drawback that the gates are not completely parallelizable, i.e., are applicable only to an asymptotically vanishing fraction of the logical qubits at a time.
As a result, sequential gate implementation is imposed, leading to a polynomially growing time overhead~\cite{gottesman2014faulttolerant,8555154}.
A key open problem in the field of FTQC, originally raised in Ref.~\cite{gottesman2014faulttolerant}, is whether we can resolve this apparent trade-off between the space and time overheads in FTQC within the law of quantum mechanics.
Simultaneously with the constant space overhead, it would be critical to achieve a strictly less time overhead than polynomials of arbitrarily small degree, so as not to ruin a large class of useful quantum accelerations including polynomial ones as well as exponential ones.

The establishment of a low-overhead protocol by resolving such a trade-off has been challenging as long as we use existing techniques.
While the existing constant-space-overhead protocol~\cite{PhysRevA.87.020304,gottesman2014faulttolerant,8555154} implements gates by the gate teleportation,
error suppression requires a large code block, and its non-parallelizability arises from the fact that state preparation required for the gate teleportation has been hard for such a large code without relying on conventional concatenated codes after all~\cite{gottesman2014faulttolerant}.
Without the concatenated codes, non-fault-tolerant state preparation, e.g., by protocols in Refs.~\cite{PhysRevA.95.032339,PhysRevA.97.032331,Zheng_2020}, would suffer from more errors as the code becomes larger, which is infeasible on large scales.
But with the concatenated codes incurring the growing space overhead, the complete parallelism in the fault-tolerant state preparation has been impossible within the constant space overhead~\cite{gottesman2014faulttolerant}.
Another gate-implementation method for the quantum LDPC codes may be to use code deformation~\cite{PhysRevX.11.011023}, but it is unknown whether such implementation can be faster than the gate teleportation due to the extra runtime of the code deformation.
A more recent method based on lattice surgery requires many auxiliary qubits for complete parallelization over all the logical qubits, which ruins the constant space overhead~\cite{cohen}.
Note that for another family of quantum LDPC codes, i.e., hyperbolic toric codes~\cite{freedman2002z2}, parallel gate implementation may be possible~\cite{Lavasani2019universallogical,NikolasB}; however, it is unknown whether these codes can feasibly realize FTQC due to lack of an efficient decoder to decide how to recover from many errors within a feasible runtime~\cite{gottesman2014faulttolerant,DBLP:journals/qic/Hastings14}.
Linear-distance quantum LDPC codes with non-vanishing rates have been developed more recently~\cite{10.1145/3519935.3520017,9996782,https://doi.org/10.48550/arxiv.2206.07750}, but no time-efficient gate-implementation method is known for these families of codes.

Even more problematically, to prove the existence of a threshold for fault tolerance, the analysis of the existing constant-space-overhead protocol assumes that classical data processing, which is used in the decoder and the gate teleportation for example, can be performed instantaneously in zero time~\cite{gottesman2014faulttolerant}.
In practice, physical experiments toward realizing FTQC are indeed challenged by the fact that implementation of classical computation has nonzero runtime that grows on large scales~\cite{9586326,Bourassa2021blueprintscalable}.
However, with finite classical computational resources incurring such growing runtime,
the time overhead and even the existence of a threshold of the existing constant-space-overhead protocol are still unknown.

To address these problems, we here develop an alternative fault-tolerant protocol that simulates a $O(\poly(M))$-size circuit within the constant space overhead $O(1)$ and only a quasi-polylogarithmic time overhead $\exp(O(\polylog(\log(M))))$.
This time overhead is substantially smaller than polynomials for arbitrarily small degrees, i.e., that shown for the existing constant-space-overhead protocol~\cite{PhysRevA.87.020304,gottesman2014faulttolerant,8555154}.
This advantage is important in realizing useful polynomial quantum speedups without polynomially large slowdown.
Remarkably, our analysis of time overhead takes into account the waiting time for the nonzero-time classical computation during FTQC\@.
The novelty of our protocol is to use a concatenated code with a non-vanishing rate constructed from a sequence of different quantum codes, rather than using a quantum LDPC code.
In the following, we show a non-vanishing rate of our code, an efficient decoder, an implementation of universal quantum computation, the existence of a threshold, and the space and time overheads, followed by the conclusion.

\begin{figure}[t]
  \centering
  \includegraphics[width=3.4in]{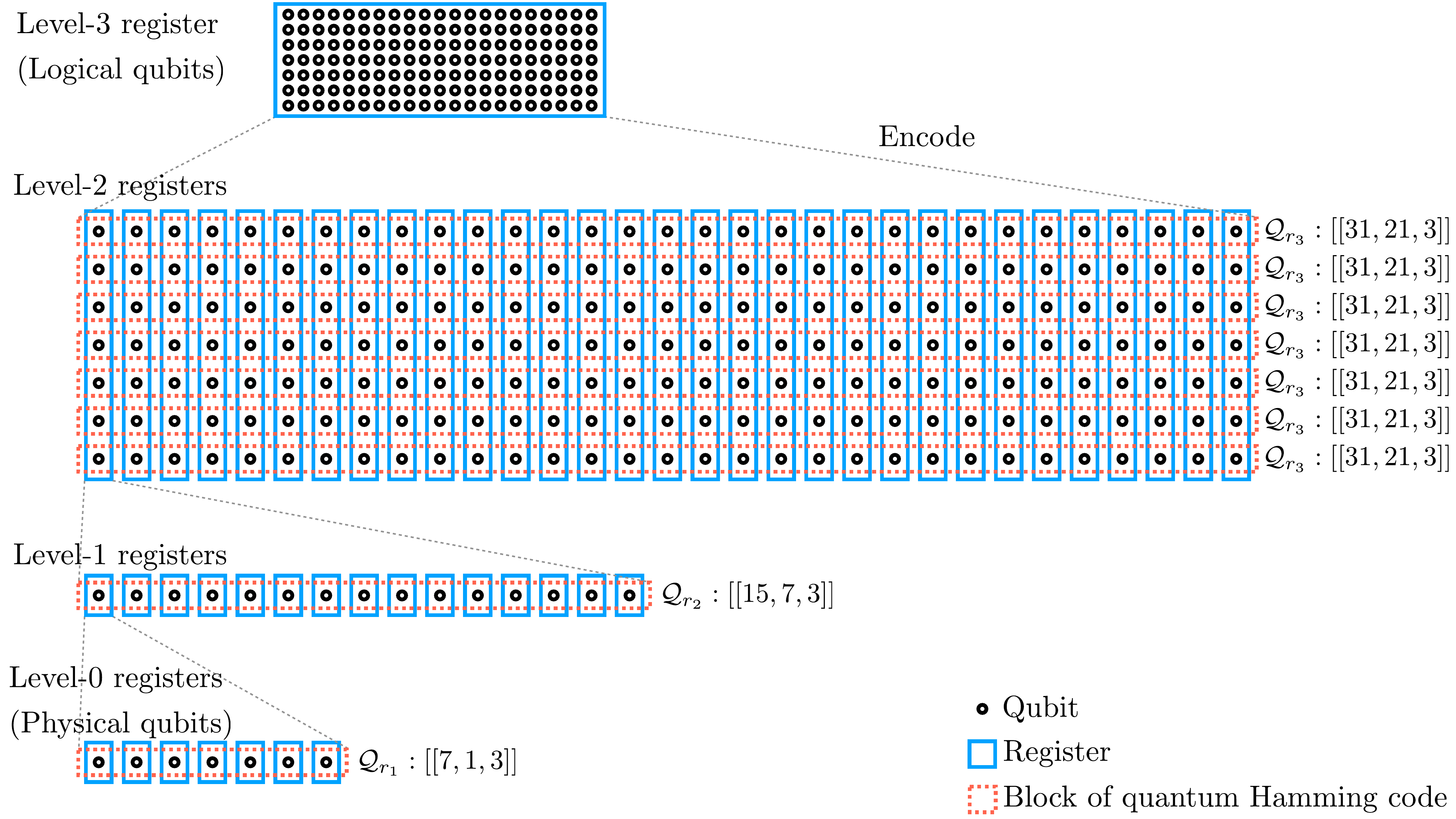}
  \caption{\label{fig:code}\textbf{Construction of the concatenated code $\Q^{(L)}$.} We concatenate a sequence of $L$ different quantum Hamming codes $\Q_{r_1},\Q_{r_2},\ldots,\Q_{r_L}$ ($L=3$ in the figure). For each concatenation level $l\in\{L, L-1,\ldots,1\}$, recursively, qubits (circles) in a level-$l$ register (blue rectangle) are encoded into those of $N_{r_l}$ level-$(l-1)$ registers using $K^{(l-1)}$ code blocks of $\Q_{r_l}$ (rectangles with red dotted border) in a grid pattern as shown in the figure. The distance-$3$ code $\Q_{r_l}$ can correct one error per code block, and this pattern is designed so that we can recover the encoded level-$l$ register even if all qubits in one of the level-$(l-1)$ registers suffer from correlated errors.}
\end{figure}

\begin{figure}[t]
  \centering
  \includegraphics[width=3.4in]{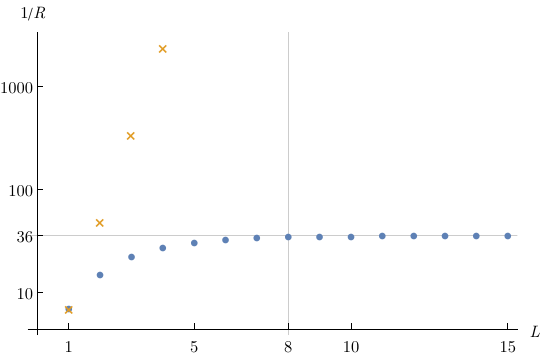}
  \caption{\label{fig:overhead}\textbf{Comparison of space overheads between our code $\Q^{(L)}$ and the concatenated $7$-qubit code.} The figure shows the overhead factor, i.e., the inverse of the rate $R(L)$, of the concatenated code $\Q^{(L)}$ developed here (blue circles) and that of the $[[7^L,1,3^L]]$ concatenated $7$-qubit code (orange $\times$) at concatenation level $L$. The overhead factor of $\Q^{(L)}$ converges to a constant $\eta_\infty<36$, saturated around $L=8$, while that of the concatenated $7$-qubit code diverges to infinity exponentially in $L$.}
\end{figure}

\paragraph*{Concatenated code at non-vanishing rate}
The crucial technique here for achieving the constant space overhead is to construct a concatenated code $\Q^{(L)}$ with a non-vanishing rate from a sequence of $L$ different quantum codes, where $L$ denotes the total number of concatenations.

We introduce the code $\Q^{(L)}$ as follows (see also Fig.~\ref{fig:code}).
With $N_r\coloneqq 2^r-1$, let $\C_r$ ($r=2,3,\ldots$) denote the family of $[N_r,N_r-r,3]$ Hamming codes~\cite{6772729}, which have $N_r$-bit block length, $(N_r-r)$-bit dimension, and distance $3$~\cite{10.5555/552386}.
Quantum Hamming codes $\Q_r$ ($r=3,4,\ldots$)~\cite{PhysRevA.54.4741} are Calderbank-Shor-Steane (CSS) codes~\cite{N4,PhysRevA.54.1098,S3} of $\C_r$ over its dual code $\C_r^\perp$, which are in a family of $[[N_r,K_r,3]]$ codes having $N_r=2^r-1$ physical qubits (i.e., code size $N_r$), $K_r\coloneqq N_r-2r$ logical qubits, and distance $3$.
We use $\Q_{r_l}$ with parameter $r_l\coloneqq l+2$ for the concatenation at each level $l\in\{1,\ldots,L\}$, which leads to the sequence $\Q_{3},\Q_{4},\ldots$ of quantum Hamming codes starting from Steane's $7$-qubit code $\Q_3$~\cite{PhysRevLett.77.793,S3} at level $1$, with rate converging to $\nicefrac{K_{r_l}}{N_{r_l}}\to 1$ as $l\to\infty$.

We construct $\Q^{(L)}$ recursively by defining $\Q^{(l)}$ for $l=L,L-1,\ldots,1$.
Let $K^{(l)}$ and $N^{(l)}$ denote the numbers of logical and physical qubits of $\Q^{(l)}$, respectively, which turn out to be $K^{(l)}=\prod_{l^{\prime}=1}^{l}K_{r_{l^{\prime}}}=\exp({O(l^2)})$, $N^{(l)}=\prod_{l^{\prime}=1}^{l}N_{r_{l^{\prime}}}=\exp({O(l^2)})$.
We define a collection of $K^{(l)}$ logical qubits of $\Q^{(l)}$ as a \textit{level-$l$ register},
where a physical qubit is referred to as a \textit{level-$0$ register} (or level-$0$ qubit).
The recursive relation between $\Q^{(l)}$ and $\Q^{(l-1)}$ is presented in Fig.~\ref{fig:code}.
In particular, we divide the $K^{(l)}=K^{(l-1)}\times K_{r_l}$ qubits in the level-$l$ register for $\Q^{(l)}$ into $K^{(l-1)}$ blocks of $K_{r_l}$ qubits.
Then, for each $k^{(l-1)}\in\{1,\ldots, K^{(l-1)}\}$, picking the $k^{(l-1)}$th qubit from each of the $N_{r_l}$ level-$(l-1)$ registers for $\Q^{(l-1)}$, we encode the $K_{r_l}$ qubits in the $k^{(l-1)}$th block into the picked $N_{r_l}$ qubits as the logical qubits of the quantum Hamming code $\Q_{r_l}$.
We design this concatenated code so that even if all qubits in one of the level-$(l-1)$ registers suffer from correlated errors, we can still recover the encoded level-$l$ register.
In this way, we obtain a $[[N^{(L)},K^{(L)},3^L]]$ concatenated code $\Q^{(L)}$.
It should be noted that, unlike quantum LDPC codes, the ability of our code to suppress errors is not fully characterized by the code distance due to its concatenated structure.

We analytically prove that $\Q^{(L)}$ has an asymptotically non-vanishing rate $R(L)\coloneqq\nicefrac{K^{(L)}}{N^{(L)}}\xrightarrow{L\to\infty}\prod_{l=1}^{\infty}\nicefrac{K_{r_l}}{N_{r_l}}\geqq\nicefrac{1}{\eta_\infty}>0$ with a finite asymptotic overhead factor $\eta_\infty$.
The overhead factor, i.e., the inverse of the rate, would diverge to infinity for the conventional concatenated and quantum LDPC codes, such as the surface code.
By contrast, we numerically show $\eta_\infty<36$ in Fig.~\ref{fig:overhead}.
Note that $\eta_\infty$ is not optimized here; e.g., the factor could be almost $1$ by starting the concatenations from $\Q_{r_{8}}$ rather than $\Q_3$.
Even at lower concatenation levels $L\leqq4$, the advantage of $\Q^{(L)}$ over the $[[7^L,1,3^L]]$ concatenated $7$-qubit code in the overhead factor can be orders of magnitude. Nevertheless, both codes can suppress the logical error rate doubly exponentially as $L\to\infty$, as shown later with our threshold analysis.
See Supplemental Information Section~B for details of our construction and analysis.

Note that the above construction of concatenated codes with non-vanishing rate is in principle applicable to other sequences of codes with rates approaching $1$ sufficiently fast; e.g., our results hold for concatenation of quantum Hamming codes with parameter $r_l\approx\log(l)$, which would be beneficial for reducing time overhead compared to $r_l\coloneqq l+2$ while a constant factor in the space overhead may become larger (see Supplemental Information Section~F for detail).
Even more generally, concatenation of a growing sequence of, e.g., $[[2^{r_l}-1, 2^{r_l}-1-2t{r_l},2t+1]]$ quantum Bose-Chaudhuri-Hocquenghem (BCH) codes ($t\geqq 1$)~\cite{PhysRevA.54.4741}, i.e., CSS codes of classical BCH codes over their dual codes, can also provide a family of concatenated codes with non-vanishing rate using the same code parameter $r_l$ as ours, which can have larger distance than quantum Hamming codes and thus have a potential advantage in faster error suppression as $L$ increases.
But apart from constructing the code, our crucial contribution will be the explicit construction of the overall fault-tolerant protocol for $\Q^{(L)}$ and the analysis of the threshold and the runtime, as shown in the following.
The general protocol construction for other concatenated codes with a non-vanishing rate is left for future work, but our results provide a fundamental design principle for such protocols.

\paragraph*{Efficient decoder}
Remarkably, $\Q^{(L)}$ has an efficient decoder even though $\Q^{(L)}$ is not a quantum LDPC code.
Efficient decoders are vital for the feasibility of FTQC yet challenging to construct in general; after all, no candidate among quantum LDPC codes for constant-space-overhead FTQC had such an efficient decoder until later research has constructed a sufficiently efficient decoder for the quantum expander code~\cite{PhysRevA.87.020304,gottesman2014faulttolerant,8555154}.

In our protocol, for each level $l$, $\Q_{r_l}$ has an efficient decoder; in particular, the efficient decoder for the classical Hamming code $\C_{r_l}$~\cite{6772729} can be used to correct one bit-flip error and one phase-flip error from syndromes of $Z$ and $X$ stabilizer generators, respectively.
The decoder for $\Q^{(L)}$ runs efficiently by recursively using the decoder for $\Q_{r_l}$ as in the conventional hard-decision decoder for concatenated codes.
With parallel classical computation, we can run this decoder in $O(\log(N_{r_l}))$ time at each level $l$, which will turn out to be polylogarithmic in the problem size $M$ and hence small compared to the $O(\poly(M))$ depth of the original circuit.
Remarkably, our threshold analysis shows that even if the decoder has this nonzero runtime, our fault-tolerant protocol provably has a threshold.

\begin{figure*}[t]
  \centering
  \includegraphics[width=7.0in]{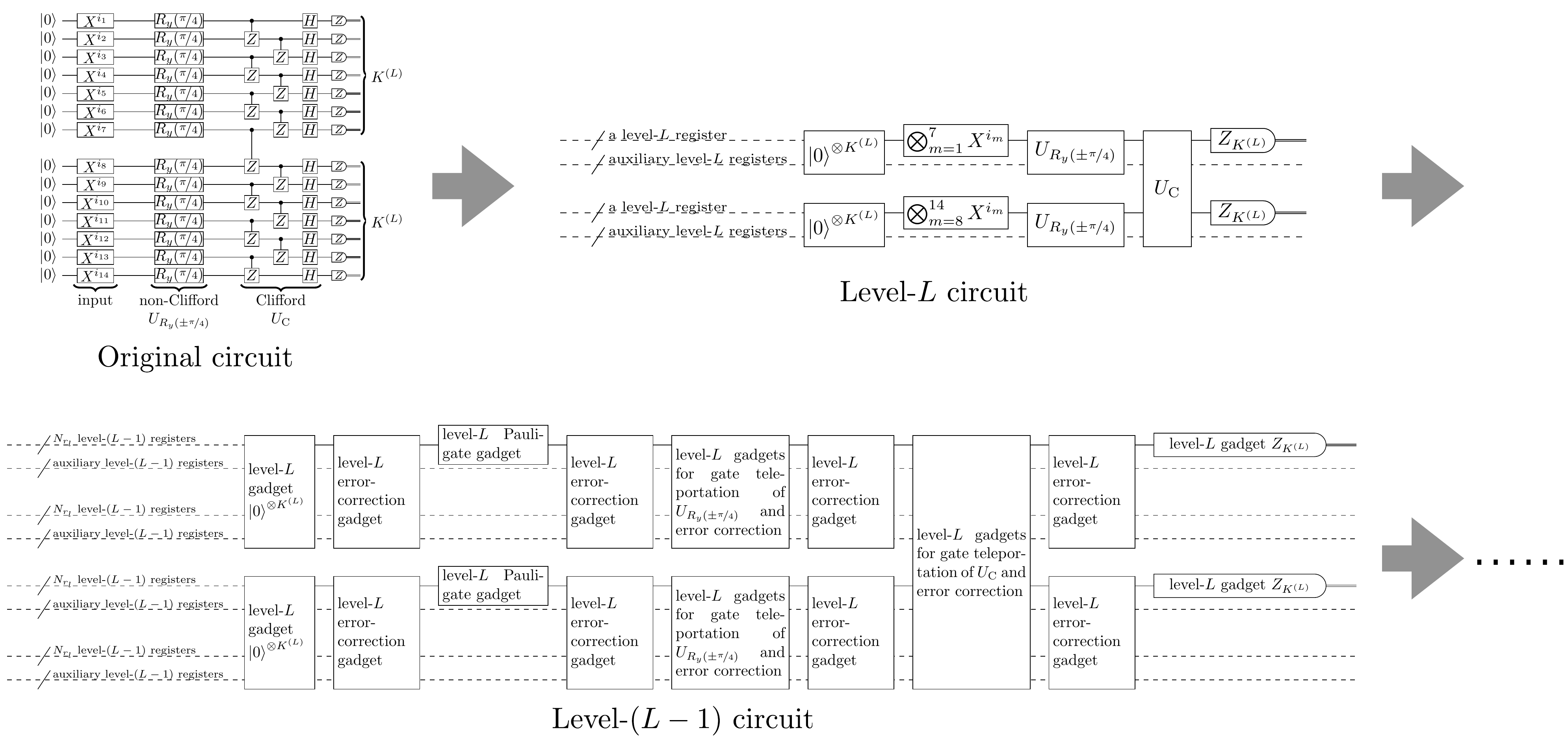}
  \caption{\label{fig:compiled_circuit}\textbf{Compilation in our fault-tolerant protocol.} We compile the original circuit into a level-$L$ circuit composed of level-$L$ elementary operations acting on level-$L$ registers. Two-register Clifford gates and one-register $R_y(\pm\nicefrac{\pi}{4})$ gates are implemented by a combination of elementary operations via gate teleportation, which are abbreviated as white boxes indicating $U_\mathrm{C}$ and $U_{R_y(\pm\nicefrac{\pi}{4})}$. For each $l=L, L-1,\ldots,1$, we further compile the level-$l$ circuit into the corresponding level-$(l-1)$ circuit by substituting each level-$l$ elementary operation into the corresponding level-$l$ gadget and inserting level-$l$ error-correction gadgets in between. Unlike conventional fault-tolerant protocols with concatenated codes, the number of level-$(l-1)$ elementary operations in level-$l$ gadgets may depend on $l$. We design our protocol in such a way that the level-$0$ circuit on physical qubits achieves the constant space overhead and the quasi-polylogarithmic time overhead compared to the original circuit.}
\end{figure*}

\paragraph*{Fault-tolerant protocol}
Using this concatenated code $\Q^{(L)}$ with the non-vanishing rate, we construct a fault-tolerant protocol.
To solve a family of computational problems with inputs represented by an $M$-bit string $(i_1,\ldots,i_M)\in{\{0,1\}}^M$, we use a $W(M)$-qubit $D(M)$-depth original circuit, where the width and the depth are polynomially bounded, i.e., $W(M)=O(\poly(M))$ and $D(M)=O(\poly(M))$.
To input $(i_1,\ldots,i_M)$ to the original circuit, we use an $M$-qubit initial state $\bigotimes_{m=1}^M\Ket{i_m}=(\bigotimes_{m=1}^M X^{i_m})\Ket{0}^{\otimes M}$, and the rest of the original circuit is determined by $M$ independently of the input.
The original circuit is written in terms of a gate set of Clifford gates $X$, $Y$, $Z$, $H$, $S$, $\textsc{CNOT}$, and $CZ$, and non-Clifford gates $R_y(\pm\nicefrac{\pi}{4})$~\cite{N4}, starting from $(\bigotimes_{m=1}^M\Ket{i_m})\otimes\Ket{0}^{\otimes (W(M)-M)}$ and ending with measurements of all qubits in $Z$ basis.
Our task is to simulate the original circuit by sampling from its output probability distribution within a given error $\epsilon$ in the total variation distance.
Based on the original circuit, our protocol recursively defines a level-$l$ circuit ($l\in\{L, L-1,\ldots,0\}$) composed of \textit{level-$l$ elementary operations} acting on level-$l$ registers, where a level-$0$ circuit is a circuit on physical qubits.
The set of level-$l$ elementary operations consists of a measurement operation, $H$-, \textsc{CNOT}-, $CZ$-, and Pauli-gate operations, initial-, Clifford-, and magic-state preparation operations, and a wait operation (see Methods).
By combining these elementary operations, our protocol performs Clifford unitaries on two registers and non-Clifford $R_y(\pm\nicefrac{\pi}{4})$ gates on each register via gate teleportation.

Figure~\ref{fig:compiled_circuit} illustrates the recursive construction of the circuits.
We first compile the original circuit into a level-$L$ circuit, where the required concatenation level $L$ is determined depending on $M$ and $\epsilon$.
For each $l=L,L-1,\ldots,1$, we further compile the level-$l$ circuit into the corresponding level-$(l-1)$ circuit using level-$l$ gadgets, i.e., level-$(l-1)$ circuits to implement level-$l$ elementary operations.
The construction here is different from the conventional protocol with concatenated codes~\cite{G} in that the circuits here are composed of elementary operations acting on registers rather than qubits, and that the procedure of converting a level-$l$ circuit to a level-$(l - 1)$ circuit depends on $l$.
Unlike the conventional protocol,
errors in the multiple qubits belonging to the same register may be highly correlated in our construction, and thus, we use the register as a unit in place of the qubit.
The gadgets used in this compilation are designed to satisfy appropriate conditions for fault tolerance.
To keep the overall space overhead constant, we design each level-$l$ gadget to use only a constant number of additional auxiliary level-$(l-1)$ registers per encoded level-$l$ register.
See Methods for details of our construction.

\paragraph*{Provable existence of a threshold}
The novelty of our protocol is to use the concatenated code $\Q^{(L)}$ constructed from a sequence of different codes $\Q_{r_1},\ldots,\Q_{r_L}$.
However, the growth of code sizes in this sequence may make the existence of a threshold nontrivial.
Conventional proofs of the threshold theorem for concatenated codes assume concatenation of the same code~\cite{G};
in contrast, a level-$l$ register for $\Q^{(l)}$ is encoded into a growing number $N_{r_l}$ of level-$(l-1)$ registers.
We nevertheless prove that a threshold exists even if the number of level-$(l-1)$ elementary operations per level-$l$ gadget grows polynomially $O(\poly(N_{r_l}))$, which is the case in our gadgets.
Remarkably, our analysis deals with the setting where the runtime of classical computation in the decoder may also grow.
The threshold analysis of the existing constant-space-overhead protocol has assumed that the decoder must run in zero time at arbitrarily large scales~\cite{gottesman2014faulttolerant},
and whether constant-space-overhead FTQC is possible with such growth of the nonzero runtime of classical computation has been unknown.
After all, large-scale FTQC needs a large-size quantum LDPC code for error suppression, but if we wait for a growing runtime of the decoder for the large-size quantum LDPC code, physical qubits suffer from more errors during waiting.
This situation violates the essential assumption for the existence of a threshold: having a constant physical error rate between performing the error corrections.
As a result, the protocols for the quantum LDPC codes may fail to correct a general class of errors on large scales (see Supplemental Information Section~E for details).
Note that the problem of requiring the non-growing runtime of classical computation persists even in the conventional protocols using quantum LDPC codes such as the $2$D surface code, where decoders must process the stream of syndrome data at the rate it is received even if the code distance grows~\cite{10.1109/ISCA45697.2020.00053,https://doi.org/10.48550/arxiv.2208.01178,https://doi.org/10.48550/arxiv.2209.08552,https://doi.org/10.48550/arxiv.2209.09219,bombin2023modular}.
By contrast, our protocol based on the concatenated code $\Q^{(L)}$ establishes how we can perform constant-space-overhead FTQC even with finite computational resources.

In our setting, we assume that the circuit on physical qubits undergoes a conventional local stochastic error model, where adversarial and correlated errors may occur at faulty locations of operations in the level-$0$ circuit at a physical error rate $p_0>0$, but the probability of $s$ locations simultaneously having the errors is bounded by $p_0^s$~\cite{G,gottesman2014faulttolerant}.
We call each level-$l$ elementary operation in a level-$l$ circuit a level-$l$ location.
Then, our analysis proves that faults at level-$L$ locations in the level-$L$ circuit of our protocol also occur according to the local stochastic error model; moreover, we prove the following proposition on error suppression, which scales doubly exponentially in $L$ in the same way as the conventional concatenated codes~\cite{G}.
(See Methods for details.)
\begin{proposition}[Error suppression with concatenation of quantum Hamming codes]
Under the local stochastic error model, we have an explicit construction of a fault-tolerant protocol using the concatenated code $\mathcal{Q}^{(L)}$ with a nonzero threshold constant $p_\mathrm{th}>0$ such that,
if the physical error rate is below the threshold $p_0<p_\mathrm{th}$,
then our protocol using the concatenated quantum Hamming code $\Q^{(L)}$ can suppress the logical error rate as $p_L=\exp(-O(2^L))$.
\end{proposition}

A practical threshold is achievable with minor modifications to our protocol.
For example, the same threshold as the surface code is achievable by encoding each level-$0$ qubit of $\Q^{(L)}$ for our protocol into the logical qubit of a \textit{constant-size} surface code at a logical error rate below the threshold constant $p_\mathrm{th}$ here, which can take the advantage of the surface code in tolerating biased noise on physical qubits~\cite{PhysRevLett.120.050505,PhysRevX.9.041031,PhysRevLett.124.130501,ataides2021xzzx}.
Note that even if the noise on physical qubits is biased, the logical error is not necessarily biased, and thus we do not have to modify our protocol for biased noise in this case.
Compared to conventional protocols, the concatenation of the surface code and our code $\Q^{(L)}$ has merits due to constant space overhead, potential speedup of decoders on large scales, and provable existence of a threshold even with nonzero-time decoders.

\paragraph*{Space and time overheads}
The significance of our protocol is to simultaneously achieve the constant space overhead and the quasi-polylogarithmic time overhead compared to the original circuit, as shown in the following proposition.
(See Methods for details.)

\begin{proposition}[Overhead achieved by concatenation of quantum Hamming codes]
Under the local stochastic error model, we have an explicit construction of a fault-tolerant protocol using the concatenated code $\mathcal{Q}^{(L)}$ with a concatenation level $L=\Theta(\log(\log(\nicefrac{M}{\epsilon})))$ to simulate any $W(M)$-qubit $D(M)$-depth original circuit within error $\epsilon>0$ in total variation distance using at most $W(M)\times O(1)$ physical qubits and $D(M)\times\exp(O(\log^2(\log(\nicefrac{M}{\epsilon}))))$ runtime in terms of the depth of the fault-tolerant circuit, where $W(M)=O(\poly(M))$ and $D(M)=O(\poly(M))$.
\end{proposition}
These overheads include those for preparing the auxiliary states required for the gate teleportation and the error correction.
Unlike the previous analysis of the existing constant-space-overhead protocol~\cite{PhysRevA.87.020304,gottesman2014faulttolerant,8555154},
the runtime also includes wait operations to wait for nonzero-time classical computations such as ones for the decoder and the gate teleportation.

As long as we use the existing techniques for the constant-space-overhead protocol~\cite{PhysRevA.87.020304,gottesman2014faulttolerant,8555154}, it was challenging to achieve these space and time overheads simultaneously.
After all, the existing protocol~\cite{PhysRevA.87.020304,gottesman2014faulttolerant,8555154} relies on conventional concatenated codes in preparing the auxiliary states for gate teleportation and hence cannot achieve the parallel gate implementation on all logical qubits within constant space overhead; then, sequential gate implementation incurs a polynomially large time overhead in implementing parallel gates of the original circuit.
By contrast, our protocol is designed to attain complete parallelizability in the gate teleportation to apply the gates to all logical qubits of $\Q^{(L)}$ at a time.
All the required auxiliary states for the gate teleportation can be prepared in parallel within the constant space overhead owing to the non-vanishing rate of $\Q^{(L)}$.
Consequently, our protocol is advantageous in terms of the time overhead compared to the existing constant-space-overhead protocol~\cite{PhysRevA.87.020304,gottesman2014faulttolerant,8555154}.

\paragraph*{Tolerance for architectural overheads}
Remarkably, our analysis also shows that a threshold would exist even with any polynomially growing architectural time overhead $O(\poly(N^{(l)}))$ in the code size at each concatenation level $l$, which may be imposed by restrictions such as nearest-neighbor interactions on $2$D geometry, limited classical computational resources for the decoder, and insufficient parallelization in preparing auxiliary states used for gate teleportation and error correction.
The suppression of the logical error rate is much faster than the growth of the code size, and thus, the concatenated codes can tolerate the time overhead at higher concatenation levels by just waiting by performing identity gates, as in our protocol.
This unique property of the concatenated code contrasts with the fact that quantum LDPC codes have to be decoded at least once in a constant time to avoid the accumulation of errors.
Thus, our protocol is expected to be implementable on various architectures with minor adaptation.

For example, one can rewrite fault-tolerant protocols with concatenated codes into those respecting the $2$D (or even $1$D) geometry by well-established procedures in Refs.~\cite{10.1137/S0097539799359385,doi:10.1080/09500340008244046,PhysRevA.72.022317,DBLP:journals/qic/SvoreDT07}, and our analysis shows that a threshold exists even with a polynomial time overhead in such rewriting.
Note that when the original circuit includes two-qubit gates on arbitrary pairs of qubits, any protocol on such $2$D architectures to simulate the original circuit unavoidably incurs a polynomially long time overhead (see Supplemental Information Section~A for detail); moreover, the constant space overhead would not be achievable on a single fully $2$D chip~\cite{https://doi.org/10.48550/arxiv.2302.04317}.
Hence, it is essential to investigate architectures with multiple $2$D layers, such as that in Ref.~\cite{PhysRevLett.129.050504}, or with full connectivity, such as photonics.
We leave the investigation of practical architectures to implement our protocol for future research.

We also leave exact evaluation of the threshold $p_\mathrm{th}$ for future research, which requires numerical simulation with taking the architectural overhead into account as the analytical bound is not usually tight.
In such numerical simulation, comparison with architectural proposals for the constant-space-overhead protocol with quantum LDPC codes~\cite{PhysRevLett.129.050504} may also be an interesting direction.
Yet importantly, the above merit of our protocol in tolerating architectural constraints such as non-constant-time decoders is not known to hold for the existing constant-space-overhead protocol with the quantum LDPC codes.
We expect that the numerical simulation can also clarify the effects of the architectural constraints that appear in practice.

\paragraph*{Conclusion and outlook}
We have constructed a protocol for FTQC achieving constant space overhead and quasi-polylogarithmic time overhead simultaneously.
A crucial technical development is to use a concatenated code constructed from a growing sequence of quantum Hamming codes.
Our technique leads to a non-vanishing rate, the existence of an efficient decoder, the space-saving and fast protocol for simulating universal quantum computation, and the provable existence of a threshold for doubly exponential error suppression as we increase the concatenation level.
Progressing beyond previous studies of the existing constant-space-overhead protocol based on quantum LDPC codes~\cite{PhysRevA.87.020304,gottesman2014faulttolerant,8555154}, we take into account nonzero runtime of classical computation in proving these results.
Our results are fundamental for realizing FTQC feasibly within constant space overhead and yet in short time overhead with parallelization.
Remarkably, this achievement is made possible with the technique of code concatenation, which opens a promising way for the low-overhead FTQC\@.

\section*{Methods}

We here summarize our notation and then present construction of our fault-tolerant protocol, derivation of the existence of a threshold, and analysis of the space and time overheads.
See Supplemental Information for further detail.

\paragraph*{Notation}
For a qubit $\mathbb{C}^2$,
the $Z$ basis is denoted by $\{\Ket{0},\Ket{1}\}$, and the $X$ basis by $\{\Ket{\pm}\coloneqq(\nicefrac{1}{\sqrt{2}})(\Ket{0}\pm\Ket{1})\}$.
Matrix elements are represented in terms of the $Z$ basis.
By convention of Ref.~\cite{N4}, we use the following notation:
\begin{align}
  X&=\left(\begin{matrix}
      0 & 1\\
      1 & 0
  \end{matrix}\right),\\
      Z&=\left(\begin{matrix}
          1 & 0\\
          0 & -1
      \end{matrix}\right),\\
      Y&=\left(\begin{matrix}
          0 & -\mathrm{i}\\
          \mathrm{i} & 0
      \end{matrix}\right)\propto XZ,\\
          H&=\frac{1}{\sqrt{2}}\left(\begin{matrix}
              1 & 1\\
              1 & -1
          \end{matrix}\right),\\
          S&=\left(\begin{matrix}
              1 & 0\\
              0 & \mathrm{i}
          \end{matrix}\right),\\
              \textsc{CNOT}&=\left(\begin{matrix}
                  1 & 0 & 0 & 0\\
                  0 & 1 & 0 & 0\\
                  0 & 0 & 0 & 1\\
                  0 & 0 & 1 & 0
          \end{matrix}\right),\\
              CZ&=\left(\begin{matrix}
                  1 & 0 & 0 & 0\\
                  0 & 1 & 0 & 0\\
                  0 & 0 & 1 & 0\\
                  0 & 0 & 0 & -1
          \end{matrix}\right),\\
                  R_y(\theta)&=\left(\begin{matrix}
                      \cos(\nicefrac{\theta}{2}) & -\sin(\nicefrac{\theta}{2})\\
                      \sin(\nicefrac{\theta}{2}) & \cos(\nicefrac{\theta}{2})
          \end{matrix}\right).
\end{align}
The identity operator is denoted by
\begin{equation}
\mathbbm{1}=\left(\begin{matrix}
                      1 & 0\\
                      0 & 1
          \end{matrix}\right).
\end{equation}

In the same way as referring to a running time $\exp(O(\log^c(M)))$ for fixed $c>0$ as a quasi-polynomial time in $M$,
we call
\begin{equation}
  \exp(O(\log^c(\log(M))))
\end{equation}
a quasi-polylogarithmic time.
A quasi-polylogarithmic time with $c>1$ may be larger than a polylogarithmic time $\exp(O(\log(\log(M))))=O(\polylog(M))$.
However, a quasi-polylogarithmic time for any $c>0$ is much smaller than a polynomial time, i.e., $\exp(O(\log(M)))=O(\poly(M))=O(M^\alpha)$, even for an arbitrarily small degree $\alpha>0$ of the polynomial.

\paragraph*{Construction of fault-tolerant protocol}
In our fault-tolerant protocol, we use level-$l$ elementary operations to write a level-$l$ circuit for each $l\in\{L,L-1,\ldots,0\}$.
The set of level-$l$ elementary operations consist of a level-$l$ measurement operation, level-$l$ $H$-, \textsc{CNOT}-, $CZ$-, and Pauli-gate operations, level-$l$ initial-, Clifford-, and magic-state preparation operations, and a level-$l$ wait operation.
The measurement operation implements measurements in $Z$ basis of all qubits in a level-$l$ register.
The $H$-, $\textsc{CNOT}$-, $CZ$-, and Pauli-gate operations implement $H$, $\textsc{CNOT}$, and $CZ$ gates on all qubits in level-$l$ registers and tensor product of any combination of Pauli gates on the qubits in a level-$l$ register.
The initial-state preparation operation prepares a level-$l$ register in $\Ket{0}^{\otimes K^{(l)}}$.
To assist implementing any given two-register Clifford unitary $U_\mathrm{C}$ on the $2K^{(l)}$ qubits in two level-$l$ registers,
the Clifford-state preparation operation prepares four level-$l$ registers $B_1,B_2,B_3,B_4$ in
\begin{equation}
  \label{eq:Clifford_state}
  (\mathbbm{1}^{B_1B_2}\otimes U_\mathrm{C}^{B_3B_4})\ket{\Phi^{(l)}}^{B_1B_2B_3B_4},
\end{equation}
where $\ket{\Phi^{(l)}}^{B_1B_2B_3B_4}=\ket{\Phi}^{B_1B_3}\otimes\ket{\Phi}^{B_2B_4}$, and $\ket{\Phi}^{B_j B_{j^\prime}}\propto\sum_{m=0}^{2^{K^{(l)}}-1}\Ket{m}^{B_j}\otimes\Ket{m}^{B_{j^\prime}}$ for $(j,j^\prime)\in\{(1,3),(2,4)\}$ is a maximally entangled state between $B_j$ and $B_{j^\prime}$.
To assist implementing any given unitary $U_{R_y(\pm\nicefrac{\pi}{4})}$ in the form of a tensor product of $R_y(\nicefrac{\pi}{4})$, $R_y(-\nicefrac{\pi}{4})$, and $\mathbbm{1}$ on the $K^{(l)}$ qubits in a level-$l$ register,
the magic-state preparation operation prepares two level-$l$ registers in
\begin{equation}
  {(R_y(\nicefrac{\pi}{4})\Ket{0})}^{\otimes K^{(l)}}\otimes{(R_y(\nicefrac{\pi}{4})\Ket{0})}^{\otimes K^{(l)}}.
\end{equation}
A wait operation is a Pauli-gate operation of $\mathbbm{1}^{\otimes K^{(l)}}$.

Using these operations in combination, we implement $U_\mathrm{C}$ and $U_{R_y(\pm\nicefrac{\pi}{4})}$ for a level-$l$ circuit by gate teleportation.
Note that $H$-, \textsc{CNOT}-, $CZ$-gate operations in our protocol perform the gates on all qubits in the registers simultaneously, and we use $U_\mathrm{C}$ for implementing all the other Clifford gates, e.g., a single-qubit $H$ gate in one of the two registers (while acting as the identity gate on the other register), and a \textsc{CNOT} gate on a specific pair of qubits in the two registers.
Indeed, $U_\mathrm{C}$ is not limited to a one- or two-qubit Clifford gate but may represent an arbitrary sequence of Clifford gates acting on the $2K^{(l)}$ qubits in two level-$l$ registers as a single Clifford unitary; similarly, $U_{R_y(\pm\nicefrac{\pi}{4})}$ is not necessarily a single-qubit non-Clifford gate but can apply non-Clifford gates on any number of qubits in a level-$l$ register.
In particular, a level-$l$ two-register Clifford gate $U_\mathrm{C}$ in our protocol is implemented by means of gate teleportation~\cite{gottesmanchuang1999,K5,K6}, assisted by the auxiliary state $(\mathbbm{1}^{B_1B_2}\otimes U_\mathrm{C}^{B_3B_4})\ket{\Phi^{(l)}}^{B_1B_2B_3B_4}$ prepared by the level-$l$ Clifford-state preparation operation, along with other level-$l$ gate and measurement operations.
The correction of byproducts in the gate teleportation is performed by level-$l$ Pauli-gate operations.
Regarding level-$l$ $U_{R_y(\pm\nicefrac{\pi}{4})}$,
the gate teleportation for $U_{R_y(\pm\nicefrac{\pi}{4})}$ is assisted by an auxiliary magic state ${(R_y(\nicefrac{\pi}{4})\Ket{0})}^{\otimes K^{(l)}}$ prepared by the level-$l$ magic-state preparation operation, and also an auxiliary state ${\Ket{0}}^{\otimes K^{(l)}}$ prepared by the level-$l$ initial-state preparation operation, along with other level-$l$ gate and measurement operations.
To apply $R_y(\pm\nicefrac{\pi}{4})$ in $U_{R_y(\pm\nicefrac{\pi}{4})}$ to a desired subset of qubits in a register, we prepare the required auxiliary state in the tensor product of $R_y(\nicefrac{\pi}{4})\Ket{0}$ and ${\Ket{0}}$ by applying single-qubit \textsc{SWAP} gates between ${(R_y(\nicefrac{\pi}{4})\Ket{0})}^{\otimes K^{(l)}}$ and ${\Ket{0}}^{\otimes K^{(l)}}$ using the level-$l$ two-register Clifford gate.
Then, assisted by this auxiliary state, we perform the gate teleportation.
The correction of byproducts is single-qubit Clifford gates on a level-$l$ register, performed by the level-$l$ two-register Clifford gate acting trivially on another auxiliary level-$l$ register.

To simulate a level-$l$ circuit at each level $l\in\{L,\ldots,1\}$, we construct a level-$l$ gadget corresponding to each level-$l$ elementary operation, i.e., a level-$(l-1)$ circuit for simulating the elementary operation on encoded level-$l$ registers.
Apart from these level-$l$ gadgets, we use a level-$l$ error-correction gadget, a level-$(l-1)$ circuit for correcting errors on one of the $N_{r_l}$ level-$(l-1)$ registers for an encoded level-$l$ register.
For the existence of a threshold, each level-$l$ gadget must be \textit{fault-tolerant}; that is, roughly speaking, even if one of the level-$(l-1)$ locations in the gadget has a fault, the resulting error should be correctable using the decoder of $\Q_{r_l}$ at the end of the gadget (See Supplemental Information for precise definition).
This definition of fault-tolerant gadgets in our protocol is a suitable modification of conventional definition for the concatenated codes~\cite{G}, so that we can prove the existence of a threshold by applying the conventional argument in Ref.~\cite{G} to our protocol.
Using the fault-tolerant level-$l$ gadgets, we convert a level-$l$ circuit into the corresponding level-$(l-1)$ circuit by replacing each level-$l$ elementary operation with the corresponding level-$l$ gadget, followed by inserting the level-$l$ error-correction gadgets between all pairs of adjacent level-$l$ elementary operations.
Repeating this conversion recursively for $l\in\{L,\ldots,1\}$ yields a level-$0$ circuit,
which leads to a fault-tolerant circuit on physical qubits to simulate the original circuit.

In the following, we sketch our construction of level-$l$ gadgets used for the fault-tolerant protocol.
See Supplemental Information for further detail.
Note that gate implementations for some classes of CSS codes with multiple logical qubits have also been discussed in Refs.~\cite{gottesman2014faulttolerant,PhysRevA.72.052335}, but the main contribution of our work is to present the gadgets explicitly for our code $\mathcal{Q}^{(l)}$ so that we can prove the existence of a threshold and bound the space and time overheads rigorously for our fault-tolerant protocol.

We implement the level-$l$ measurement gadget by performing level-$(l-1)$ measurement operations for all the level-$(l-1)$ registers and then calculating bit values of the outcome by a decoder, using the logical $Z$ operator for each of the $K^{(l)}$ logical qubits in the encoded level-$l$ register.
We let $Z_{K^{(l)}}$ label this $Z$-basis measurement with the $K^{(l)}$-bit outcome.
The fault tolerance follows from transversality.

We implement the $H$-, $\textsc{CNOT}$-, and $CZ$-gate gadgets by applying level-$(l-1)$ $H$-, $\textsc{CNOT}$-, and $CZ$-gate operations, respectively, to all level-$(l-1)$ registers transversally.
We implement the Pauli-gate gadget by level-$(l-1)$ Pauli-gate operations to apply the tensor product of Pauli gates representing the logical Pauli operators to the level-$(l-1)$ registers transversally.
The wait gadget is a special case of the Pauli-gate gadget to apply the identity gate.
The fault tolerance follows from transversality.

The level-$l$ initial-state preparation gadget is implemented by transforming states $\Ket{0}^{\otimes K^{(l-1)}}$ prepared by the level-$(l-1)$ initial-state preparation operations into logical $\Ket{0}^{\otimes K^{(l)}}$ by a level-$(l-1)$ stabilizer circuit in a non-fault-tolerant way~\cite{PhysRevA.56.76,10.5555/2481569.2481579,Steane_2002}, followed by verification with post-selection.
For the verification, we prepare another logical $\Ket{0}^{\otimes K^{(l)}}$, and using this auxiliary $\Ket{0}^{\otimes K^{(l)}}$, we measure the logical $Z$ operators and the $Z$ stabilizer generators of $\Q_{r_l}$.
If no logical $X$ error is detected from this measurement on the logical $\Ket{0}^{\otimes K^{(l)}}$ prepared in this first run, i.e., in the case of success in the verification, then the gadget outputs this state.
Otherwise, the gadget discards the prepared state and repeats the same level-$(l-1)$ stabilizer circuit to output the logical $\Ket{0}^{\otimes K^{(l)}}$ prepared in this second run without verification.
This repetition makes the gadget fault-tolerant while it at most doubles the depth of the gadget.

Assisted by the registers in logical states ${\Ket{0}}^{\otimes K^{(l)}}$ prepared by the level-$l$ initial-state preparation gadgets,
the level-$l$ error-correction gadget is implemented here in a fault-tolerant way by Knill's error correction~\cite{K5,K6} based on quantum teleportation~\cite{PhysRevLett.70.1895}.
The fault tolerance follows from transversality.
Unlike the quantum LDPC codes using an auxiliary physical qubit per extracting each syndrome bit, the weight of stabilizer generators does not matter for feasibility of error correction with the concatenated codes; in particular, using the above technique for the concatenated codes, we can prepare encoded codewords $\Ket{0}^{\otimes K^{(l)}}$ in a fault-tolerant way, and using this fault-tolerant state preparation to perform Knill's error correction, we can obtain all the syndrome bits simultaneously from the measurement outcomes for quantum teleportation, without using the auxiliary physical qubit per syndrome.

Note that we could also use Steane's error correction here~\cite{G}, but Knill's error correction may have merits in our protocol since Knill's error correction can be implemented in the same way as gate teleportation used for implementing the level-$l$ two-register Clifford gates.
An additional advantage of Knill's error correction over Steane's error correction is its ability to correct leakage errors~\cite{K6} while the error model in our analysis does not explicitly consider the leakage errors for simplicity.

The level-$l$ Clifford-state preparation gadget is implemented by non-fault-tolerant state preparation followed by verification, similar to the level-$l$ initial-state preparation gadget.
In particular, we first transform logical states ${\Ket{0}}^{\otimes 4 K^{(l)}}$ prepared by the level-$l$ initial-state preparation gadgets into logical $(\mathbbm{1}\otimes U_\mathrm{C})\ket{\Phi^{(l)}}$ by a level-$(l-1)$ stabilizer circuit in a non-fault-tolerant way~\cite{PhysRevA.56.76}.
Then, we perform verification with post-selection.
In the verification, we let the state be in the code space of $\Q_{r_l}$ using the level-$l$ error-correction gadgets.
Then, as $(\mathbbm{1}\otimes U_\mathrm{C})\ket{\Phi^{(l)}}$ is a stabilizer state, we measure the logical stabilizer operators for $(\mathbbm{1}\otimes U_\mathrm{C})\ket{\Phi^{(l)}}$ (i.e., multiqubit Pauli operators) using the controlled Pauli gates implemented by the gate teleportation.
To make the gadget fault-tolerant,
we design the gadget in such a way that an error on the auxiliary registers used as the control qubits for these controlled Pauli gates should not propagate to $(\mathbbm{1}\otimes U_\mathrm{C})\ket{\Phi^{(l)}}$ conditioned on the post-selection, using the technique of flag qubits~\cite{PhysRevLett.121.050502}.
Since we concatenate the distance-$3$ quantum Hamming codes, the verification can be made fault-tolerant with adding one flag qubit per extraction of logical stabilizer operators as in Ref.~\cite{PhysRevLett.121.050502}.
Note that flag-qubit techniques in Refs.~\cite{Chamberland2018flagfaulttolerant,PRXQuantum.1.010302} may also be used for potential generalization to concatenating higher-distance codes.
If no error is detected from measuring the logical stabilizer operators and the flag qubits, i.e., in the case of success in the verification, then the gadget outputs the logical $(\mathbbm{1}\otimes U_\mathrm{C})\ket{\Phi^{(l)}}$ prepared in this first run.
Otherwise, the gadget discards the prepared state and repeats the same level-$(l-1)$ stabilizer circuit to output the logical $(\mathbbm{1}\otimes U_\mathrm{C})\ket{\Phi^{(l)}}$ prepared in this second run without verification.
In the same way as the initial-state preparation gadget, the repetition at most doubles the depth of the gadget.

The level-$l$ magic-state preparation gadget is also implemented by non-fault-tolerant state preparation followed by verification.
First, we prepare states ${(R_y(\nicefrac{\pi}{4})\Ket{0})}^{\otimes 2K^{(l)}}$ and ${\Ket{0}}^{\otimes 2(N^{(l)}-K^{(l)})}$ by the level-$(l-1)$ magic- and initial-state preparation operations, respectively, and transform them into logical ${(R_y(\nicefrac{\pi}{4})\Ket{0})}^{\otimes K^{(l)}}\otimes{(R_y(\nicefrac{\pi}{4})\Ket{0})}^{\otimes K^{(l)}}$ by a level-$(l-1)$ stabilizer circuit for encoding, i.e., for transforming the magic states into the same logical states in a non-fault-tolerant way~\cite{PhysRevA.56.76}.
Then, we perform the verification with post-selection via ensuring the state in the code space of $\Q_{r_l}$ and measuring the logical stabilizer operators.
This magic state preparation does not use magic state distillation~\cite{PhysRevA.71.022316,https://doi.org/10.48550/arxiv.quant-ph/0402171}
but instead uses the verification to reduce errors.
In particular, since ${(R_y(\nicefrac{\pi}{4})\Ket{0})}$ is stabilized by $H$, i.e.,  $H{(R_y(\nicefrac{\pi}{4})\Ket{0})}={R_y(\nicefrac{\pi}{4})\Ket{0}}$,
we implement controlled $H$ gates for measuring the logical stabilizer operators for ${(R_y(\nicefrac{\pi}{4})\Ket{0})}^{\otimes K^{(l)}}\otimes{(R_y(\nicefrac{\pi}{4})\Ket{0})}^{\otimes K^{(l)}}$, using techniques similar to state-of-the-art low-overhead magic-state-preparation protocols in Refs.~\cite{yamasaki2020polylogoverhead,G6,Chamberland2019faulttolerantmagic}.
To make the gadget fault-tolerant, similar to the level-$l$ Clifford-state preparation gadget,
the level-$l$ magic-state preparation gadget is also designed in such a way that an error on the auxiliary registers used as the control qubits for the controlled $H$ gates should not propagate to ${(R_y(\nicefrac{\pi}{4})\Ket{0})}^{\otimes K^{(l)}}\otimes{(R_y(\nicefrac{\pi}{4})\Ket{0})}^{\otimes K^{(l)}}$ conditioned on the post-selection, using the technique of flag qubits~\cite{PhysRevLett.121.050502}.
If no error is detected from measuring the logical stabilizer operators and the flag qubits, i.e., in the case of success in the verification, then the gadget outputs the logical ${(R_y(\nicefrac{\pi}{4})\Ket{0})}^{\otimes K^{(l)}}\otimes{(R_y(\nicefrac{\pi}{4})\Ket{0})}^{\otimes K^{(l)}}$ prepared in this first run.
Otherwise, the gadget discards the prepared state and repeats the same level-$(l-1)$ circuit to output the logical ${(R_y(\nicefrac{\pi}{4})\Ket{0})}^{\otimes K^{(l)}}\otimes{(R_y(\nicefrac{\pi}{4})\Ket{0})}^{\otimes K^{(l)}}$ prepared in this second run without verification.
In the same way as the initial- and Clifford-state preparation gadgets, the repetition at most doubles the depth of the gadget.

With synthesis of stabilizer circuits~\cite{PhysRevA.70.052328,10.5555/2011763.2011767}, we show that all the level-$l$ gadgets here have at most $O(\poly(N_{r_l}))$ depths including the wait operations to wait for classical computation.
Consequently, each level-$l$ gadget has at most $O(\poly(N_{r_l}))$ locations, even if we take into account wait operations to wait for nonzero-time classical computation such as ones in the decoder and the gate teleportation.

\paragraph*{Analysis of threshold existence and improvement}
We sketch the proof of the existence of a threshold in our fault-tolerant protocol and discuss how to achieve a practical threshold with minor protocol modifications.
See Supplemental Information for further detail.

As in the conventional proof for the concatenated code,
the proof of the existence of a threshold in our protocol is given by bounding a logical error rate at a higher concatenation level by that at a lower level, based on counting the number of locations in extended rectangles (ExRecs)~\cite{G}.
(See also the figure illustrating ExRecs in Supplementary Information.)
Given a level-$l$ circuit for $l\in\{1,\ldots,L\}$, a level-$l$ ExRec refers to a part of the corresponding level-$(l-1)$ circuit that includes a level-$l$ gadget at each level-$l$ location and its adjacent level-$l$ error-correcting gadgets~\cite{G}.
For the distance-$3$ code such as the code used here, a level-$l$ ExRec is said to be good if the ExRec contains at most one faulty level-$(l-1)$ location and bad otherwise.
Intuitively, a good ExRec can implement the logical operation correctly, but a bad ExRec may not.
Thus, to bound the logical error rate, it suffices to evaluate the probability of having a bad ExRec using the number of locations therein.

In particular, let $A(l)$ be the maximum number of pairs of level-$(l-1)$ locations in a level-$l$ ExRec, where we take the maximum over all the possible level-$l$ ExRecs.
Since all the level-$l$ gadgets used in our protocol has $O(\poly(N_{r_l}))$ level-$(l-1)$ locations, we have $A(l)=O(\poly(N_{r_l}))=\exp({O(l)})$.
For simplicity of presentation, let $\alpha>0$ denote a constant factor such that
\begin{equation}
  \label{eq:A_l}
  A(l)\leqq 2^{\alpha l}
\end{equation}
for all $l\geqq 1$.
Crucially, our definition of a gadget being fault-tolerant is made analogous to the conventional definition in Ref.~\cite{G}, so that the same argument as Ref.~\cite{G} is applicable to our protocol.
When a level-$(l-1)$ circuit simulates a level-$l$ circuit, this argument leads to the fact that if the level-$(l-1)$ circuit undergoes the local statistic error model, the level-$l$ circuit also does.
Then, let $p_0$ be the physical error rate of level-$0$ locations, and $p_l$ denote the logical error rate of level-$l$ locations at each level $l$.
The conventional argument for the threshold theorem proves that the logical error rates are upper bounded by the probability of having two errors in an ExRec, i.e., $p_l\leqq A(l)p_{l-1}^2$ for each $l$~\cite{G}, which leads to $p_l\leqq 2^{\alpha l}p_{l-1}^2$.
Using this bound recursively, we can prove that the logical error rate $p_L$ is bounded by $p_L\leqq{(2^{2\alpha}p_0)}^{2^L}$, as shown in Supplemental Information.
This shows the existence of a threshold $p_\mathrm{th}\geqq 2^{-2\alpha}>0$ such that the logical error rate $p_L\leqq{(\nicefrac{p_0}{p_\mathrm{th}})}^{2^L}p_\mathrm{th}$ can be suppressed doubly exponentially in $L$ if the physical error rate satisfies $p_0<p_\mathrm{th}$.
Note that the same argument as ours for the threshold existence holds even in cases where the exponent $\alpha l$ of $2^{\alpha l}$ in~\eqref{eq:A_l} is replaced with $\poly(l)$; e.g., even if the gadgets had $O(\poly(N^{(l)}))$ depths due to architectural overhead or insufficient parallelization, a threshold would still exist.

Remarkably, a practical threshold is also achievable with minor modifications of our protocol.
Any quantum code with one logical qubit can be concatenated with $\Q^{(L)}$ by using the logical qubit of the code in place of each level-$0$ qubit of $\Q^{(L)}$, as long as the code can implement required operations for our fault-tolerant protocol at level $0$, namely, preparation of a qubit in $\Ket{0}$, a single-qubit measurement in the $Z$ basis, and the $H$, $S$, $\textsc{CNOT}$, $CZ$, Pauli, and $R_y(\pm\nicefrac{\pi}{4})$ gates.
For example, we can concatenate the surface code and $\Q^{(L)}$ and replace physical operations for our fault-tolerant protocol at concatenation level $0$ with the corresponding logical operations on the surface code.
Indeed, the surface code has well-established procedures for implementing logical operations for universal quantum computation~\cite{Horsman_2012,Litinski2019gameofsurfacecodes}; thus, we can use the logical qubit of a constant-size surface code in place of each physical qubit of $\Q^{(L)}$ in our protocol.
With this modification, we can achieve the same threshold as that of the surface code, and at the same time attain the constant overhead asymptotically.
See also Supplemental Information regarding further options of protocol modifications for a better threshold.

We also remark that the above lower bound $2^{-2\alpha}$ of the nonzero threshold value $p_\mathrm{th}$, derived here for the rigorous proof of its existence, is not necessarily close to $p_\mathrm{th}$; thus, it would be misleading to use $2^{-2\alpha}$ as an estimate of $p_\mathrm{th}$.
To estimate $p_\mathrm{th}$, one needs to perform more precise numerical simulation, which is essential for finding out which part of the protocol is a bottleneck to be modified for further improvement.
In addition to the threshold value itself, the achievable logical error rate at a finite concatenation level may also be of interst.
After all, what matters to FTQC in practice is the overall balance of the protocol, depending on specific settings of the error model and the architectural constraint.
We leave the numerical simulation of the protocols based on concatenating quantum Hamming codes for future work, but our theoretical contribution is fundamental for the research toward such a practical direction.

\paragraph*{Analysis of space and time overhead}
We sketch the analysis of space and time overheads of our fault-tolerant protocol.
See Supplemental Information for further detail.

To achieve the constant space overhead, our protocol uses the code $\Q^{(L)}$ with a non-vanishing rate of logical qubits per physical qubit.
However, it is still nontrivial to achieve the constant space overhead since the protocol may additionally use auxiliary level-$(l-1)$ registers in level-$l$ gadgets for implementability.
Crucially, we design each level-$l$ gadget to use only a constant number of auxiliary level-$(l-1)$ registers per encoded level-$l$ register, so as to keep the overall space overhead constant
\begin{equation}
  O(1)\quad\text{as $M\to\infty$},
\end{equation}
including physical qubits used for the auxiliary registers.

To save the time overhead, it is essential to realize gates acting on all the level-$L$ registers in parallel.
At the same time, it is also crucial to keep the code size for the sufficient error suppression as small as possible; after all, a smaller code size leads to a faster preparation of auxiliary states for gate teleportation and thus a smaller time overhead in implementing each gate acting on the level-$L$ registers.
As our threshold analysis shows, the suppression of the logical error rate $p_L=\exp(-O(2^L))$ in our protocol is exponentially faster than the growth of the code size $N^{(L)}=\exp(O(L^2))$ of $\mathcal{Q}^{(L)}$.
By choosing $L=\Theta(\log\log(\nicefrac{M}{\epsilon}))$,
we can reduce the overall error in simulating the original circuit to $\epsilon$.
With this choice, the size of each code block $\mathcal{Q}^{(L)}$ becomes only quasi-polylogarithmic
$N^{(L)}=\exp(O(L^2))=\exp(O(\log^2(\log(\nicefrac{M}{\epsilon}))))$.
On the other hand, each gadget in our protocol is designed to be implementable within at most polynomial time in the code size.
Therefore, this code size leads to the quasi-polylogarithmically small time overhead
\begin{equation}
  \exp(O(\log^2(\log(\nicefrac{M}{\epsilon}))))
\end{equation}
in implementing the gates and thus in simulating the original circuit.
Significantly, this time overhead includes that for preparing auxiliary states for gate teleportation and error correction, and also that for waiting for the nonzero-time classical computation during the protocol, such as ones required for the decoder and the gate teleportation.

\section*{Data availability}

Data sharing is not applicable since no datasets were generated or analyzed during this study.

\section*{Code availability}

The codes used in this study are available from the corresponding author upon reasonable request.

\begin{acknowledgments}
  H.Y.\ acknowledges Keisuke Fujii and Yasunari Suzuki for comments in the meeting of JST PRESTO\@.
  This work was supported by JSPS Overseas Research Fellowships, JST PRESTO Grant Number JPMJPR201A, and JST [Moonshot R\&D][Grant Number JPMJMS2061].
\end{acknowledgments}

\section*{Author contributions}

Both authors contributed to the conception of the work, the analysis and interpretation in the work, and the preparation and revision of the manuscript.

\section*{Competing interests}

The authors declare no competing interests.

\section*{Additional information}

Supplementary Information is available for this paper.
Correspondence and requests for materials should be addressed to Hayata Yamasaki.

\newpage
\appendix
\counterwithin{figure}{section}

\section*{Supplemental Information: Time-Efficient Constant-Space-Overhead Fault-Tolerant Quantum Computation}

Supplemental Information on ``Time-Efficient Constant-Space-Overhead Fault-Tolerant Quantum Computation'' is organized as follows.
In Sec.~\ref{sec:setting}, we present the setting of fault-tolerant quantum computation (FTQC).
In Sec.~\ref{sec:hamming}, the definition of quantum codes used in our fault-tolerant protocol and the proof of the non-vanishing rate of our quantum code are presented in detail.
In Sec.~\ref{sec:gadget}, we explain how our fault-tolerant protocol compiles the original circuit into the fault-tolerant circuit using gadgets.
In Sec.~\ref{sec:fault_tolerant_gadget}, we show the construction of the gadgets.
In Sec.~\ref{sec:threshold_proof}, we prove the existence of a threshold in our protocol.
In Sec.~\ref{sec:overhead}, we analyze the space and time overheads of our protocol.

\section{\label{sec:setting}Setting}

We present the setting of fault-tolerant quantum computation (FTQC) in our work.
Quantum computation aims to solve a family of computational problems using a quantum circuit.
We call this circuit the original circuit~\cite{G}.
A classical input of the computational problems is represented by an $M$-bit string
\begin{equation}
  \label{eq:input}
  (i_1,\ldots,i_M)\in{\{0,1\}}^M.
\end{equation}

The original circuit is composed of the following operations: preparations, gates, measurements, and waits performing the identity gate $\mathbbm{1}$.
Each operation in a circuit is called a location.
The width and the depth of the original circuit are denoted by $W(M)$ and $D(M)$, respectively.
The original circuit is assumed to have a polynomial size in $M$;
that is, $W(M)$ and $D(M)$ are bounded by
\begin{align}
  \label{eq:W}
  W(M)&=O(\poly(M)),\,W(M)\geqq M,\\
  D(M)&=O(\poly(M)),\,D(M)\geqq 0.
\end{align}
The original circuit starts with the preparation of all the $W(M)$ qubits in $\Ket{0}^{\otimes W(M)}$.
Then, we input the $M$-bit string $(i_1,\ldots,i_M)$ in~\eqref{eq:input} to the original circuit, using an $M$-qubit input state prepared at the initial part of the original circuit as
\begin{equation}
  \label{eq:input_part}
  \bigotimes_{m=1}^{M}\Ket{i_m}=\left(\bigotimes_{m=1}^{M}X^{i_m}\right)\Ket{0}^{\otimes M},
\end{equation}
where $X^0=\mathbbm{1}$.
The part of the original circuit for preparing~\eqref{eq:input_part} is called the \textit{input part}.
The original circuit holds $(\bigotimes_{m=1}^{M}\Ket{i_m})\otimes\Ket{0}^{\otimes (W(M)-M)}$ at the end of the input part.
After the input part, we apply gates to the qubits.
We write the original circuit in terms of a gate set of Clifford gates $X$, $Y$, $Z$, $H$, $S$, $\textsc{CNOT}$, and $CZ$, and non-Clifford gates $R_y(\pm\nicefrac{\pi}{4})$~\cite{N4}.
The gates in the original circuit can be performed on all the qubits in parallel, within the constraint that at most one gate can be performed on each qubit at each time step; that is, the depth of the original circuit refers to the number of time steps under this constraint.
Two-qubit gates in the original circuit may be performed on arbitrary pairs of the qubits.
Finally, we perform measurements of all the $W(M)$ qubits in the $Z$ basis, which achieves sampling of an $W(M)$-bit string that is returned as a result of quantum computation.
In our setting, the original circuit does not include measurements in the middle of computation and thus does not have feed-forward operations conditioned on measurement outcomes; that is, the width of the original circuit (i.e., the number of qubits in the original circuit) remains constant for each time step.
Except for the input part, the original circuit is fixed, i.e., is independent of $(i_1,\ldots,i_M)$ and depends only on $M$.
Note that in this setting, we can perform computation controlled by $(i_1,\ldots,i_M)$ with controlled gates, using the initial state~\eqref{eq:input_part} for the state of the control qubits.
We will show an example of original circuits in Fig.~\ref{fig:circuit_compilation} of Sec.~\ref{sec:gadget}.

The goal of FTQC is to simulate the original circuit using a circuit on physical qubits that may suffer from errors.
Prior to starting the execution of quantum computation, we have the $O(\poly(M))$-size classical description of the original circuit.
For FTQC, we compile the original circuit into a circuit on physical qubits, which we call a \textit{fault-tolerant circuit}.
The fault-tolerant circuit is composed of the following operations: preparation of a physical qubit in $\Ket{0}$, Clifford gates $X$, $Y$, $Z$, $H$, $S$, $\textsc{CNOT}$, and $CZ$, non-Clifford gates $R_y(\pm\nicefrac{\pi}{4})$, measurement of each physical qubit in the $Z$ basis, and wait, i.e., performing the identity gate $\mathbbm{1}$.
Our model is as follows.
\begin{enumerate}
  \item \textbf{Local stochastic error model}: We use a conventional but sufficiently general error model, a \textit{local stochastic error model}~\cite{G,gottesman2014faulttolerant}. We say that a circuit undergoes a local statistical error model if the faults occurring in the circuit satisfy the following~\cite{G}: (i) a set $S$ of faulty locations in the circuit is chosen randomly with probability $p(S)$, and errors occur at the locations in $S$ in such a way that the operations at the locations in $S$ are replaced with an arbitrary quantum channel $\mathcal{E}$ that is consistent with the causal order of the locations in $S$ (i.e., $\mathcal{E}$ may be adversarial and correlated); (ii) each location $i$ in the circuit has a parameter $p_{0,i}$ such that for any set $R$ of locations, the probability $\Pr\{S\supseteqq R\}$ of having faults at every location in $R$ (i.e., the probability of the randomly chosen set $S$ including $R$) is at most
  \begin{equation}
    \label{eq:local_stochastic_error_model_physical}
    \prod_{i\in R} p_{0,i}.
  \end{equation}
  We assume that the fault-tolerant circuit undergoes the local stochastic error model.
  \item \textbf{Parallel quantum operation}: The operations on physical qubits can be performed in parallel, where each qubit is involved in at most one operation at a time.
  \item \textbf{Faultless classical computation}: Conditioned on the outcome of measurements on physical qubits, we can perform classical computation to change the subsequent operations on the qubits adaptively. For simplicity of analysis of overhead, we assume that classical computation is faultless.
    Note that our threshold analysis in Sec.~\ref{sec:threshold_proof} will prove that our fault-tolerant protocol has a threshold even with a polynomially large architectural time overhead, and thus even if classical computation is performed by faulty circuits with the overhead of using a classical error-correcting code~\cite{185425,10.1007/11672142_44}, a threshold still exists.
  \item \textbf{Parallel classical computation but nonzero runtime}: Similar to the operations on physical qubits, classical computation is assumed to be performed in parallel so that the depth of classical circuits determines the runtime of classical computation.
    The number of parallel processes of classical computation is limited to the same order as the number of physical qubits up to a quasi-polylogarithmic factor, as given later by~\eqref{eq:parallel_process_number}.
    In the analysis of FTQC, it has been conventional to ignore the runtime of classical computation, i.e., to assume that the classical computation runs instantaneously in zero time that never grows on large scales~\cite{G,gottesman2014faulttolerant}.
    However, in practice, the required runtime of classical computation in FTQC such as that for the decoder may become longer on larger scales. In view of this practical requirement, we do not ignore the runtime of classical computation;
    that is, we take into account the runtime of classical computation in our analysis by performing wait operations on physical qubits during the runtime of classical computation.
    In our setting, errors may also occur on these wait operations according to the local stochastic error model.
  \item \textbf{Allocation of qubits and bits}: We allocate qubits by the preparation and deallocate the qubits by the measurement. The measurement allocates bits for classical computation to process the outcome of the measurement. After using the measurement outcomes, we discard the outcomes and deallocate the bits. The number of qubits and bits at a time is that which has already been allocated before and is not yet deallocated at the time.
  \item \textbf{No geometrical constraint}: Following the convention of the previous works~\cite{gottesman2014faulttolerant,PhysRevA.87.020304,8555154}, we assume that two-qubit gates are applicable to any pair of physical qubits without geometrical constraint as in photonic systems~\cite{yamasaki2020polylogoverhead,PhysRevResearch.2.023270,PhysRevLett.123.200502,PhysRevLett.112.120504,doi:10.1063/1.5100160} and distributed architectures~\cite{Y5,Wehnereaam9288,7562346}.
    Classical computation can also be performed without such geometrical constraints. This assumption is essential for avoiding polynomially large time overhead in simulating the original circuit; for example, if one could use only $2$D local interactions on $m$ qubits aligned on an $O(\sqrt{m})\times O(\sqrt{m})$ square lattice, a realization of each long-range two-qubit gate would require $O(\sqrt{m})$ operations to mediate, which we can avoid with our assumption.
    Note that our threshold analysis in Sec.~\ref{sec:threshold_proof} will prove that our fault-tolerant protocol still has a threshold even with such polynomial time overhead arising from the $2$D nearest-neighbor interactions; in such a case, we can incorporate the techniques in Refs.~\cite{10.1137/S0097539799359385,doi:10.1080/09500340008244046,PhysRevA.72.022317,DBLP:journals/qic/SvoreDT07} for fault-tolerant implementation of the gates within the geometrical constraints.
\end{enumerate}

A fault-tolerant protocol aims to simulate the $W(M)$-qubit $D(M)$-depth original circuit within any given target error $\epsilon>0$, i.e., to output a $W(M)$-bit string sampled from a probability distribution close to the output probability distribution of the original circuit within error in the total variation distance at most $\epsilon$.
To achieve this simulation, the fault-tolerant protocol replaces the qubits in the original circuit with logical qubits of a quantum error-correcting code, to obtain the fault-tolerant circuit using multiple physical qubits per logical qubit.
Let $W_\mathrm{FT}(M)$ denote the maximum of the total number of physical qubits allocated at a time, where the maximum is taken over all the time steps in the fault-tolerant circuit.
The space overhead refers to $W_\mathrm{FT}(M)$ divided by the number $W(M)$ of qubits in the original circuit, i.e.,
\begin{equation}
  \label{eq:space_overhead}
  \frac{W_\mathrm{FT}(M)}{W(M)}.
\end{equation}
Apart from the physical qubits, for classical computation, we allow the fault-tolerant protocol to use a classical memory with a size of
\begin{equation}
\label{eq:classical_memory}
  O(\poly(M))~\text{bits};
\end{equation}
after all, just to store the classical description of the original circuit, we may need to use $O(\poly(M))$ bits.
On the other hand, we limit the number of parallel processes of classical computation to be of the same order as the number $W(M)$ of qubits up to a quasi-polylogarithmic factor, i.e.,
\begin{equation}
  \label{eq:parallel_process_number}
  W(M)\times\exp(O(\polylog(\log(M)))).
\end{equation}
The time overhead refers to the depth $D_\mathrm{FT}(M)$ of the fault-tolerant circuit (including wait operations to wait for classical computation) divided by $D(M)$ of the original circuit, i.e.,
\begin{equation}
  \label{eq:time_overhead}
  \frac{D_\mathrm{FT}(M)}{D(M)}.
\end{equation}
Note that, due to the limitation~\eqref{eq:classical_memory} of the number of bits, it is prohibited to use an exponentially large number of parallel processes of classical computation to classically simulate the original circuit in a polynomial time in terms of the circuit depth.
For fixed $\epsilon$, the fault-tolerant protocol is said to achieve a constant space overhead if the space overhead is $O(1)$ as $M\to\infty$.

The classical process for converting the original circuit into a fault-tolerant circuit is called compilation.
We perform the compilation prior to starting the execution of quantum computation, i.e., prior to knowing the input $(i_1,\ldots,i_M)$ in~\eqref{eq:input} and allocating physical qubits.
By convention, we do not include the runtime of compilation in the time overhead.
Note that this setting may potentially justify solving an NP-hard problem of circuit optimization heuristically during compilation, but our fault-tolerant protocol will not require any process to solve such an NP-hard problem in our compilation; in particular, we will clarify in Secs.~\ref{sec:gadget} and~\ref{sec:fault_tolerant_gadget} that the compilation can be performed feasibly using techniques for converting the stabilizer circuits.
Also, due to the limitation~\eqref{eq:classical_memory} of the number of bits, it is prohibited to store the result of classical simulation of the original circuit for all possible inputs~\eqref{eq:input} by performing an exponentially long-time classical computation during the compilation.

\section{\label{sec:hamming}Concatenated code constructed from a sequence of quantum Hamming codes}

In this section, we summarize the notations on quantum codes and define the quantum error-correcting code used in our fault-tolerant protocol.
For details of quantum codes, see also Refs.~\cite{G,P,D,T10,lidar_brun_2013,N4}.

A quantum code $\Q$ on $N$ physical qubits is a subspace of ${(\mathbb{C}^{2})}^{\otimes N}$, where $\mathbb{C}^2=\spn\{\Ket{0},\Ket{1}\}$ represents a qubit.
A stabilizer code $\Q(\mathcal{S})$ is a quantum code represented by a stabilizer $\mathcal{S}\subset \mathcal{P}_N$, i.e., an Abelian subgroup of the Pauli group $\mathcal{P}_N$ on $N$ qubits generated by $\{\mathrm{i}\mathbbm{1},X_1,Z_1,\ldots,X_N,Z_N\}$ with $-\mathbbm{1}\not\in \mathcal{S}$, where $X_n$ and $Z_n$ for each $n\in\{1,\ldots,N\}$ are Pauli-$X$ and $Z$ operators, respectively, acting on the $n$th qubit.
A centralizer $C(\mathcal{S})$ of $\mathcal{S}$ is the set of Pauli operators in $\mathcal{P}_N$ that commute with all the elements in $\mathcal{S}$.
Suppose that $\mathcal{S}$ is generated by $N-K$ independent generators.
The code space $\Q(\mathcal{S})$ is the $2^K$-dimensional subspace of $N$ qubits spanned by all $N$-qubit states invariant under the action of $\mathcal{S}$, which defines $K$ logical qubits.
Logical operators are elements in $C(\mathcal{S})/\mathcal{S}$, which is isomorphic to the Pauli group $\mathcal{P}_K$  on $K$ qubits~\cite{G}.
In particular, for each $k\in\{1,\ldots,K\}$, the logical $X$ and $Z$ operators of the $k$th logical qubit of $\Q(\mathcal{S})$ are defined as the elements in $C(\mathcal{S})/\mathcal{S}$ that are identified with $X_k$ and $Z_k$, respectively, in $\mathcal{P}_K$ acting on the $k$th qubit, under this isomorphism.
The rate of $\Q(\mathcal{S})$ is $\nicefrac{K}{N}$.
The distance $D$ of $\Q(\mathcal{S})$ is the minimal weight of nontrivial logical operators, where the weight of an $N$-qubit operator is the number of qubits on which the operator acts nontrivially (rather than $\mathbbm{1}$).
A stabilizer code with $N$ physical qubits, $K$ logical qubits, and distance $D$ is written as an $[[N,K,D]]$ code in double square brackets.
We may call $N$ the size of the code block.

\begin{table*}[tpb]
  \centering
  \caption{\label{table:hamming}The parity-check matrix $H_{r}$ of the Hamming codes $\C_r$ with $r\in\{2,3,\ldots\}$. The $i$th row of $H_{r}$ is denoted by $(b_{i,1},b_{i,2},\ldots,b_{i,N_r})$ and given in the table, where $i\in\{1,\ldots,r\}$ and $N_r=2^r-1$.}
  \begin{tabular}{@{}c|cccccccccccccccccccc@{}}
    \toprule
    $i$ & $b_{i,1}$ & $b_{i,2}$ & \multicolumn{1}{c|}{$b_{i,3}$} & $b_{i,4}$ & $b_{i,5}$ & $b_{i,6}$ & \multicolumn{1}{c|}{$b_{i,7}$} & $b_{i,8}$ & $b_{i,9}$ & $b_{i,10}$ & $b_{i,11}$ & $b_{i,12}$ & $b_{i,13}$ & $b_{i,14}$ & \multicolumn{1}{c|}{$b_{i,15}$} & $b_{i,16}$ & $b_{i,17}$ & $b_{i,18}$ & $b_{i,19}$ &  $\cdots$ \\
\midrule
    $1$ & $1$ & $0$ & \multicolumn{1}{c|}{$1$} & $0$ & $1$ & $0$ & \multicolumn{1}{c|}{$1$} & $0$ & $1$ & $0$ & $1$ & $0$ & $1$ & $0$ & \multicolumn{1}{c|}{$1$} & $0$ & $1$ & $0$ & $1$ \\
    $2$ & $0$ & $1$ & \multicolumn{1}{c|}{$1$} & $0$ & $0$ & $1$ & \multicolumn{1}{c|}{$1$} & $0$ & $0$ & $1$ & $1$ & $0$ & $0$ & $1$ & \multicolumn{1}{c|}{$1$} & $0$ & $0$ & $1$ & $1$ \\
    \cmidrule{1-4}
    $3$ & $0$ & $0$ & $0$ & $1$ & $1$ & $1$ & \multicolumn{1}{c|}{$1$} & $0$ & $0$ & $0$ & $0$ & $1$ & $1$ & $1$ & \multicolumn{1}{c|}{$1$} & $0$ & $0$ & $0$ & $0$ \\
    \cmidrule{1-8}
    $4$ & $0$ & $0$ & $0$ & $0$ & $0$ & $0$ & $0$ & $1$ & $1$ & $1$ & $1$ & $1$ & $1$ & $1$ & \multicolumn{1}{c|}{$1$} & $0$ & $0$ & $0$ & $0$ \\
    \cmidrule{1-16}
    $5$ & $0$ & $0$ & $0$ & $0$ & $0$ & $0$ & $0$ & $0$ & $0$ & $0$ & $0$ & $0$ & $0$ & $0$ & $0$ & $1$ & $1$ & $1$ & $1$ \\
    \cmidrule{1-20}
    $\vdots$\\
    \bottomrule
  \end{tabular}
\end{table*}

A Calderbank-Shor-Steane (CSS) code provides a way of constructing a class of stabilizer codes from two classical linear codes.
A classical linear code $\C$ of block length $N$ is a linear subspace of $\mathbb{F}_2^{N}$, where $\mathbb{F}_2=\{0,1\}$ is the finite field of order $2$ representing a bit~\cite{10.5555/552386}.
The linear code $\C$ is characterized by an $(N-K)\times N$ parity-check matrix $H$ such that the inner product of any codeword $v\in\C$ and any row of $H$ is zero, i.e., $\C=\{v\in\mathbb{F}_2^{N}:Hv^T=0\}$, where the codeword $v$ of $\C$ is represented as a row vector, and thus the transpose $v^T$ of $v$ is a column vector.
Each row of $H$ is used as a parity check to detect and correct errors.
The dimension of $\C$ is $K\coloneqq\log_2(\dim \C)$ in bits.
The rate of $\C$ is $\nicefrac{K}{N}$.
The distance $D$ of $\C$ is the minimal Hamming distance between any distinct codewords $u,v\in \C$,
where the Hamming distance of $u$ and $v$ is the weight, i.e., the number of $1$s in the $N$ elements, of $u+v$.
A classical linear code of block length $N$, dimension $K$, and distance $D$ is written as an $[N,K,D]$ code in single square brackets.
To obtain a CSS code, an $[N,K_Z,D_Z]$ code $\C_Z$ and an $[N,K_X,D_X]$ code $\C_X$ of the same block length $N$ satisfying $\C_X^\perp\subseteqq \C_Z$ are used, where $\C_X^\perp$ is the dual code of $\C_X$ spanned by $(N-K_X)$ rows of the parity-check matrix of $\C_X$, i.e., $\C_X^\perp=\{u\in\mathbb{F}_2^{N}:\forall v\in \C_X,\Braket{u|v}=0\}$.
Let $H_Z$ and $H_X$ denote the parity-check matrices of $\C_Z$ and $\C_X$, respectively; then, the condition $\C_X^\perp\subseteqq \C_Z$ is equivalent to $H_Z H_X^T=0$, where $H_X^T$ is the transpose of $H_X$.
The CSS code of $\C_Z$ over $\C_X^\perp$~\cite{N4} is a quantum code denoted by $\css(\C_Z,\C_X)\coloneqq\spn\{\sum_{w\in \C_X^\perp}\Ket{u+w}:u\in \C_Z\}$.
The code $\css(\C_Z,\C_X)$ is a class of stabilizer codes with stabilizer generators given only in $X$s or only in $Z$s.
In particular, for any row vector $(b_1,\ldots,b_N)\in\mathbb{F}_2^N$ given by a row of $H_Z$, the stabilizer of $\css(\C_Z,\C_X)$ has a $Z$ generator $\bigotimes_{n=1}^{N}Z_n^{b_n}$, where $Z^{1}=Z$ and $Z^0=\mathbbm{1}$.
Also, for any row $(b_1,\ldots,b_N)\in\mathbb{F}_2^N$ of $H_X$, the stabilizer of $\css(\C_1,\C_2)$ has an $X$ generator $\bigotimes_{n=1}^{N}X_n^{b_n}$, where $X^{1}=X$ and $X^0=\mathbbm{1}$.
As a whole, $\css(\C_Z,\C_X)$ has $(N-K_Z)$ stabilizer generators in $Z$ and $(N-K_X)$ stabilizer generators in $X$.
The condition $\C_X^\perp\subseteqq \C_Z$, i.e., $H_Z H_X^T=0$, requires that each pair of $Z$ and $X$ generators should commute with each other, so that the stabilizer of $\css(\C_Z,\C_X)$ should be Abelian.
The CSS code $\css(\C_Z,\C_X)$ is an $[[N,K_Z+K_X-N,D]]$ stabilizer code, where the distance is lower bounded by $D\geqq\min\{D_Z,D_X\}$.

Hamming codes $(\C_r:r=2,3,\ldots)$ are a family of linear classical codes such that the parity-check matrix
\begin{equation}
  \label{eq:parity_check_matrix}
  H_{r}
\end{equation}
of $\C_r$ is an $r\times N_r$ matrix where $N_r\coloneqq 2^r-1$, and the $i$th row for each $i\in\{1,\ldots,r\}$ has $1$s between $(2^{i-1}+2^i j)$th--$(2^{i}-1+2^i j)$th columns for all $j\in\{0,1,\ldots\}$ and $0$s otherwise~\cite{6772729}.
For example, Table~\ref{table:hamming} gives the $i$th row $(b_{i,1},b_{i,2},\ldots,b_{i,N_r})$ of $H_{r}$.
The code $\C_r$ has block length $N_r$, dimension $N_r-r$, and distance $3$; that is, $\C_r$ is an $[N_r,N_r-r,3]$ code.

Quantum Hamming codes $(\Q_r:r=3,4,\ldots)$ are defined as CSS codes of $\C_r$ over $\C_r^\perp$~\cite{PhysRevA.54.4741}.
Since $\C_r$ is weakly self-dual, i.e., $H_{r}H_{r}^T=0$, the condition for CSS codes is satisfied.
For each $i\in\{1,\ldots,r\}$ to specify the $i$th row $(b_{i,1},\ldots,b_{i,N_r})\in{\{0,1\}}^{N_r}$ of the parity-check matrix $H_r$ of $\C_r$, the stabilizer of $\Q_r$ has a $Z$ generator and an $X$ generator given respectively by
\begin{align}
  \label{eq:Z_generator}
  \bigotimes_{n=1}^{N_r}Z_n^{b_{i,n}},\\
  \label{eq:X_generator}
  \bigotimes_{n=1}^{N_r}X_n^{b_{i,n}}.
\end{align}
As a whole, the stabilizer of $\Q_r$ has $r$ $Z$ generators and $r$ $X$ generators.
A code block of $\Q_r$ has
\begin{align}
  \label{eq:N_r}
  N_r&=2^r-1~\text{physical qubits},\\
  \label{eq:K_r}
  K_r&\coloneqq N_r-2r~\text{logical qubits},
\end{align}
and distance $3$; that is, $\Q_r$ is an $[[N_r,K_r,3]]$ code.
For example, $\Q_3$ is a $[[7,1,3]]$ code also known as Steane's $7$-qubit code~\cite{PhysRevLett.77.793,S3}, $\Q_4$ is $[[15,7,3]]$, $\Q_5$ is $[[31,21,3]]$, $\Q_6$ is $[[63,51,3]]$, $\Q_7$ is $[[127,113,3]]$, $\Q_8$ is $[[255,239,3]]$, and $\Q_9$ is $[[511,493,3]]$.
The rate of $\Q_r$ converges to
\begin{equation}
  \label{eq:rate_Q_r_l}
  \lim_{r\to\infty}\frac{K_r}{N_r}=1,
\end{equation}
which is necessary for constructing a family of concatenated codes with a non-vanishing rate in the limit of large concatenation levels.
For each $k\in\{1,\ldots,K_r\}$,
the $k$th logical qubit of $\Q_r$ has the logical $Z$ and $X$ operators given respectively by
\begin{align}
  \label{eq:kth_logical}
  &\bigotimes_{n=1}^{N_r}Z_n^{b_n^{(k)}},\nonumber\\
  &\bigotimes_{n=1}^{N_r}X_n^{b_n^{(k)}},
\end{align}
in terms of the same bit string $(b_1^{(k)},\ldots,b_{N_r}^{(k)})\in{\{0,1\}}^{N_r}$ representing the $k$th logical qubit, which can be calculated by techniques in Refs.~\cite{Gottesman1997,PhysRevA.79.062322}.
Although the choice of this bit string $(b_1^{(k)},\ldots,b_{N_r}^{(k)})$ for each $r$ and $k$ is not unique, it is assumed here and henceforth that a fixed choice of the bit string is adopted.
We let
\begin{equation}
  \label{eq:generator_matrix}
  G_r
\end{equation}
denote a $K_r\times N_r$ matrix where the $k$th row for $k\in\{1,\ldots,K_r\}$ is $(b_1^{(k)},\ldots,b_{N_r}^{(k)})$.
Note that we have $G_r H_r^T=0$ as with a generator matrix $G$ and a parity-check matrix $H$ of a classical linear code satisfying $GH^T=0$.

Note that, for a family of CSS codes, each code in the family is called a quantum low-density parity-check (LDPC) code if the parity-check matrices $H_Z$ and $H_X$ of $\C_Z$ and $\C_X$ for the CSS codes in the family have $O(1)$ nonzero elements in each column and row as the code size increases; that is, the $Z$ and $X$ stabilizer generators of the quantum LDPC codes have only $O(1)$ weights, and only a constant number $O(1)$ of stabilizer generators act nontrivially on each physical qubit of the quantum LDPC codes.
By definition, quantum Hamming codes are not quantum LDPC codes.

The following observations are essential for implementing logical gates~\cite{G}.
\begin{itemize}
  \item  Since $\Q_r$ is a CSS code obtained from a weakly self-dual linear code $\C_r$, transversal \textsc{CNOT} and $CZ$ gates between all the physical qubits between two code blocks of $\Q_r$ implement logical \textsc{CNOT} and $CZ$ gates between all the logical qubits between the code blocks. Also, transversal $H$ gates acting on all the $N_{r}$ physical qubits of $\Q_r$, i.e., $H^{\otimes N_{r}}$, implement a logical operator $H^{\otimes K_{r}}$ on all the $K_{r}$ logical qubits of $\Q_r$.
  \item Since $\Q_r$ is a stabilizer code, any logical Pauli operation in the Pauli group $\mathcal{P}_{K_r}$ corresponds to a physical Pauli operation in the Pauli group $\mathcal{P}_{N_r}$, and is thus implemented by applying a suitable Pauli gate ($X$, $Z$, $Y$, or $\mathbbm{1}$) on each of the $N_r$ physical qubits.
\end{itemize}

For example, consider the $[[7,1,3]]$ code $\Q_3$.
The stabilizer is generated by
\begin{align}
  &Z\otimes\mathbbm{1}\otimes Z\otimes\mathbbm{1}\otimes Z\otimes\mathbbm{1}\otimes Z,\\
  &X\otimes\mathbbm{1}\otimes X\otimes\mathbbm{1}\otimes X\otimes\mathbbm{1}\otimes X,\\
  &\mathbbm{1}\otimes Z\otimes Z\otimes \mathbbm{1}\otimes\mathbbm{1} \otimes Z\otimes Z,\\
  &\mathbbm{1}\otimes X\otimes X\otimes \mathbbm{1}\otimes\mathbbm{1} \otimes X\otimes X,\\
  &\mathbbm{1}\otimes \mathbbm{1}\otimes\mathbbm{1} \otimes Z\otimes Z \otimes Z\otimes Z,\\
  &\mathbbm{1}\otimes \mathbbm{1}\otimes\mathbbm{1} \otimes X\otimes X \otimes X\otimes X.
\end{align}
The logical $Z$ and $X$ operators on the logical qubit are, respectively,
\begin{align}
  &Z\otimes Z\otimes Z\otimes \mathbbm{1}\otimes \mathbbm{1}\otimes \mathbbm{1}\otimes \mathbbm{1},\\
  &X\otimes X\otimes X\otimes \mathbbm{1}\otimes \mathbbm{1}\otimes \mathbbm{1}\otimes \mathbbm{1}.
\end{align}
The logical $H$ on the logical qubit is given by
\begin{align}
  H\otimes H\otimes H\otimes H\otimes H\otimes H\otimes H.
\end{align}

Here the construction of the concatenated code $\Q^{(L)}$ is described in full detail.
For concatenation, we use a sequence of quantum Hamming codes $(\Q_{r_l}:1,2,\ldots, L)$, where we choose the parameter $r_l$ as
\begin{equation}
  \label{seq:r_l}
  r_l\coloneqq l+2.
\end{equation}
Note that, due to~\eqref{eq:N_r} and~\eqref{eq:K_r}, this choice of $r_l$ leads to
\begin{align}
\label{eq:N_r_scaling}
N_{r_l}&=\exp(O(l)),\\
\label{eq:K_r_scaling}
K_{r_l}&=\exp(O(l)),
\end{align}
as $l\to\infty$.
For $l\in\{L,L-1,\ldots,1,0\}$, the construction here introduces a collection of $K^{(l)}$ qubits called a \textit{level-$l$ register},
where $K^{(l)}$ is explicitly given by
\begin{align}
  \label{eq:K_(l)}
  K^{(l)}\coloneqq\prod_{l^\prime=1}^{l}K_{r_{l^\prime}}=\prod_{l^\prime=1}^{l}(2^{r_{l^\prime}}-2r_{l^\prime}-1)
\end{align}
and $K^{(0)}\coloneqq 1$.
Every qubit in a level-$l$ register is distinguished by a label
\begin{equation}
  k^{(l)} \in \{1 ,\ldots, K^{(l)}\}.
\end{equation}

The code $\Q^{(L)}$ is constructed recursively for $l=L, L-1, ,\ldots, 1$ by encoding the $K^{(l)}=K_{r_l}\times K^{(l-1)}$ qubits in a level-$l$ register into $N_{r_l}\times K^{(l-1)}$ qubits in a set of $N_{r_l}$ level-$(l-1)$ registers, using $K^{(l-1)}$ blocks of the quantum Hamming code $\Q_{r_l}$.
The level-$L$ register represents the set of logical qubits of $\Q^{(L)}$,
and a level-$0$ register (or a level-$0$ qubit) refers to a physical qubit of $\Q^{(L)}$.
Encoding into lower levels of qubits involves all the qubits in a register, and consequently, the qubits in the register may suffer from a common fault at the same time.
In contrast, qubits belonging to distinct registers should have independent statistics of their errors if physical errors at level $0$ occur independently.

To specify the assignment of each qubit in a level-$l$ register to a code block according to its index $k^{(l)}$, define a map $ k^{(l)} \mapsto (k,k^{(l-1)})$ by the relation
\begin{equation}
  \label{eq:k_Q_l}
  k^{(l)}= (k-1) K^{(l-1)} +  k^{(l-1)},
\end{equation}
where $k^{(l-1)} \in \{1,\ldots, K^{(l-1)} \}$ and $k\in \{1,\ldots, K_{r_l}\}$.
The $k^{(l)}$th qubit in the level-$l$ register is then assigned as the $k$th logical qubit of the $k^{(l-1)}$th block of $\Q_{r_l}$.
Each code block encodes $K_{r_l}$ logical qubits into $N_{r_l}$.
Hence,
the $K^{(l-1)}$ code blocks have $K_{r_l}\times K^{(l-1)}=K^{(l)}$ logical qubits in total to cover all the qubits in a level-$l$ register,
which are encoded into $N_{r_l}\times K^{(l-1)}$ qubits in the set of $N_{r_l}$ level-$(l-1)$ registers as a whole.
These qubits in the set of $N_{r_l}$ level-$(l-1)$ registers are naturally indexed by
\begin{equation}
  \label{eq:n_Q_r_l_k_Q_l_1}
  (n,k^{(l-1)})
\end{equation}
with $n\in\{1,\ldots,N_{r_l}\}$,
indicating the $n$th physical qubit of the $k^{(l-1)}$th block of $\Q_{r_l}$.
Note that the indexes of the logical and physical qubits of $\Q_{r_l}$ used here should follow~\eqref{eq:kth_logical} with a fixed choice of the bit strings.

Finally, these $N_{r_l} K^{(l-1)}$ qubits are sorted into $N_{r_l}$ level-$(l-1)$ registers.
This is simply done by assigning the qubit indexed by $(n,k^{(l-1)})$ to the $k^{(l-1)}$th qubit in the $n$th level-$(l-1)$ register.
Now the whole procedure of encoding a level-$l$ register into $N_{r_l}$ level-$(l-1)$ registers is specified, which can be recursively applied to define the concatenated code $\Q^{(L)}$ that encodes $K^{(L)}$ qubits in a level-$L$ register into level-$0$ physical qubits.
In the above rule of assignment, every physical qubit of a code $\Q_{r_l}$ is assigned to a distinct level-$(l-1)$ register. This also means that every qubit in a level-$(l-1)$ register is  assigned to a distinct block of $\Q_{r_l}$.
As mentioned above, qubits in the same register may suffer from errors simultaneously. The construction here is thus intended to work analogously to interleaving techniques widely used for burst error correction.

The number $N^{(L)}$ of physical qubits of $\Q^{(L)}$ is computed from the fact that a level-$l$ register is encoded into $N_{r_l}$ level-$(l-1)$ registers, and that a level-$0$ register is a single physical qubit by definition; thus, it holds that
\begin{equation}
  \label{eq:n_L}
  N^{(L)}=\prod_{l=1}^{L}N_{r_l}=\prod_{l=1}^{L}(2^{r_l}-1).
\end{equation}
The number of logical qubits of $\Q^{(L)}$ is $K^{(L)}$ by definition, i.e.,
\begin{equation}
  \label{eq:k_L}
  K^{(L)}=\prod_{l=1}^{L}K_{r_l}=\prod_{l=1}^{L}(2^{r_l}-2r_l-1).
\end{equation}
Since quantum Hamming codes are distance-$3$ codes, the distance of $\Q^{(L)}$ obtained by concatenating the quantum Hamming codes $L$ times is $3^L$.
Therefore, $\Q^{(L)}$ is a $[[N^{(L)},K^{(L)},3^L]]$ code.
Note that the ability of our code to suppress errors is not fully characterized by the code distance since our protocol suppresses error by increasing the concatenation level $L$, unlike quantum LDPC codes increasing distance for error suppression.
For the choice of~\eqref{seq:r_l}, we have
\begin{align}
  \label{eq:n_L_scaling}
  N^{(L)}&=\exp(O(L^2)),\\
  \label{eq:k_L_scaling}
  K^{(L)}&=\exp(O(L^2)).
\end{align}

We here prove that $\Q^{(L)}$ has an asymptotically non-vanishing rate.
Let
\begin{equation}
  R(L)\coloneqq\frac{K^{(L)}}{N^{(L)}}
\end{equation}
denote the rate of $\Q^{(L)}$,
where $K^{(L)}$ and $N^{(L)}$ are given by~\eqref{eq:k_L} and~\eqref{eq:n_L}, respectively.
Then, it suffices to derive a lower bound of
\begin{align}
  \frac{K^{(L)}}{N^{(L)}}&=\prod_{l=1}^{L}\frac{K_{r_l}}{N_{r_l}}\\
      &=\prod_{l=1}^{L}\frac{2^{r_l}-2r_l-1}{2^{r_l}-1}\\
  \label{eq:n}
      &=\prod_{l=1}^{L}\left(1-\frac{2(l+2)}{2^{l+2}-1}\right).
\end{align}
Since $\nicefrac{K^{(L)}}{N^{(L)}}$ monotonically decreases as $L$ increases,
it holds for any $L$ that
\begin{equation}
  \label{eq:infinite_product}
  \frac{K^{(L)}}{N^{(L)}}\geqq\prod_{l=1}^{\infty}\frac{K_{r_l}}{N_{r_l}}=\prod_{l=1}^{\infty}\left(1-\frac{2(l+2)}{2^{l+2}-1}\right).
\end{equation}
Here, an infinite sum
\begin{equation}
  \label{eq:infinite_sum}
  \sum_{l=1}^{\infty}\left|\frac{N_{r_l}-K_{r_l}}{N_{r_l}}\right|=\sum_{l=1}^{\infty}\left|\frac{2(l+2)}{2^{l+2}-1}\right|
\end{equation}
converges to a finite constant,
which proves that the infinite product on the right-hand side of~\eqref{eq:infinite_product} converges to a positive constant~\cite{Agarwal2011}.
Therefore, for any $L$, the rate $R(L)$ of $\Q^{(L)}$ is lower bounded by a positive constant, i.e.,
\begin{equation}
  \label{eq:rate}
  R(L)=\frac{K^{(L)}}{N^{(L)}}\geqq\lim_{l\to\infty}\frac{K^{(l)}}{N^{(l)}}=\frac{1}{\eta_\infty}>0,
\end{equation}
where $\eta_\infty$ is given by
\begin{equation}
  \eta_\infty\coloneqq\lim_{l\to\infty}\frac{N^{(l)}}{K^{(l)}},
\end{equation}
and called an asymptotic overhead factor.
The value of the asymptotic overhead factor $\eta_\infty$ of $\Q^{(L)}$ can be obtained by evaluating~\eqref{eq:n} numerically.
In particular, the numerical evaluation yields
\begin{equation}
  \eta_\infty<36,
\end{equation}
which contrasts with the fact that such overhead factors of conventional concatenated and topological codes would diverge to infinity.

We remark that the above proof of non-vanishing rate for concatenated codes is also applicable to other sequences of codes as long as the sequence of the codes satisfies the condition in~\eqref{eq:infinite_sum}.
In particular, the above argument leads to the following theorem.
\begin{theorem}[\label{thm:general_rate}Concatenated code with non-vanishing rate]
  For any sequence of $[[n_l,k_l,d]]$ quantum codes ($l\in\{1,\ldots,L\}$),
  the above construction leads to a $[[\prod_{l=1}^{L} n_l,\prod_{l=1}^{L} k_l, d^L]]$ concatenated code,
  and if it holds that
  \begin{equation}
    \sum_{l=1}^{\infty}\left|\frac{n_l-k_l}{n_l}\right|<\infty,
  \end{equation}
  then the concatenated code has a non-vanishing rate
  \begin{equation}
    \frac{\prod_{l=1}^{L} k_l}{\prod_{l=1}^{L} n_l}>\prod_{l=1}^{\infty}\frac{k_l}{n_l}>0.
  \end{equation}
\end{theorem}

For example, as we will discuss later in~\eqref{eq:r_l_log}, concatenation of quantum Hamming codes with parameter $r_l=\lceil\beta\log_2(l)\rceil$ for $\beta>1$ in place of~\eqref{seq:r_l} still achieves a non-vanishing rate and can be used for constant-space-overhead FTQC\@.
In the same way, concatenation of a growing sequence of, e.g., $[[2^{r_l}-1, 2^{r_l}-1-2t{r_l},2t+1]]$ quantum Bose-Chaudhuri-Hocquenghem (BCH) codes ($t\geqq 1$)~\cite{PhysRevA.54.4741} can also provide a family of concatenated codes with non-vanishing rate using the same $r_l$ as~\eqref{seq:r_l}.
This code family has a larger distance than a quantum Hamming code of the same size and thus may have a potential advantage in faster error suppression as $L$ increases.
Nevertheless, our analysis in the following sections focuses on an explicit construction of the fault-tolerant protocol for $\Q^{(L)}$.
We leave construction and analysis of general fault-tolerant protocols for other code families for future work, but our explicit construction of the fault-tolerant protocol for $\Q^{(L)}$ clarifies a fundamental design principle for such protocols.

\section{\label{sec:gadget}Compilation of original circuit into fault-tolerant circuit}

In this section, we present our fault-tolerant protocol that compiles the $W(M)$-qubit $D(M)$-depth original circuit into the fault-tolerant circuit to achieve the target error $\epsilon$.
In place of the $W(M)$ qubits of the original circuit, we will use the qubits in level-$L$ registers of the quantum error-correcting code $\Q^{(L)}$, where $L$ is the required concatenation level depending on $M$ and $\epsilon$.
In particular, for given $M$ and $\epsilon$, $L$ will be given by~\eqref{eq:L} in Sec.~\ref{sec:threshold_proof}.
We here compile the original circuit on the $W(M)$ qubits into a level-$L$ circuit using $O(\nicefrac{W(M)}{K^{(L)}})$ level-$L$ registers.
As in the conventional fault-tolerant protocol for concatenated codes in Ref.~\cite{G}, the actual implementation of the intended level-$L$ circuit by a level-$0$ circuit, i.e., the fault-tolerant circuit, is derived by recursively constructing the corresponding level-$(l-1)$ circuit from that at level $l$ ($l=L, L-1, \ldots, 1$).
But the construction here is different from that of Ref.~\cite{G} in that the circuits here are composed of elementary operations acting on registers rather than qubits, and that the procedure of converting a level-$l$ circuit to a level-$(l-1)$ circuit depends on $l$.

More concretely, for each level $l=0, \ldots, L$, we will define a set of elementary operations (preparations, gates, a measurement, and wait) acting on level-$l$ registers.
We require that a level-$l$ circuit should be written in terms only of the level-$l$ elementary operations in the set.
In a level-$l$ circuit that satisfies this requirement, each of the level-$l$ elementary operations is called a level-$l$ \textit{location}.
For each level-$l$ elementary operation, a corresponding level-$l$ \textit{gadget} will be defined, which is a level-$(l-1)$ circuit intended to carry out the logical elementary operation on the encoded level-$l$ registers.
We will also define a level-$l$ error-correction gadget, which is a level-$(l-1)$ circuit intended to carry out quantum error correction for an encoded level-$l$ register.
To implement some of the gates required for universal quantum computation, our protocol may use gate teleportation that combines multiple level-$l$ elementary operations to implement a gate.
For simplicity of presentation, we may represent these multiple level-$l$ elementary operations collectively as a level-$l$ \textit{abbreviation}, which is a level-$l$ circuit intended to carry out the gate on the level-$l$ registers.
A level-$l$ circuit that includes level-$l$ abbreviations is identified with that composed only of level-$l$ elementary operations obtained by expanding all the level-$l$ abbreviations.
In the following, we list up the set of level-$l$ elementary operations, gadgets, and abbreviations, while the precise definitions will be given later in Sec.~\ref{sec:fault_tolerant_gadget}.

We use the following set of level-$l$ elementary operations:
\begin{align}
  \label{eq:measurement}
  &\text{measurement:}\nonumber\\&\quad\includegraphics{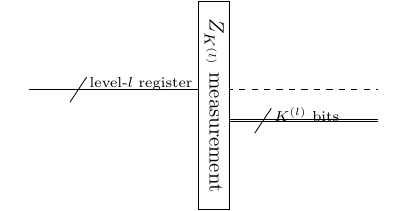},\\
  \label{eq:h_gate}
  &\text{$H$ gate:}\nonumber\\&\quad\includegraphics{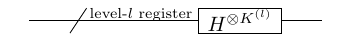},\\
  \label{eq:cnot_gate}
  &\text{\textsc{CNOT} gate:}\nonumber\\&\quad\includegraphics{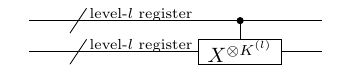},\\
  \label{eq:cz_gate}
  &\text{$CZ$ gate:}\nonumber\\&\quad\includegraphics{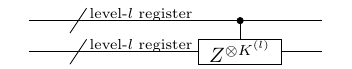},\\
  \label{eq:pauli_gate}
  &\text{Pauli gate:}\nonumber\\&\quad\includegraphics{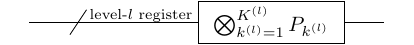},\\
  \label{eq:initial_state_preparation}
  &\text{initial-state preparation:}\nonumber\\&\quad\includegraphics{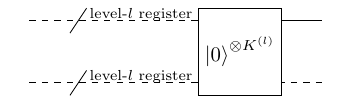},\\
  \label{eq:clifford_state_preparation}
  &\text{Clifford-state preparation:}\nonumber\\&\quad\includegraphics{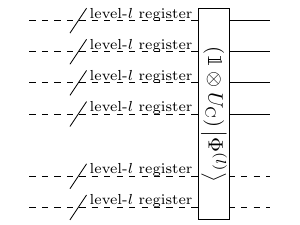},\\
  \label{eq:magic_state_preparation}
  &\text{magic-state preparation:}\nonumber\\&\quad\includegraphics{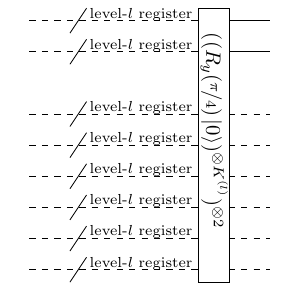},\\
  \label{eq:wait}
  &\text{wait:}\nonumber\\&\quad\includegraphics{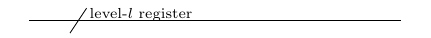}.
\end{align}
In these notations,
a dashed input wire turning into a solid output wire implies the allocation of a level-$l$ register.
A solid wire turning into a dashed one implies deallocation.
When an input wire and an output wire are both dashed, the corresponding level-$l$ register is used as a workspace in implementing the elementary operation.
Double-line wires represent $K^{(l)}$ bits.
The level-$l$ measurement operation performs measurements in $Z$ basis $\{\Ket{0},\Ket{1}\}$ of all the $K^{(l)}$ qubits in a level-$l$ register, deallocating the level-$l$ register and outputting the $K^{(l)}$-bit measurement outcome,
where we let $Z_{K^{(l)}}$ denote a label for representing this measurement.
The level-$l$ $H$-, \textsc{CNOT}-, and $CZ$-gate operations perform $H^{\otimes K^{(l)}}$, $\textsc{CNOT}^{\otimes K^{(l)}}$, and $CZ^{\otimes K^{(l)}}$ gates acting on all the qubits in level-$l$ registers.
The level-$l$ Pauli-gate operation performs a tensor product of arbitrary Pauli gates
\begin{equation}
  \label{eq:pauli_l}
  \bigotimes_{k^{(l)}=1}^{K^{(l)}}P_{k^{(l)}},
\end{equation}
where $P_{k^{(l)}}\in\{X, Z, Y,\mathbbm{1}\}$ is a single-qubit Pauli (or identity) gate acting on the ${k^{(l)}}$th qubit with labeling~\eqref{eq:k_Q_l}, and the global phase is ignored.
The level-$l$ initial-state preparation operation allocates two level-$l$ registers $B_1,B_2$ and prepares a state $\Ket{0}^{\otimes K^{(l)}}$ of $K^{(l)}$ qubits in $B_1$, followed by deallocating the level-$l$ register $B_2$ used as a workspace.
The level-$l$ Clifford-state preparation operation allocates six level-$l$ registers $B_1,B_2,B_3,B_4,B_5,B_6$ aligned from top to bottom in~\eqref{eq:clifford_state_preparation} and prepares a state $\left(\mathbbm{1}^{B_1 B_2}\otimes U_\mathrm{C}^{B_3 B_4}\right)\ket{\Phi^{(l)}}^{B_1 B_2 B_3 B_4}$, where $U_\mathrm{C}^{B_3 B_4}$ is an arbitrary Clifford unitary operator acting on qubits in $B_3$ and $B_4$,
\begin{align}
  \label{eq:phi}
  &\ket{\Phi^{(l)}}^{B_1 B_2 B_3 B_4}=\ket{\Phi}^{B_1 B_3}\otimes\ket{\Phi}^{B_2 B_4},
\end{align}
and
\begin{align}
  &\ket{\Phi}^{B_j B_{j^\prime}}=\frac{1}{\sqrt{2^{K^{(l)}}}}\sum_{m=0}^{2^{K^{(l)}}-1}\Ket{m}^{B_j}\otimes\Ket{m}^{B_{j^\prime}},\nonumber\\
  &\quad(j,j^\prime)\in\{(1,3),(2,4)\}.
\end{align}
Here, the superscripts represent the registers to which states and operators belong.
The two auxiliary level-$l$ registers $B_5$ and $B_6$ are deallocated at the end of the implementation.
The level-$l$ magic-state preparation operation allocates eight level-$l$ registers $B_1,B_2,B_3,B_4,B_5,B_6,B_7,B_8$ aligned from top to bottom in~\eqref{eq:magic_state_preparation}, prepares a state ${(R_y(\nicefrac{\pi}{4})\ket{0})}^{\otimes K^{(l)}}$ of $K^{(l)}$ qubits in each of $B_1$ and $B_2$, and deallocates $B_3,B_4,B_5,B_6,B_7,B_8$ at the end of the implementation.
To avoid potential confusion, in our presentation, we always write the registers $B_1,B_2,\ldots$ in~\eqref{eq:initial_state_preparation},~\eqref{eq:clifford_state_preparation}, and~\eqref{eq:magic_state_preparation} from top to bottom in a circuit.
However, under our assumption of no geometrical constraint, we can arbitrarily bend, cross, and permute the wires, which we do not consider as elementary operations.
The level-$l$ wait operation performs the identity operator on a level-$l$ register, which will be regarded as a special case of Pauli gates henceforth and will not be explained explicitly.

We may omit dashed wires output from the operations while we do not omit those input into the operations.
It is then possible to determine the number of omitted output wires by comparing the total numbers of the input and the explicitly shown output wires.
For example, the measurement operation~\eqref{eq:measurement} with omitting the output dashed wire may be denoted by
\begin{align}
  &\text{measurement:}\nonumber\\&\quad\includegraphics{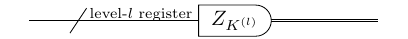}.
\end{align}
The notations on elementary operations may also be used to show the corresponding gadgets; in particular, whenever a level-$l$ elementary operation is depicted as the one acting on $N_{r_l}$ level-$(l-1)$ registers instead of each level-$l$ register, this notation represents the corresponding level-$l$ gadget.

Apart from these elementary operations and the corresponding gadgets, the level-$l$ error-correction gadget is denoted by
\begin{align}
  \label{eq:error_correction}
  &\text{error correction:}\nonumber\\
  &\quad\includegraphics{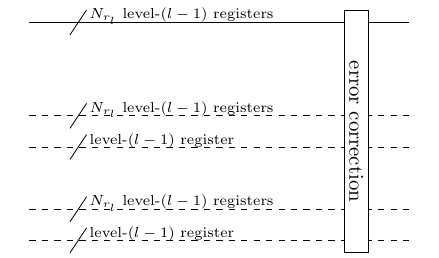}.
\end{align}
The level-$l$ error-correction gadget performs quantum error correction on an encoded level-$l$ register for the first set of $N_{r_l}$ level-$(l-1)$ registers (solid wire), temporarily using the other $2(N_{r_l}+1)$ level-$(l-1)$ registers (dashed wires).
In addition, the following level-$l$ abbreviations are used:
\begin{align}
  \label{eq:clifford_abbreviation}
  &\text{two-register Clifford gate:}\nonumber\\&\quad\includegraphics{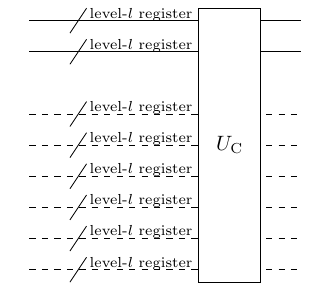},\\
  \label{eq:r_y_abbreviation}
  &\text{$R_y(\pm\nicefrac{\pi}{4})$ gate:}\nonumber\\&\quad\includegraphics{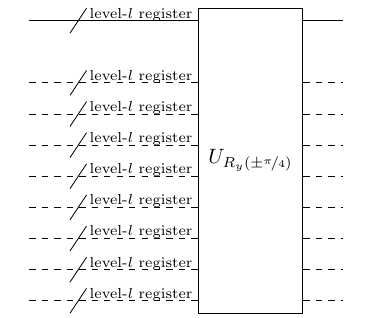},
\end{align}
which are abbreviated notations of combinations of level-$l$ elementary operations with functions described as follows.
The level-$l$ two-register Clifford-gate abbreviation applies an arbitrary Clifford unitary $U_\mathrm{C}$ on the first two level-$l$ registers (solid wires), temporarily using the other six auxiliary level-$l$ registers (dashed wires) during its implementation.
The level-$l$ $R_y(\pm\nicefrac{\pi}{4})$-gate abbreviation applies $U_{R_y(\pm\nicefrac{\pi}{4})}$ gates on the first level-$l$ register (solid wire), temporarily using the other eight auxiliary level-$l$ registers (dashed wires) during its implementation, where $U_{R_y(\pm\nicefrac{\pi}{4})}$ is an arbitrary tensor product of $R_y(\nicefrac{\pi}{4})$, $R_y(-\nicefrac{\pi}{4})$, and $\mathbbm{1}$.
Similar to the elementary operations, the output dashed wires in these notations may be omitted.

\begin{turnpage}
\begin{figure*}[t]
  \centering
  \includegraphics[width=9.4in]{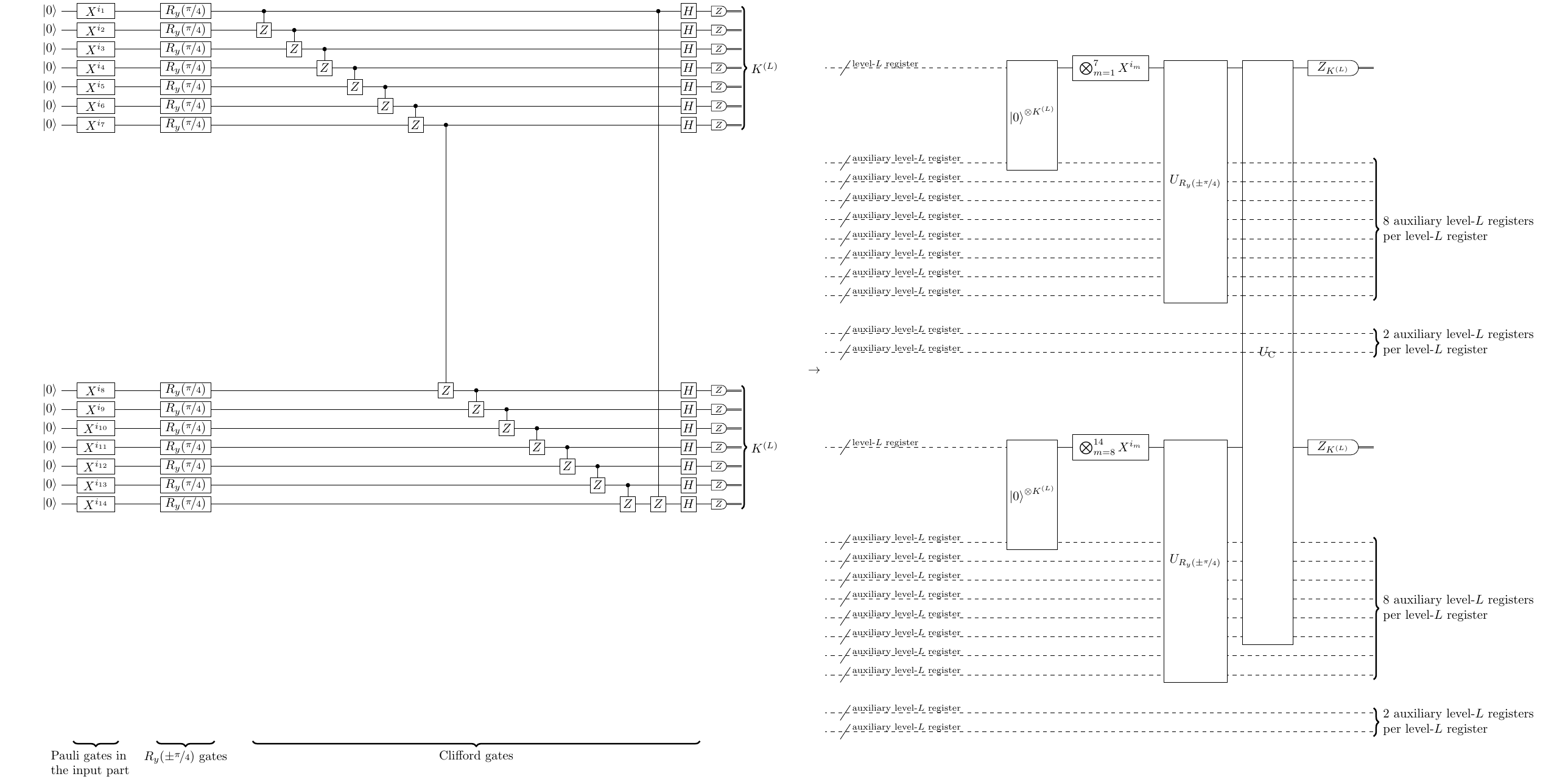}
  \caption{\label{fig:circuit_compilation}An example of $14$-qubit original circuits on the left ($W(M)=14$ with $M=14$) and the corresponding level-$2$ circuit obtained from this original circuit on the right ($L=2$ and $K^{(L)}=7$). The original circuit is composed of a gate set of Clifford gates $X$, $Y$, $Z$, $H$, $S$, $\textsc{CNOT}$, and $CZ$, and non-Clifford gates $R_y(\pm\nicefrac{\pi}{4})$. We represent the $W(M)$ qubits in the original circuit by the qubits in the $\lceil\nicefrac{W(M)}{K^{(L)}}\rceil$ level-$L$ registers and add $8+2=10$ auxiliary level-$L$ registers per level-$L$ register. Among these ten, eight are used for the level-$L$ abbreviations and the level-$L$ elementary operations as shown in the figure, and the other two are used for error correction as will be explained in Fig.~\ref{fig:circuit_conversion}. Pauli gates in the input part~\eqref{eq:input_part} of the original circuit are compiled into level-$L$ Pauli-gate operations. Then, a multiple-depth part of the original circuit composed only of the Clifford gates is compiled into a single use of the level-$L$ two-register Clifford-gate abbreviation $U_\mathrm{C}$, and a one-depth part of the original circuit with $R_y(\pm\nicefrac{\pi}{4})$ gates into a single use of the level-$L$ $R_y(\pm\nicefrac{\pi}{4})$-gate abbreviation $U_{R_y(\pm\nicefrac{\pi}{4})}$. The abbreviations acting on all the level-$L$ registers can be performed in parallel.}
\end{figure*}
\end{turnpage}

\begin{turnpage}
\begin{figure*}[t]
  \centering
  \includegraphics[width=9.4in]{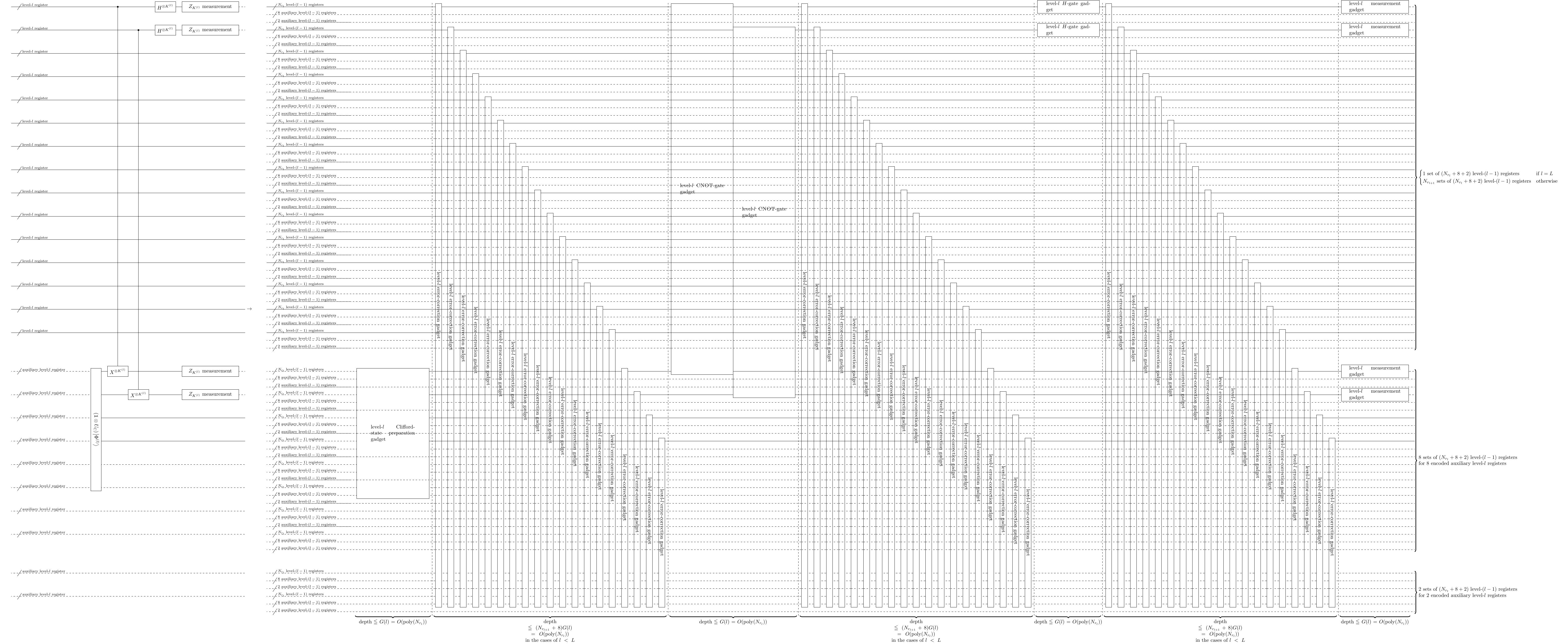}
  \caption{\label{fig:circuit_conversion}The recursive procedure to compile a level-$l$ circuit (left) into a level-$(l-1)$ circuit (right) for $l=L,\ldots,1$, where we replace each level-$l$ location of the level-$l$ elementary operation with the corresponding level-$l$ gadget and insert the level-$l$ error-correction gadgets in between. The figure shows the case of $l=1<L$, i.e., $N_{r_l}=7$ and $N_{r_{l+1}}=15$. Note that the figure omits the double-line wires of measurement operations~\eqref{eq:measurement} for representing the measurement outcome. The level-$L$ circuit is given by the procedure of Fig.~\ref{fig:circuit_compilation}. Then for $l=L,\ldots,1$, in place of each level-$l$ register in the level-$l$ circuit (e.g., each of the solid and dashed wires on the right of Fig.~\ref{fig:circuit_compilation}), we use a set of $N_{r_l}$ level-$(l-1)$ registers and add $8+2=10$ auxiliary level-$(l-1)$ registers per set in the corresponding level-$(l-1)$ circuit. Among these ten auxiliary level-$(l-1)$ registers, eight are used for level-$(l-1)$ elementary operations and level-$(l-1)$ abbreviations, and the other two are level-$(l-1)$ dormant registers that are never used explicitly in the level-$(l-1)$ circuit but will be used for level-$(l-1)$ error-correction gadgets. The level-$l$ elementary operations are replaced with the corresponding level-$l$ gadgets, which are level-$(l-1)$ circuits to perform the elementary operation on the encoded level-$l$ registers. Then, level-$l$ error-correction gadgets are inserted in between for all the allocated encoded level-$l$ registers in a synchronized way; that is, during each time period of performing the level-$l$ error-correction gadgets on all the encoded level-$l$ registers, we do not perform the other level-$l$ gadgets, as shown in the figure. For the synchronization, we may insert wait operations before and after each level-$l$ gadget to wait for the timing of the error correction. The maximal depth among all the level-$l$ gadgets is denoted by $G(l)$, which satisfies $G(l)=O(\poly(N_{r_l}))$ in our protocol. Then, the sum of the depth of each gadget and the wait operations is upper bounded by $G(l)$. In each time period of the error correction, we perform the error-correction gadgets sequentially by reusing the workspace of the two auxiliary encoded level-$l$ registers. In the case of $l=L$, for allocated registers among each encoded level-$L$ register and its eight auxiliary level-$L$ registers for elementary operations, we perform (at most) $1+8=9$ level-$L$ error-correction gadgets sequentially. The depth of each part of the error correction in the level-$(L-1)$ circuit is bounded by $9G(l)=O(\poly(N_{r_l}))$. In the cases of $l<L$, for allocated registers among each set of $N_{r_{l+1}}$ encoded level-$l$ registers and their eight auxiliary level-$l$ registers for elementary operations, we perform (at most) $N_{r_{l+1}}+8$ level-$l$ error-correction gadgets sequentially. In these cases, the depth of each part of the error correction in the level-$(l-1)$ circuit is bounded by $(N_{r_{l+1}}+8)G(l)=O(\poly(N_{r_l}))$.}
\end{figure*}
\end{turnpage}

Using these elementary operations and abbreviations, we compile the original circuit into a level-$L$ circuit as shown in Fig.~\ref{fig:circuit_compilation}.
In particular, we use $\lceil\nicefrac{W(M)}{K^{(L)}}\rceil$ level-$L$ registers and represent the $W(M)$ qubits in the original circuit as the qubits in these level-$L$ registers,
where
\begin{equation}
   \lceil x \rceil
\end{equation}
is the ceiling function representing the smallest integer larger than or equal to $x$.
Additionally, for each of these level-$L$ registers, we allocate $8+2=10$ auxiliary level-$L$ registers.
Among these ten, eight auxiliary level-$L$ registers are used for the level-$L$ abbreviations and the level-$L$ elementary operations as explained in the following.
The other two, which we call level-$L$ \textit{dormant} registers, are never used explicitly in the level-$L$ circuit.
The level-$L$ dormant registers are used to secure a workspace for the level-$L$ error-correction gadgets appearing in the level-$(L-1)$ circuit, as will be explained later.
To obtain the level-$L$ circuit,
we replace the Pauli gates in the input part~\eqref{eq:input_part} of the original circuit with the corresponding level-$L$ Pauli-gate operations,
and then replace the Clifford and $R_y(\pm\nicefrac{\pi}{4})$ gates in the original circuit with the corresponding level-$L$ Clifford- and $R_y(\pm\nicefrac{\pi}{4})$-gate abbreviations, respectively;
in this replacement, the dashed wires in the level-$L$ abbreviations~\eqref{eq:clifford_abbreviation} and~\eqref{eq:r_y_abbreviation} are always selected from the eight out of the ten auxiliary level-$L$ registers added to the level-$L$ registers on which the abbreviations act.
We also replace preparations of $\Ket{0}$ and measurements in the $Z$ basis in the original circuit with the corresponding level-$L$ preparation and measurement operations, respectively, in the level-$L$ circuit, where each level-$L$ preparation operation uses, as the dashed wire in~\eqref{eq:initial_state_preparation}, one of the eight auxiliary level-$L$ registers added to the level-$L$ register on which the operation acts.

Since the eight auxiliary level-$L$ registers per level-$L$ register are sufficient for performing each level-$L$ abbreviation (and preparation operation), we attain complete parallelizability; that is, we can perform the level-$L$ abbreviations (and preparation operations) acting on all $\lceil\nicefrac{M}{K^{(L)}}\rceil$ level-$L$ registers in parallel to reduce the time overhead (see Fig.~\ref{fig:circuit_compilation}).
Note that $R_y(\pm\nicefrac{\pi}{4})$ gates, preparations, and measurements in a level-$L$ register can be collectively replaced with a single use of the corresponding level-$L$ abbreviation and operations.
Similarly, multiple Clifford gates acting on qubits in the same pair of level-$L$ registers can be collectively replaced with a single use of level-$L$ two-register Clifford-gate abbreviation since the level-$L$ Clifford-gate abbreviation can implement an arbitrarily long sequence of Clifford gates in the pair of level-$L$ registers.
On the other hand, if a one-depth part of the original circuit includes multiple Clifford gates acting on qubits in different pairs of level-$L$ registers, this part requires multiple level-$L$ two-register Clifford-gate abbreviations, as we will explain in Sec.~\ref{sec:overhead}.

Then, for each $l=L,L-1,\ldots,1$, we perform a recursive procedure to compile the level-$l$ circuit into a level-$(l-1)$ circuit, as shown in Fig.~\ref{fig:circuit_conversion}.
For each level-$l$ register in the level-$l$ circuit, we use, in the corresponding level-$(l-1)$ circuit, a set of $N_{r_l}$ level-$(l-1)$ registers and further add $8+2=10$ auxiliary level-$(l-1)$ registers per set.
Similar to the level-$L$ case, among these ten, eight auxiliary level-$(l-1)$ registers are used for the level-$(l-1)$ abbreviations and the level-$(l-1)$ elementary operations in the level-$l$ gadgets.
The other two, the level-$(l-1)$ dormant registers, are never used explicitly in the level-$(l-1)$ circuit and are used as the workspace for the level-$(l-1)$ error-correction gadgets.
The corresponding level-$(l-1)$ circuit is given by replacing every level-$l$ location of a level-$l$ elementary operation in the level-$l$ circuit with its corresponding level-$l$ gadget and by inserting the level-$l$ error-correction gadgets between all the adjacent pairs of the level-$l$ locations (see Fig.~\ref{fig:circuit_conversion}).
For deallocated level-$l$ registers, we do not perform error correction; that is, we insert the level-$l$ error-correction gadgets only on the sets of $N_{r_l}$ level-$(l-1)$ registers for the allocated encoded level-$l$ registers.
In particular, the dormant level-$l$ registers never require error correction,
and we use the corresponding level-$(l-1)$ registers (i.e., $(N_{r_l}+8)$ level-$(l-1)$ registers for each dormant level-$l$ register) as a workspace for the level-$l$ error correction gadgets.
The ratio of the allocated level-$l$ registers to the dormant level-$l$ registers is at most $\nicefrac{(N_{r_l}+8)}{2}$ for $l< L$ and is $\nicefrac{9}{2}$ for $l=L$.
From~\eqref{eq:error_correction}, we see that two level-$l$ dormant registers, which corresponds to $2(N_{r_l}+8)$ level-$(l-1)$ usable registers, provide an enough workspace for error correction of an encoded level-$l$ register.
The error correction of the encoded level-$l$ registers cannot be done fully in parallel; that is, the level-$l$ dormant registers are time-shared in a series of error-correction gadgets.
The length of the series is bounded by $N_{r_l}+8$ for $l< L$ and is $9$ for $l=L$.

We perform error correction in a synchronized way as shown in Fig.~\ref{fig:circuit_conversion}; that is,
we start the level-$l$ error-correction gadgets only after we complete the level-$l$ gadgets for elementary operations over all the allocated encoded level-$l$ registers.
During a time period of performing the level-$l$ error-correction gadgets, we do not start performing the other level-$l$ gadgets.
In our protocol, the depths of different level-$l$ gadgets may vary.
Thus, for the synchronization, we insert wait operations after the level-$l$ gadgets that finish earlier.
In particular, let
\begin{equation}
  \label{eq:G_l_def}
  G(l)
\end{equation}
be the maximum depth of the level-$(l-1)$ circuit for a level-$l$ gadget, where the maximum is taken over all level-$l$ gadgets including elementary operations and error correction.
Then, for each level-$l$ gadget for a level-$l$ elementary operation, the sum of the depth of the gadget itself and that of the wait operations inserted for the synchronization is bounded by $G(l)$.
As will be shown in Sec.~\ref{sec:fault_tolerant_gadget}, we construct gadgets in such a way that $G(l)=O(\poly(N_{r_l}))$.
Similarly, wait operations are filled appropriately before and after the error-correction gadgets so that the next level-$l$ gadgets for elementary operations commence synchronously.
The depth of the error-correction part of the level-$(l-1)$ circuit is thus upper bounded by $(N_{r_l+1}+8)G(l)$ for $l<L$ and by $9G(l)$ for $l=L$, which are $O(\poly(N_{r_l}))$ in both cases.
Note that to optimize the runtime further, non-synchronized time scheduling of error correction could also be considered instead of the synchronized scheduling here while we leave such optimization for future work; for example, it may be possible to perform multiple short-depth level-$l$ gadgets while waiting for another long-depth level-$l$ gadget, but our analysis does not consider such optimization.

By performing this procedure recursively, we obtain the level-$0$ circuit.
Just after performing the above recursive procedure, the level-$0$ circuit is composed of the level-$0$ abbreviations and the level-$0$ elementary operations, which may use auxiliary level-$0$ registers by definition.
However, at level $0$, we are allowed to perform operations directly on physical qubits.
Thus, we substitute the level-$0$ two-register Clifford-gate abbreviations in the level-$0$ circuit with direct applications of the two-qubit Clifford gates and the level-$0$ $R_y(\pm\nicefrac{\pi}{4})$-gate abbreviations with the one-qubit $R_y(\pm\nicefrac{\pi}{4})$ gates, using the operations on physical qubits rather than performing the gate teleportation.
After this substitution for the level-$0$ abbreviations, we also substitute the remaining level-$0$ elementary operations with the corresponding operations on the physical qubits.
As a result, we do not use the $8+2=10$ auxiliary level-$0$ registers.
Hence after these substitutions, we remove the auxiliary level-$0$ registers from the level-$0$ circuit, which yields the fault-tolerant circuit on physical qubits to be executed in our fault-tolerant protocol.

In this compilation, some of the level-$l$ elementary operations and abbreviations are assumed to be invoked with classical arguments, i.e., classical bit strings that dictate the action of the elementary operations and abbreviations.
Elementary operations and abbreviations invoked with the classical arguments are called \textit{on-demand} elementary operations and abbreviations.
In particular, the Pauli-gate operation~\eqref{eq:pauli_gate} can be invoked with the argument of a classical description of Pauli gates $\bigotimes_{k^{(l)}=1}^{K^{(l)}}P_{k^{(l)}}$, which may not be determined during the compilation but can be given during execution; then, invoked with this classical argument, the Pauli-gate operation can perform the Pauli gates designated by the argument on demand.
Also, the Clifford-state preparation operation~\eqref{eq:clifford_state_preparation} and the two-register Clifford-gate abbreviation~\eqref{eq:clifford_abbreviation} can be invoked with the argument of a classical description of the Clifford unitary $U_{\mathrm{C}}$ and can then perform, respectively, the corresponding state preparation and the corresponding Clifford gate for $U_{\mathrm{C}}$ designated by this argument on demand.
The level-$l$ abbreviations pass the classical arguments to the appropriate level-$l$ elementary operations included in the abbreviations,
and the level-$l$ gadgets for the level-$l$ elementary operations are designed to be able to translate the arguments from level $l$ to level $(l-1)$ during execution, by performing classical computation.
As discussed in Sec.~\ref{sec:setting}, the original circuit has the input part~\eqref{eq:input_part}, which is implemented by the Pauli gates designated by the input~\eqref{eq:input}.
Also, in the level-$l$ circuit, we may use gate teleportation and error correction, which depend on the outcomes of measurements to be obtained after starting quantum computation.
As a whole, on-demand elementary operations and abbreviations are required in the following cases.
\begin{enumerate}
  \item The level-$L$ Pauli-gate operations used for the input part~\eqref{eq:input_part}.
  \item The level-$l$ Pauli-gate operations used for correction in the level-$l$ two-register Clifford-gate abbreviation in Fig.~\ref{fig:inblock_clifford} and the level-$l$ error-correction gadget in Fig.~\ref{fig:error_correction}.
  \item The level-$l$ two-register Clifford-gate abbreviations used for correction in the level-$l$ $R_y(\pm\nicefrac{\pi}{4})$-gate abbreviation in Fig.~\ref{fig:interblock_T}.
  \item The level-$l$ Clifford-state preparation operations invoked in the level-$l$ on-demand Clifford-gate abbreviations.
  \item The level-$(l-1)$ two-register Clifford-state abbreviations invoked in the level-$l$ gadgets of level-$l$ on-demand Clifford-state preparations in Fig.~\ref{fig:preparation_clifford}.
\end{enumerate}
It turns out that the above requirement is fulfilled in our protocol if the following set of on-demand elementary operations and abbreviations are available.
\begin{enumerate}
  \item The level-$l$ on-demand Pauli-gate operations invoked with a $2K^{(l)}$-bit binary row vector $(x_1,z_1,\ldots,x_{K^{(l)}},z_{K^{(l)}})\in{\{0,1\}}^{2K^{(l)}}$ to represent the Pauli gates
    \begin{equation}
      \label{eq:argument_pauli}
      \bigotimes_{k^{(l)}=1}^{K^{(l)}}P_{k^{(l)}}=\bigotimes_{k^{(l)}=1}^{K^{(l)}}X^{x_{k^{(l)}}}Z^{z_{k^{(l)}}}.
    \end{equation}
  \item The level-$l$ on-demand two-register Clifford-gate abbreviations invoked with $K^{(l)}$ $4\times 4$ binary matrices representing
    \begin{equation}
      \label{eq:argument_clifford}
      U_\mathrm{C}=\bigotimes_{k^{(l)}=1}^{K^{(l)}}U_{\mathrm{C},l}^{(k^{(l)})},
    \end{equation}
    where $U_{\mathrm{C},l}^{(k^{(l)})}$ is an arbitrary two-qubit Clifford unitary acting on the $k^{(l)}$th qubit in each of the two level-$l$ registers.
    Here, each $U_{\mathrm{C},l}^{(k^{(l)})}$ is represented as a $4\times 4$ binary matrix in such a way that the conjugation of two-qubit Pauli operators by $U_{\mathrm{C},l}^{(k^{(l)})}$ can be calculated via multiplication of the $4\times 4$ binary matrix representing $U_{\mathrm{C},l}^{(k^{(l)})}$ (from the right of the $4$-bit binary row vector representing the two-qubit Pauli operators), as shown in Ref.~\cite{PhysRevA.70.052328}.
  \item The level-$l$ on-demand Clifford-state preparations invoked with the $K^{(l)}$ $4\times 4$ binary matrices representing $U_\mathrm{C}$ in the form of~\eqref{eq:argument_clifford}.
\end{enumerate}
We will construct abbreviations and gadgets in Sec.~\ref{sec:fault_tolerant_gadget} in such a way that the abbreviations and gadgets offer the above on-demand functions and are implemented by using only the above on-demand functions.
Prior to starting the execution of quantum computation, the compilation represents these parts of the fault-tolerant circuit in terms of on-demand elementary operations to be invoked with the classical arguments, and all the other parts of the circuits are given by fixed circuits without such classical arguments, referred to as fixed parts.
The fixed parts may also include the Pauli-gate operation, the Clifford-state preparation operations, and the two-register Clifford-gate abbreviations that are not invoked with the classical arguments.
In the fixed parts, the actions of all the elementary operations and the abbreviations are determined during the compilation so as to reduce the time overhead of waiting for the classical computation during execution.

\section{\label{sec:fault_tolerant_gadget}Fault-tolerant gadgets and abbreviations}

\begin{figure*}[tpb]
  \centering
  \includegraphics[width=3.4in]{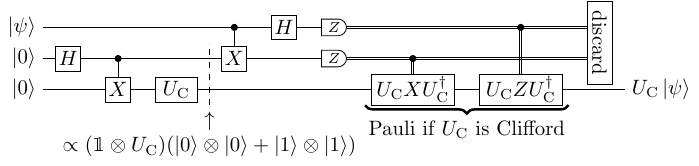}
  \includegraphics[width=7.0in]{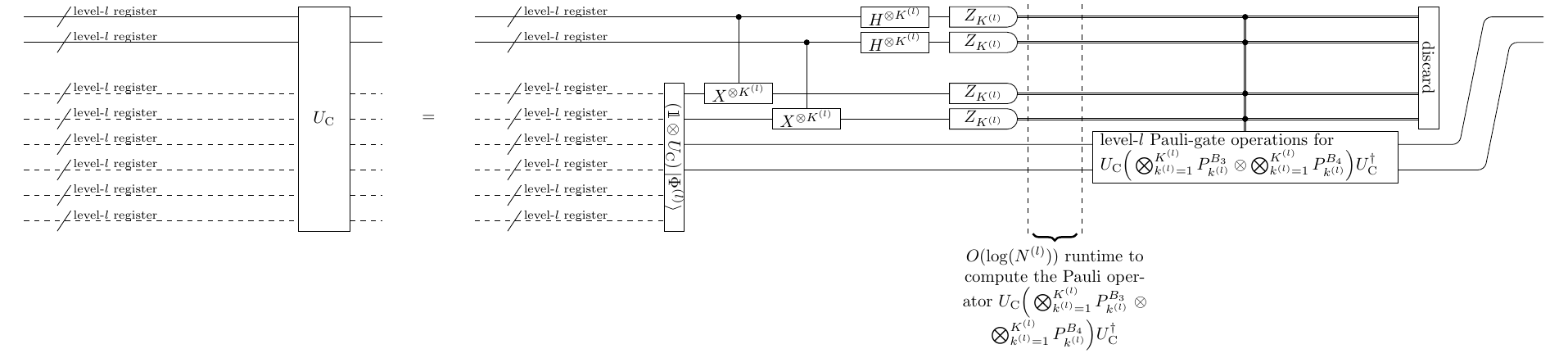}
  \caption{\label{fig:inblock_clifford}Gate teleportation for implementing a single-qubit Clifford gate at the top, and the level-$l$ two-register Clifford-gate abbreviation based on the gate teleportation for applying a Clifford gate $U_\mathrm{C}$ on two level-$l$ registers at the bottom. The implementation is assisted by auxiliary level-$l$ registers in a state $(\mathbbm{1}^{B_1 B_2}\otimes U_\mathrm{C}^{B_3 B_4})\ket{\Phi^{(l)}}^{B_1B_2B_3B_4}$, which is prepared by the Clifford-state preparation operation in~\eqref{eq:clifford_state_preparation}. In the figure, the level-$l$ registers $B_1, B_2, B_3, B_4$ are shown from top to bottom. Conditioned on the measurement outcome, a tensor product of Pauli operators $U_\mathrm{C}\left(\bigotimes_{k^{(l)}=1}^{K^{(l)}}P_{k^{(l)}}^{B_3}\otimes\bigotimes_{k^{(l)}=1}^{K^{(l)}}P_{k^{(l)}}^{B_4}\right)U_\mathrm{C}^\dag$ in~\eqref{eq:quantum_teleportation_correction_clifford} is calculated and applied as the correction in the quantum teleportation by two level-$l$ Pauli-gate operations in~\eqref{eq:pauli_gate}. If invoked with the argument dictating $U_\mathrm{C}$, the abbreviation passes the argument to the Clifford-state preparation operation in the circuit.}
\end{figure*}

\begin{turnpage}
\begin{figure*}[tpb]
  \centering
  \includegraphics[width=3.4in]{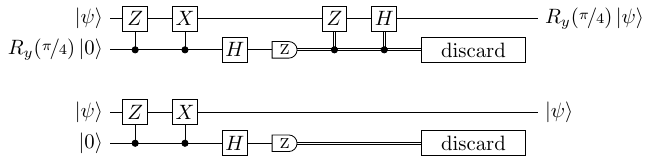}
  \includegraphics[width=9.4in]{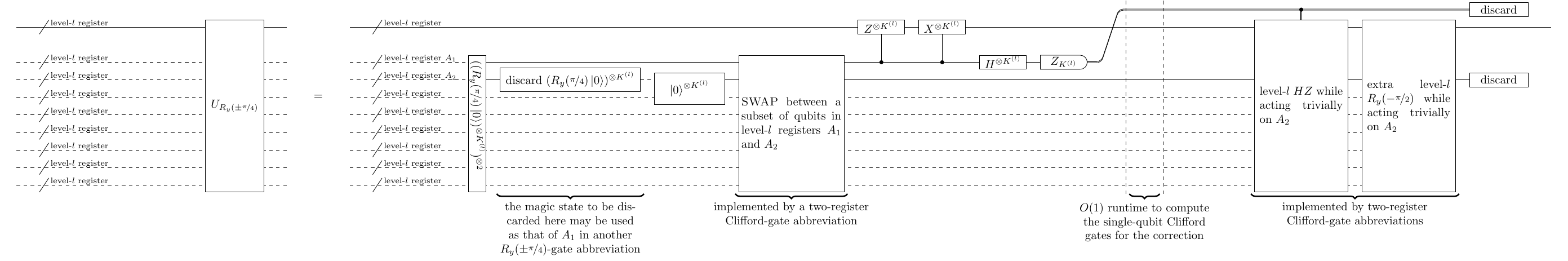}
  \caption{\label{fig:interblock_T}Gate teleportation assisted by a magic state $R_y(\nicefrac{\pi}{4})\Ket{0}$ for performing $R_y(\nicefrac{\pi}{4})$ at the top, the same circuit as the gate teleportation assisted by $\Ket{0}$ for acting trivially as $\mathbbm{1}$ in the middle, and the level-$l$ $R_y(\pm\nicefrac{\pi}{4})$-gate abbreviation based on the gate teleportation for applying $U_{R_y(\pm\nicefrac{\pi}{4})}$ gates to a level-$l$ register at the bottom, where $U_{R_y(\pm\nicefrac{\pi}{4})}$ is an arbitrary tensor product of $R_y(\nicefrac{\pi}{4})$, $R_y(-\nicefrac{\pi}{4})$, and $\mathbbm{1}$. The implementation is assisted by two types of auxiliary level-$l$ registers $A_1$ and $A_2$. The registers $A_1$ and $A_2$ are prepared in ${(R_y(\nicefrac{\pi}{4})\ket{0})}^{\otimes K^{(l)}}$ and ${\ket{0}}^{\otimes K^{(l)}}$ by the magic-state preparation operation in~\eqref{eq:magic_state_preparation} and by the initial-state preparation operation in~\eqref{eq:initial_state_preparation}, respectively. For each qubit on which $U_{R_y(\pm\nicefrac{\pi}{4})}$ acts trivially as $\mathbbm{1}$, a \textsc{SWAP} gate is applied between $A_1$ and $A_2$, so as to avoid the subsequent application of the $R_y(\pm\nicefrac{\pi}{4})$ gates. These \textsc{SWAP} gates are collectively implemented by a two-register Clifford-gate abbreviation in Fig.~\ref{fig:inblock_clifford}. Then, the gate teleportation for applying $R_y(\nicefrac{\pi}{4})$ and $\mathbbm{1}$ is performed. Conditioned on the measurement outcome, level-$l$ Clifford gates for the correction, i.e., $HZ$ gates, are applied. To perform $R_y(-\nicefrac{\pi}{4})$, we use the circuit at the top with an extra Clifford gate $R_y(-\nicefrac{\pi}{2})=ZH$ applied in the end after the correction with the $HZ$ gate. Each of the correction $HZ$ and the extra $R_y(-\nicefrac{\pi}{2})$ is implemented by a two-register Clifford-gate abbreviation in Fig.~\ref{fig:inblock_clifford} acting trivially on $A_2$ that is to be discarded in the end.}
\end{figure*}
\end{turnpage}

\begin{figure*}[tpb]
  \centering
  \includegraphics[width=7.0in]{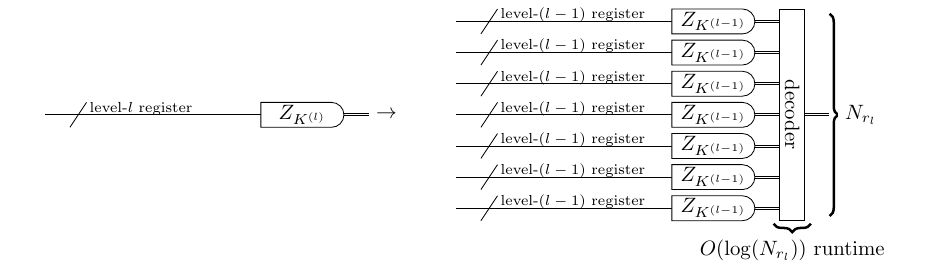}
  \caption{\label{fig:measurement}The level-$l$ measurement gadget for performing measurements in $Z$ basis for all the qubits in a level-$l$ register. The gadget is implemented by transversally performing level-$(l-1)$ measurement operations, followed by the efficient decoder. Since classical computation for this decoder has $O(\log(N_{r_l}))$ runtime, we may need to perform $O(\log(N_{r_l}))$ level-$(l-1)$ wait operations before using the measurement outcome.}
\end{figure*}

\begin{figure*}[t]
  \centering
  \includegraphics[width=7.0in]{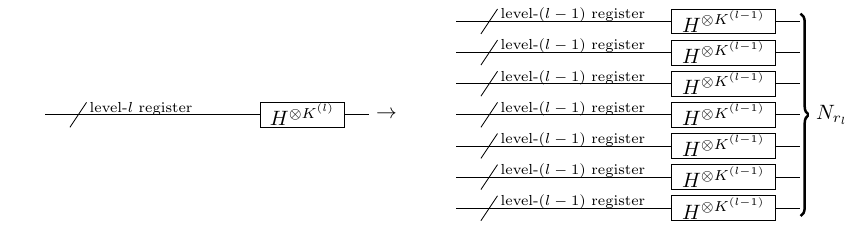}
  \includegraphics[width=7.0in]{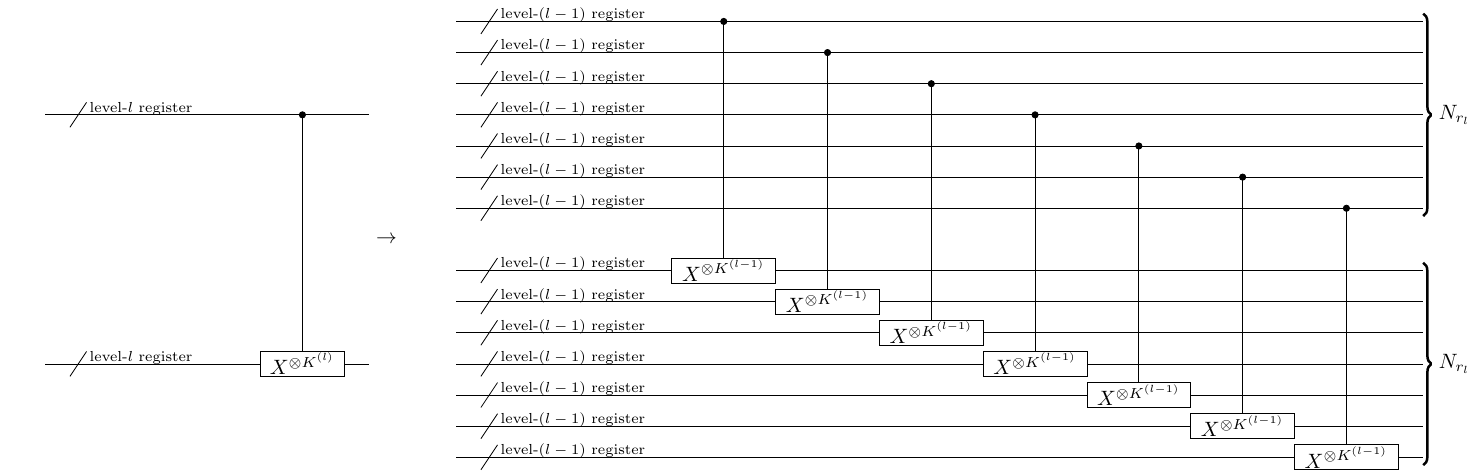}
  \includegraphics[width=7.0in]{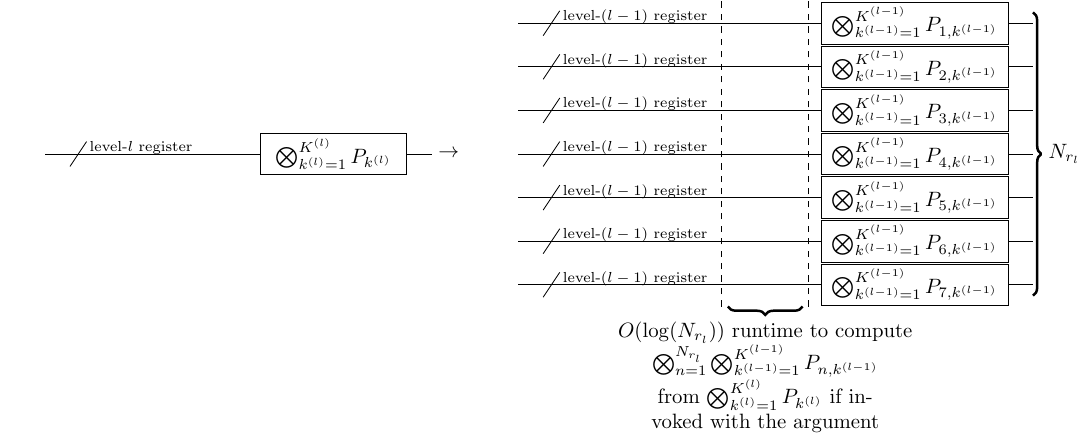}
  \caption{\label{fig:transversal_gate}The level-$l$ $H$-gate gadget for performing $H^{\otimes K^{(l)}}$ acting on all the $K^{(l)}$ qubits in a level-$l$ register at the top, the level-$l$ $\textsc{CNOT}$-gate gadget for performing $\textsc{CNOT}^{\otimes K^{(l)}}$ acting on all the $K^{(l)}$ control qubits in the upper level-$l$ register and all the $K^{(l)}$ target qubits in the lower level-$l$ register in the middle, and the level-$l$ Pauli-gate gadget for performing an arbitrary tensor product of Pauli gates $\bigotimes_{{k^{(l)}}=1}^{K^{(l)}}P_{k^{(l)}}$ for $P_{k^{(l)}}\in\{X, Z, Y,\mathbbm{1}\}$ in~\eqref{eq:pauli_l} acting on the $K^{(l)}$ qubits in a level-$l$ register at the bottom. The level-$l$ $H$- and $\textsc{CNOT}$-gate gadgets are implemented by level-$(l-1)$ $H$- and $\textsc{CNOT}$-gate operations, respectively, acting transversally on all the level-$(l-1)$ registers. In the same way as the level-$l$ $\textsc{CNOT}$-gate gadget but using $CZ$ in place of $\textsc{CNOT}$, the level-$l$ $CZ$-gate gadget for performing $CZ^{\otimes K^{(l)}}$ acting on two level-$l$ registers is defined. The level-$l$ Pauli-gate gadget can be implemented by performing Pauli gates $\bigotimes_{n=1}^{N_{r_l}}\bigotimes_{k^{(l-1)}=1}^{K^{(l-1)}}P_{n,k^{(l-1)}}$ in~\eqref{eq:pauli_l_1}, i.e., by applying level-$(l-1)$ Pauli-gate operations transversally to the $N_{r_l}$ level-$(l-1)$ registers. Note that the $O(\log(N_{r_l}))$-time classical computation shown in the Pauli-gate gadget can be performed during the compilation prior to starting execution of the quantum computation, except for the cases where the gadget is invoked with the argument to specify the gate.}
\end{figure*}

\begin{turnpage}
\begin{figure*}[t]
  \centering
  \includegraphics[width=9.4in]{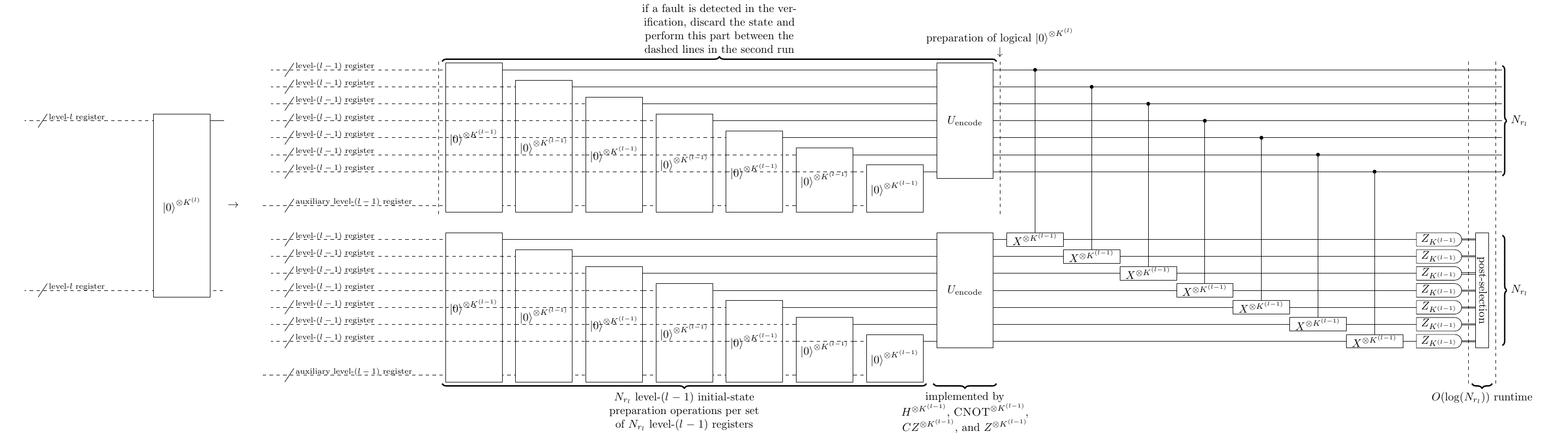}
  \caption{\label{fig:preparation_0}The level-$l$ initial-state preparation gadget for preparing a logical state $\Ket{0}^{\otimes K^{(l)}}$ in an encoded level-$l$ register. The gadget starts with a level-$(l-1)$ stabilizer circuit for preparing logical $\Ket{0}^{\otimes K^{(l)}}$ by implementing a Clifford unitary $U_\mathrm{encode}$ in~\eqref{eq:encode} for encoding. In this level-$(l-1)$ stabilizer circuit, preparation operations are performed sequentially using the auxiliary level-$(l-1)$ registers added to each set of $N_{r_l}$ level-$(l-1)$ registers on which the operations themselves act. Note that most of the auxiliary level-$(l - 1)$ registers are omitted in the figure except for those used for these preparations. Then, verification is performed by measuring the logical $Z$ operator for each of the logical qubits of $\Q_{r_l}$, and all the $Z$ generators of the stabilizer of $\Q_{r_l}$, for each of the $K^{(l-1)}$ code blocks. If no error is detected in the verification, the gadget outputs the logical $\Ket{0}^{\otimes K^{(l)}}$ prepared in this first run; otherwise, this prepared state is discarded, and the gadget performs the part between the dashed lines as shown in the figure and outputs the logical $\Ket{0}^{\otimes K^{(l)}}$ prepared in the second run without verification.}
\end{figure*}
\end{turnpage}

\begin{turnpage}
\begin{figure*}[tpb]
  \centering
  \includegraphics[width=3.4in]{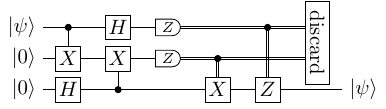}
  \includegraphics[width=9.4in]{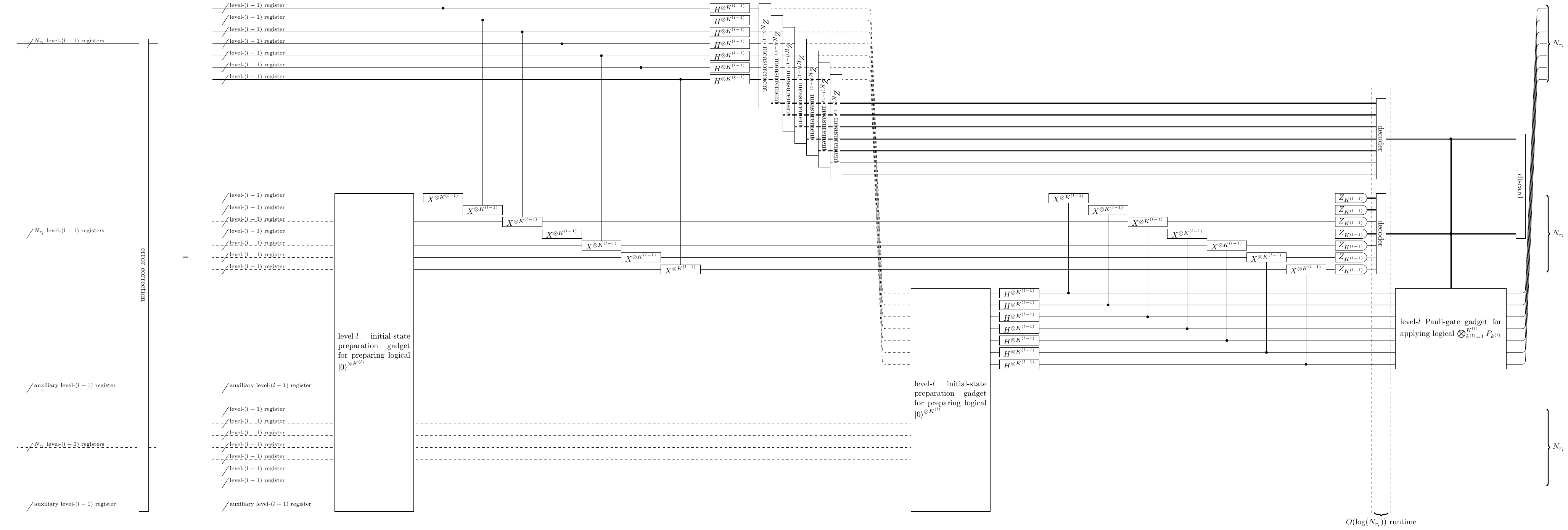}
  \caption{\label{fig:error_correction}A circuit for Knill's error correction at the top and the level-$l$ error-correction gadget based on Knill's error correction at the bottom. The gadget is assisted by $2(N_{r_l}+1)$ level-$(l-1)$ registers for preparing logical $\Ket{0}^{\otimes K^{(l)}}$. The white boxes in the figure for the level-$l$ initial-state preparation gadgets and the level-$l$ Pauli-gate gadgets are to be replaced with the corresponding level-$(l-1)$ circuits shown on the right-hand sides of Figs.~\ref{fig:preparation_0} and~\ref{fig:transversal_gate}, respectively.
  We calculate the measurement outcome in the same way as the measurement gadget in Fig.~\ref{fig:measurement}.
  Given the measurement outcome, we calculate a logical operator $\bigotimes_{k^{(l)}=1}^{K^{(l)}}P_{k^{(l)}}$ in~\eqref{eq:quantum_teleportation_correction} in the form of a tensor product of Pauli gates $P_{k^{(l)}}$ according to correction in quantum teleportation. In the level-$(l-1)$ circuit, logical $\bigotimes_{k^{(l)}=1}^{K^{(l)}}P_{k^{(l)}}$ for the correction in the quantum teleportation is implemented by invoking the level-$l$ Pauli-gate gadgets in Fig.~\ref{fig:transversal_gate} with an argument to specify the gate.}
\end{figure*}
\end{turnpage}

\begin{turnpage}
\begin{figure*}[tpb]
  \centering
  \includegraphics[width=9.4in]{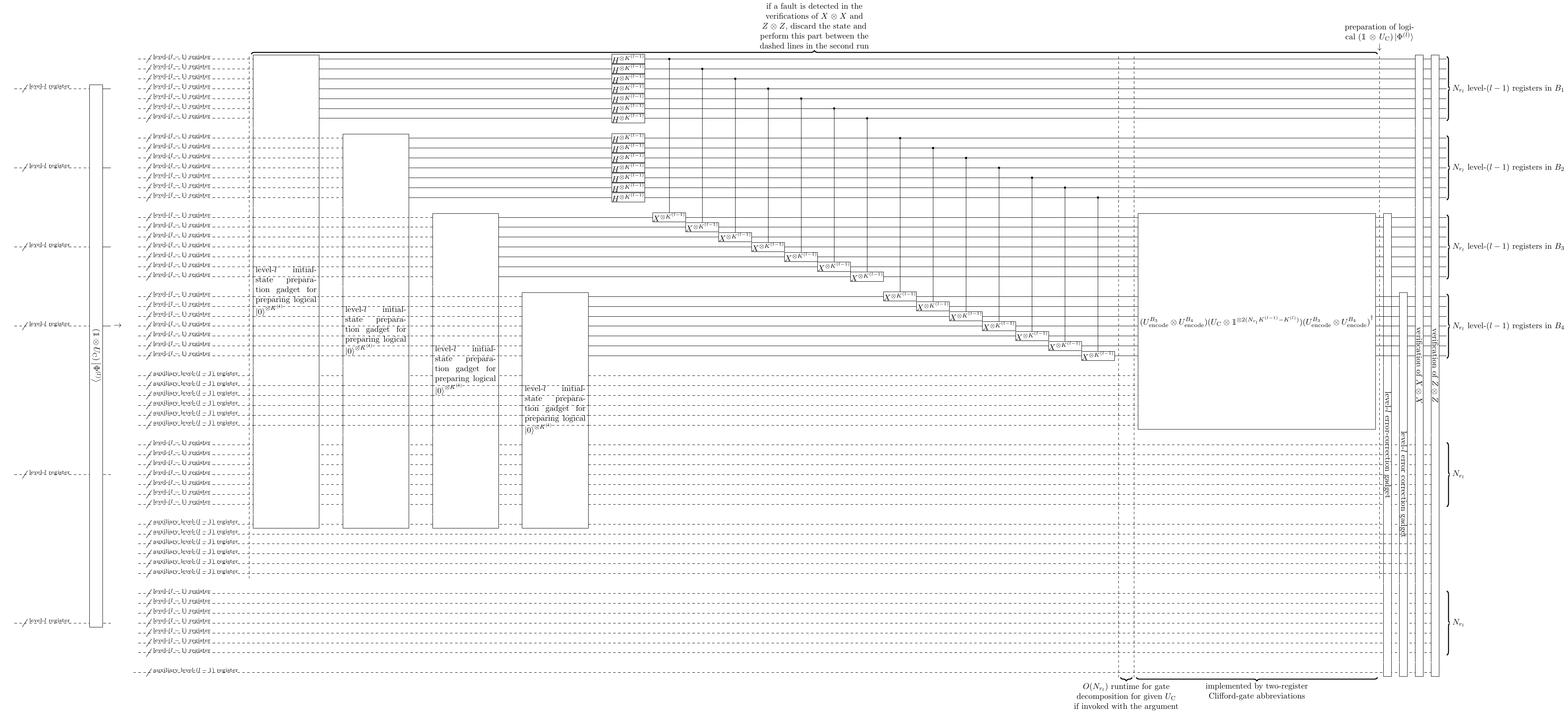}
\caption{\label{fig:preparation_clifford}The level-$l$ Clifford-state preparation gadget for preparing a logical state $(\mathbbm{1}^{B_1 B_2}\otimes U_\mathrm{C}^{B_3 B_4})\ket{\Phi^{(l)}}^{B_1 B_2 B_3 B_4}$, where each of $B_1,B_2,B_3,B_4$ is $N_{r_l}$ level-$(l-1)$ registers for an encoded level-$l$ register aligned from top to bottom in the level-$(l-1)$ circuit. The white boxes of the level-$l$ initial-state preparation gadgets and the level-$l$ error-correction gadgets are replaced with the level-$(l-1)$ circuits on the right-hand side of Figs.~\ref{fig:preparation_0} and~\ref{fig:error_correction}, respectively. The white boxes of verification of $X\otimes X$ and $Z\otimes Z$ are replaced with the level-$(l-1)$ circuit on the right-hand side of Fig.~\ref{fig:verification_clifford} and the same level-$(l-1)$ circuit with substituting $X$ with $Z$, respectively. The gadget starts with a level-$(l-1)$ stabilizer circuit for preparing logical $(\mathbbm{1}\otimes U_\mathrm{C})\ket{\Phi^{(l)}}$ by implementing~\eqref{eq:Clifford}. Then, verification is performed by using the level-$l$ error-correction gadgets in Fig.~\ref{fig:error_correction} so as to ensure that the state is in the code space of $\Q_{r_l}$, followed by measuring the logical stabilizer operators of logical $(\mathbbm{1}\otimes U_\mathrm{C})\ket{\Phi^{(l)}}$, i.e., all the logical operators in~\eqref{eq:xx} and~\eqref{eq:zz}, to detect logical errors. A post-selection is made according to the verifications of $X\otimes X$ and $Z\otimes Z$. If no error is detected in these verifications, the gadget outputs the logical $(\mathbbm{1}^{B_1 B_2}\otimes U_\mathrm{C}^{B_3 B_4})\ket{\Phi^{(l)}}^{B_1 B_2 B_3 B_4}$ prepared in this first run; otherwise, this prepared state is discarded, and the gadget again performs the part between the dashed lines as shown in the figure and outputs the logical $(\mathbbm{1}^{B_1 B_2}\otimes U_\mathrm{C}^{B_3 B_4})\ket{\Phi^{(l)}}^{B_1 B_2 B_3 B_4}$ prepared in this second run without verification. Note that classical computation for the gate decomposition of the gadget can be performed during the compilation prior to starting execution of the quantum computation, except for the cases where the gadget is invoked with the argument to specify the gate as discussed with~\eqref{eq:runtime_gate_decomposition_clifford}.}
\end{figure*}
\end{turnpage}

\begin{turnpage}
\begin{figure*}[tpb]
  \centering
  \includegraphics[width=9.4in]{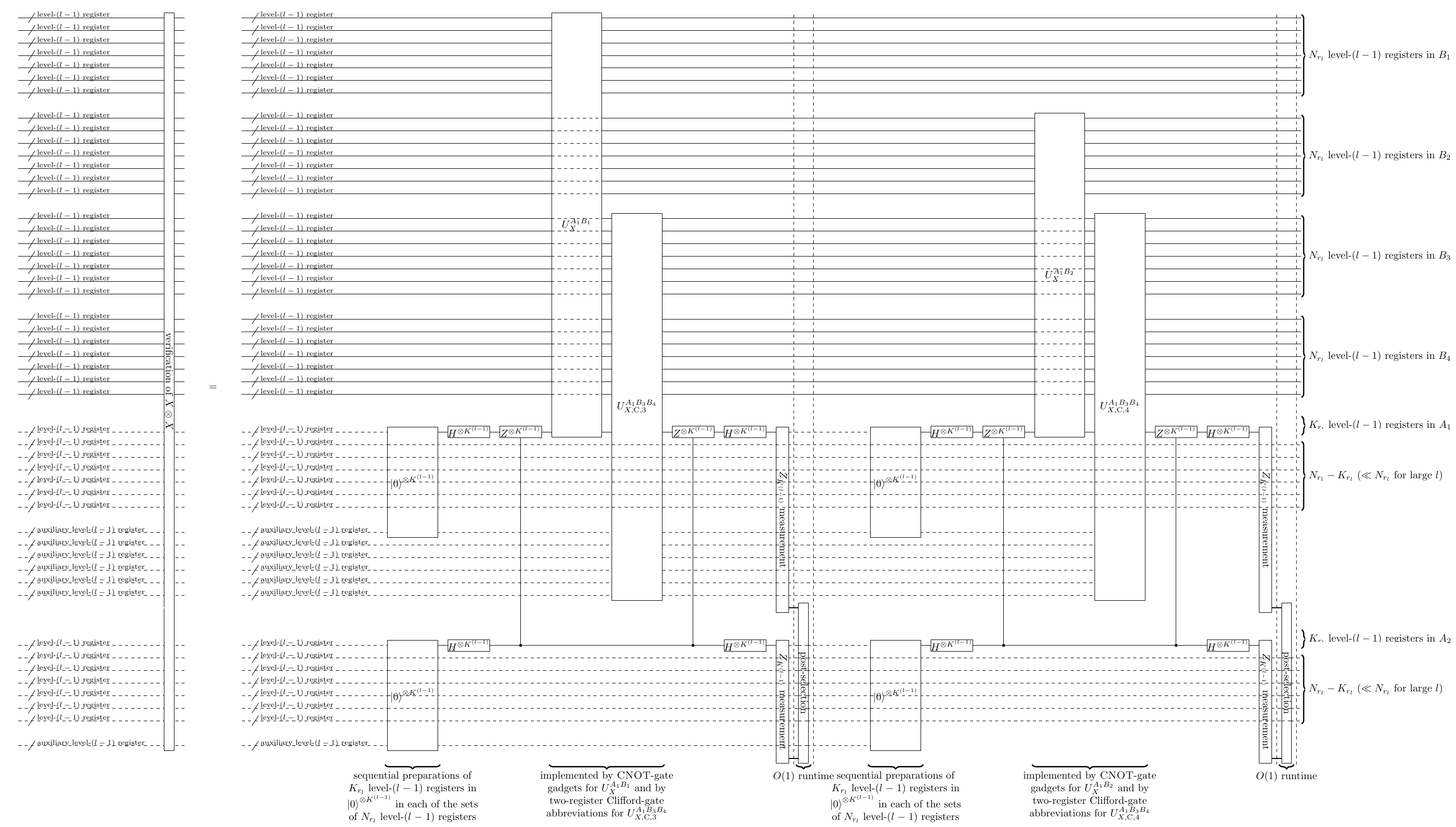}
  \caption{\label{fig:verification_clifford}A circuit for measuring the logical stabilizer operators of logical $(\mathbbm{1}^{B_1 B_2}\otimes U_\mathrm{C}^{B_3 B_4})\ket{\Phi^{(l)}}^{B_1 B_2 B_3 B_4}$. The circuit for extracting all the logical operators in~\eqref{eq:xx} to a set of $K_{r_l}$ level-$(l-1)$ registers is performed by controlled Pauli gates $U_X^{A_1B_1}$, $U_X^{A_2B_2}$, $U_{X,\mathrm{C}}^{A_3B_3}$, and $U_{X,\mathrm{C}}^{A_4B_4}$ defined as~\eqref{eq:U_X} and~\eqref{eq:U_XC}, while that for measuring all the logical operators in~\eqref{eq:zz} is the same as this circuit with substituting $X$ with $Z$. In extracting these logical operators, another set of $K_{r_l}$ level-$(l-1)$ registers are used as flag qubits to make the verification fault-tolerant. The state is discarded unless no error is detected either from these logical stabilizer operators in~\eqref{eq:xx} and~\eqref{eq:zz} and the flag qubits.}
\end{figure*}
\end{turnpage}

\begin{turnpage}
\begin{figure*}[tpb]
  \centering
  \includegraphics[width=9.4in]{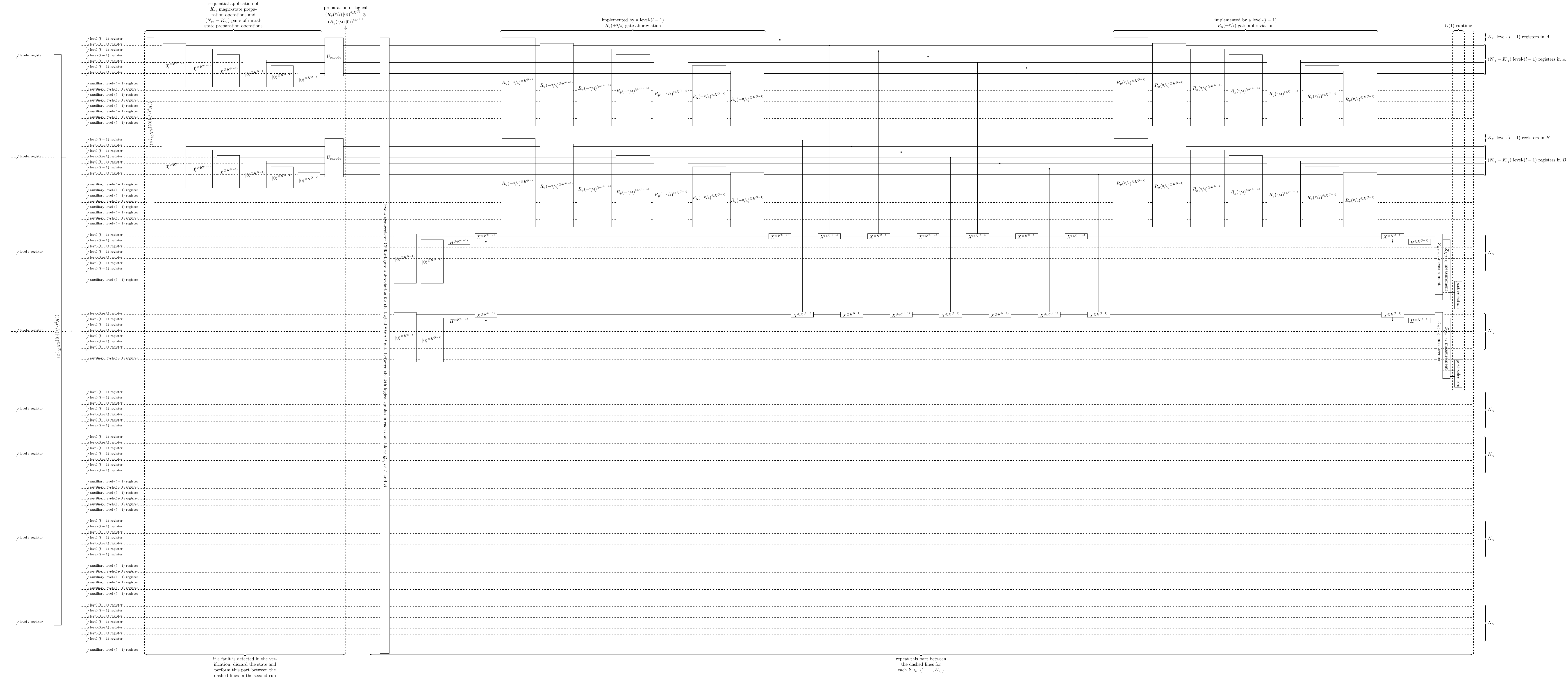}
  \caption{\label{fig:preparation_magic}The level-$l$ magic-state preparation gadget for preparing a logical state $((R_y(\nicefrac{\pi}{4})\ket{0})^{\otimes K^{(l)}})\otimes((R_y(\nicefrac{\pi}{4})\ket{0})^{\otimes K^{(l)}})$ of two encoded level-$l$ registers, where the two sets of $N_{r_l}$ level-$(l-1)$ registers for these two encoded level-$l$ registers are denoted respectively by $A$ and $B$. The white box for the level-$l$ two-register Clifford-gate abbreviation is to be replaced with the corresponding level-$(l - 1)$ circuit obtained from the level-$l$ circuit in Fig.~\ref{fig:inblock_clifford} by replacing each level-$l$ operation with the corresponding level-$l$ gadget. The gadget starts with a level-$(l-1)$ stabilizer circuit for encoding, i.e., for transforming $K_{r_l}$ level-$(l-1)$ registers in ${(R_y(\nicefrac{\pi}{4})\ket{0})}^{\otimes K^{(l-1)}}$ and $(N_{r_l}-K_{r_l})$ level-$(l-1)$ registers in $\ket{0}^{\otimes K^{(l-1)}}$ into the logical state ${(R_y(\nicefrac{\pi}{4})\ket{0})}^{\otimes K^{(l)}}$ for each of $A$ and $B$, in the same way as the initial-state preparation gadget in Fig.~\ref{fig:preparation_0}. In the verification, after ensuring that the state is in the code space of $\Q_{r_l}$, we measure the logical stabilizer operators~\eqref{eq:H_stabiliziers} of logical ${(R_y(\nicefrac{\pi}{4})\ket{0})}^{\otimes K_{r_l}}\otimes {(R_y(\nicefrac{\pi}{4})\ket{0})}^{\otimes K_{r_l}}$ for each pair of code blocks of $\Q_{r_l}$ in $A$ and $B$, where the logical stabilizer operators are given according to $H(R_y(\nicefrac{\pi}{4})\ket{0})=R_y(\nicefrac{\pi}{4})\ket{0}$. To measure each of the logical stabilizer operators iteratively, for $k=1,\ldots,K_{r_l}$, the level-$l$ two-register Clifford-gate abbreviation is used for implementing the logical \textsc{SWAP} gate between the $k$th logical qubits in each code block $\Q_{r_l}$ of $A$ and $B$, and then logical $H^{\otimes K_{r_l}}$ is measured for each code block $\Q_{r_l}$ of $A$ and $B$. For this measurement in each iteration, we extract $H=R_y(\nicefrac{\pi}{4})Z{R_y(-\nicefrac{\pi}{4})}$ to a level-$(l-1)$ register by combining the \textsc{CNOT}-gate operations with $R_y(\pm\nicefrac{\pi}{4})$ gates implemented by level-$(l-1)$ $R_y(\pm\nicefrac{\pi}{4})$-gate abbreviations in Fig.~\ref{fig:interblock_T}. In each extraction, another level-$(l-1)$ register is used as flag qubits to make the verification fault-tolerant. If no error is detected from logical stabilizer operators~\eqref{eq:H_stabiliziers} and the flag qubits in these iterations, then the gadget outputs the logical ${(R_y(\nicefrac{\pi}{4})\ket{0})}^{\otimes K_{r_l}}\otimes {(R_y(\nicefrac{\pi}{4})\ket{0})}^{\otimes K_{r_l}}$ prepared in this first run; otherwise, this prepared state is discarded, and the gadget again performs the part between the dashed lines as shown in the figure and outputs the logical ${(R_y(\nicefrac{\pi}{4})\ket{0})}^{\otimes K_{r_l}}\otimes {(R_y(\nicefrac{\pi}{4})\ket{0})}^{\otimes K_{r_l}}$ prepared in this second run without verification.}
\end{figure*}
\end{turnpage}

In this section, after defining the conditions of fault-tolerant gadgets, the detail of the construction of the gadgets and the abbreviations in our fault-tolerant protocol is explained one by one, using Figs.~\ref{fig:inblock_clifford}-\ref{fig:preparation_magic} to show the correspondence.
To show an abbreviation,
the corresponding level-$l$ circuit in terms of level-$l$ elementary operations will be given in the figures.
In particular, the two-register Clifford-gate abbreviation is given by Fig.~\ref{fig:inblock_clifford}, and the $R_y(\pm\nicefrac{\pi}{4})$-gate abbreviation by Fig.~\ref{fig:interblock_T}.
As for a gadget, the corresponding level-$(l-1)$ circuit of the gadget will be given in the figures.
In particular, the measurement gadget is given by Fig.~\ref{fig:measurement}, the $H$-, \textsc{CNOT}-, $CZ$-, and Pauli-gate gadgets by Fig.~\ref{fig:transversal_gate}, the initial-state preparation gadget by Fig.~\ref{fig:preparation_0}, the error-correction gadget by Fig.~\ref{fig:error_correction}, the Clifford-state preparation gadget by Fig~\ref{fig:preparation_clifford}, and the magic-state preparation gadget by Fig.~\ref{fig:preparation_magic}.
For explicitness, level-$(l-1)$ circuits of level-$l$ gadgets are shown using $N_{r_l}=7$ and $K_{r_l}=1$ as in the case of $l=1$.
To simplify the description of the level-$(l-1)$ circuits, gadgets may refer to other gadgets as sub-gadgets;
in such a case, the level-$(l-1)$ circuit may include a box with the name of another level-$l$ gadget, which is to be replaced with the level-$(l-1)$ circuit of the level-$l$ gadget of the box.
In addition, the level-$(l-1)$ circuit may include level-$(l-1)$ abbreviations, which should be replaced with the corresponding level-$(l-1)$ circuits.
Similar to sub-gadgets, gadgets may refer to a combination of other sub-gadgets as a sub-abbreviation;
in such a case, the level-$(l-1)$ circuit may include a box with the name of a level-$l$ abbreviation, which should be considered to be replaced with the level-$(l-1)$ circuit obtained from the level-$l$ circuit of the level-$l$ abbreviation of the box by replacing each level-$l$ elementary operation with the corresponding level-$l$ gadget.

\textbf{Conditions of fault-tolerant gadgets}:
We define the properties of gadgets required for fault tolerance.
To define such requirements, as in Ref.~\cite{G}, we introduce an ideal decoder and $r$-filters for our protocol.
In Ref.~\cite{G}, the conditions for fault-tolerant gadgets are given in the form of equivalence relations between circuits including the gadget, the $r$-filters, the ideal decoder, and the ideal intended elementary operation, conditioned on the number of faulty locations in the gadget.
Although Ref.~\cite{G} provides concrete definitions for the ideal decoder and the $r$-filters in the case of a $[[N, 1, 2t + 1]]$ code for $0\leqq r\leqq t$, the proof of the threshold theorem does not rely on specific definitions but on the equivalence relations alone.
Thus, to use the same argument as the proof of the threshold theorem in Ref.~\cite{G},
we will give appropriately modified definitions of an ideal filter and $r$-filters so that the gadgets proposed here should satisfy effectively the same set of equivalence relations as those in Ref.~\cite{G}.

An ideal decoder, a $0$-filter, and a $1$-filter in our case ($t=1$) are defined as follows.
As explained in Sec.~\ref{sec:hamming}, a level-$l$ register is encoded into $N_{r_l} K^{(l-1)}$ qubits in $N_{r_l}$ level-$(l-1)$ registers,
and the $k^{(l-1)}$th qubit of the $n$th level-$(l-1)$ register is labeled $(n,k^{(l-1)})$ with $k^{(l-1)}\in\{1,\ldots,K^{(l-1)}\}$ and $n\in\{1,\ldots,N_{r_l}\}$.
Let
\begin{equation}
  \mathcal{H}_{n,k^{(l-1)}}\cong\mathbb{C}^2
\end{equation}
be the Hilbert space of the qubit $(n,k^{(l-1)})$ in a level-$(l-1)$ register.
Define
\begin{align}
  \mathcal{H}_{k^{(l-1)}}&\coloneqq\bigotimes_{n=1}^{N_{r_l}} \mathcal{H}_{n,k^{(l-1)}},\\
  \label{eq:H}
  \mathcal{H}&\coloneqq\bigotimes_{k^{(l-1)}=1}^{K^{(l-1)}} \mathcal{H}_{k^{(l-1)}},
\end{align}
where $\mathcal{H}$ is the whole space of the $N_{r_l}$ level-$(l-1)$ registers.
The quantum Hamming code $\Q_{r_l}$ defines a code space for each $k^{(l-1)}$
\begin{equation}
  \mathcal{H}_{k^{(l-1)}}^{\mathrm{code}} \subset \mathcal{H}_{k^{(l-1)}}.
\end{equation}
Then the whole code space in the $N_{r_l}$ level-$(l-1)$ registers is given by
\begin{equation}
  \mathcal{H}^{\mathrm{code}}\coloneqq\bigotimes_{k^{(l-1)}=1}^{K^{(l-1)}} \mathcal{H}_{k^{(l-1)}}^{\mathrm{code}} \subset \mathcal{H}.
\end{equation}
The level-$l$ $0$-filter acting on the $N_{r_l}$ level-$(l-1)$ registers is defined as a projector
\begin{equation}
  \label{eq:0filter}
  \Pi^{\mathrm{code}}~\text{onto the subspace $\mathcal{H}^{\mathrm{code}}\subset\mathcal{H}$}.
\end{equation}
Suppose that a codeword $\ket{\psi_{k^{(l-1)}}} \in \mathcal{H}_{k^{(l-1)}}^{\mathrm{code}}$ suffers from an error represented by a Pauli operator $P_{k^{(l-1)}} \in \mathcal{P}_{N_{r_l}}$ acting on $\mathcal{H}_{k^{(l-1)}}$.
If the weight of $P_{k^{(l-1)}}$ is at most $1$, the code $\Q_{r_l}$ can correct the error.
As a result, if a codeword $\ket{\psi} \in \mathcal{H}^{\mathrm{code}}$ suffers from an error represented by $P=\bigotimes_{k^{(l-1)}=1}^{K^{(l-1)}} P_{k^{(l-1)}}$, the $K^{(l-1)}$ blocks of the code $\Q_{r_l}$ can correct the error $P$ as long as every $P_{k^{(l-1)}}$ has a weight at most $1$.
Such an error $P$ is said to be \textit{correctable} here.
An ideal decoder here is defined to do this correction and decoding; that is, the ideal decoder receives the $N_{r_l}$ level-$(l-1)$ registers in an input state $P\ket{\psi}$, corrects errors to recover the state $\ket{\psi}$, and converts it to the non-encoded state of a level-$l$ register.
The level-$l$ $1$-filter acting on the $N_{r_l}$ level-$(l-1)$ registers is defined as a projector onto the subspace spanned by all the states in the form $P\ket{\psi}$ with $P$ correctable and $\ket{\psi} \in \mathcal{H}^{\mathrm{code}}$.
It turns out that this subspace is the entire space $\mathcal{H}$, and the $1$-filter is trivial, i.e.,
\begin{equation}
  \label{eq:1filter}
  \mathbbm{1}^{\otimes N_{r_l}K^{(l-1)}}~\text{on $\mathcal{H}$}.
\end{equation}

With these definitions of the ideal decoder and the $r$-filters ($r=0,1$), the notations used for the condition for a gadget to be fault-tolerant are introduced as follows.
Note that the notations here are made analogous to Ref.~\cite{G} using qubits, but are different from Ref.~\cite{G} in that each wire represents one or multiple registers.
The ideal decoder is denoted by
\begin{align}
  &\text{ideal decoder:}\nonumber\\&\quad\includegraphics{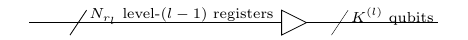},
\end{align}
where the thin wire after the ideal decoder represents a level-$l$ register composed of $K^{(l)}=K_{r_l}K^{(l-1)}$ qubits, and each of the other wires with regular thickness represents $N_{r_l}$ level-$(l-1)$ registers collectively.
The $0$-filter here reduces to $\Pi^{\mathrm{code}}$ as shown in~\eqref{eq:0filter}, and is denoted by
\begin{align}
  &\text{$0$-filter:}\nonumber\\&\quad\includegraphics{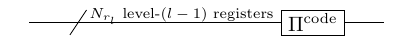}.
\end{align}
The $1$-filter is not illustrated explicitly since it is trivial as shown in~\eqref{eq:1filter}.
The variable $s$ placed in the upper right of a gadget represents the number of faulty locations in the gadget.
An operation depicted using thin lines and acting on thin wires is an ideal operation.
With these notations, every nontrivial condition for a gadget to be fault-tolerant given in Ref.~\cite{G} is translated as follows,
where dashed wires are omitted for simplicity of presentation.
\begin{widetext}
The measurement gadget is fault-tolerant if it satisfies
\begin{align}
  \label{eq:measA}
  &\text{\textbf{meas:} when $s=0$}\nonumber\\&\includegraphics{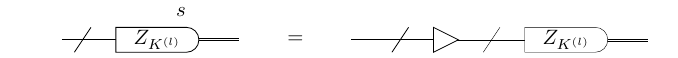},\nonumber\\
  &\quad\text{and when $s=0,1$}\nonumber\\&\includegraphics{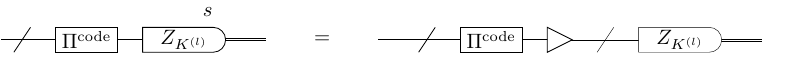}.
\end{align}
The initial-state preparation gadget is fault-tolerant if it satisfies
\begin{align}
  \label{eq:prepA}
  &\text{\textbf{prep A\@:} when $s=0$}\nonumber\\&\includegraphics{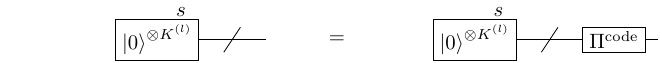},\\
  \label{eq:prepB}
  &\text{\textbf{prep B\@:} when $s=0,1$}\nonumber\\&\includegraphics{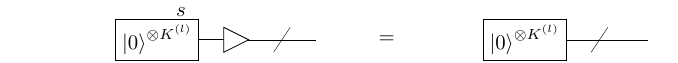}.
\end{align}
The Clifford-state preparation gadget is fault-tolerant if it satisfies
\begin{align}
  \label{eq:prepCA}
  &\text{\textbf{prep A\@:} when $s=0$}\nonumber\\&\includegraphics{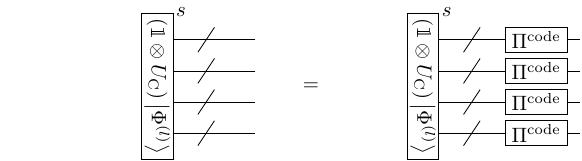},\\
  \label{eq:prepCB}
  &\text{\textbf{prep B\@:} when $s=0,1$}\nonumber\\&\includegraphics{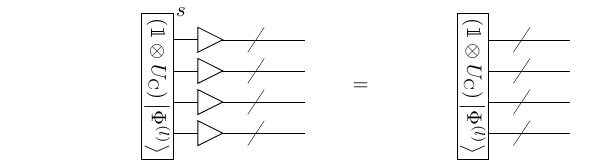}.
\end{align}
The magic-state preparation gadget is fault-tolerant if it satisfies
\begin{align}
  \label{eq:prepmA}
  &\text{\textbf{prep A\@:} when $s=0$}\nonumber\\&\includegraphics{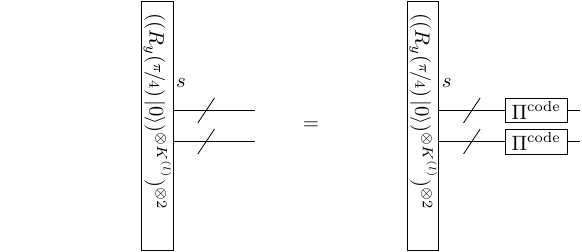},\\
  \label{eq:prepmB}
  &\text{\textbf{prep B\@:} when $s=0,1$}\nonumber\\&\includegraphics{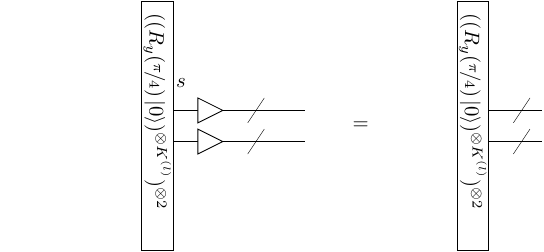}.
\end{align}
The $H$- or Pauli-gate gadget is fault-tolerant if it satisfies
\begin{align}
  \label{eq:gateA}
  &\text{\textbf{gate A\@:} when $s=0$}\nonumber\\&\includegraphics{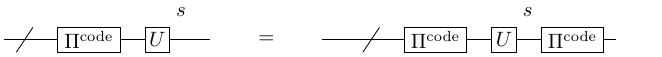},\\
  \label{eq:gateB}
  &\text{\textbf{gate B\@:} when $s=0$}\nonumber\\&\includegraphics{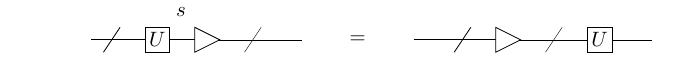},\nonumber\\
  &\quad\text{and when $s=0,1$}\nonumber\\&\includegraphics{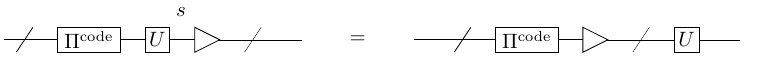},
\end{align}
where $U$ is the logical gate applied by the gadget.
The \textsc{CNOT}- or $CZ$-gate gadget is fault-tolerant if it satisfies
\begin{align}
  \label{eq:gate2A}
  &\text{\textbf{gate A\@:} when $s=0$}\nonumber\\&\includegraphics{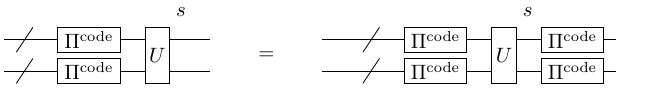},\\
  \label{eq:gate2B}
  &\text{\textbf{gate B\@:} when $s=0$}\nonumber\\&\includegraphics{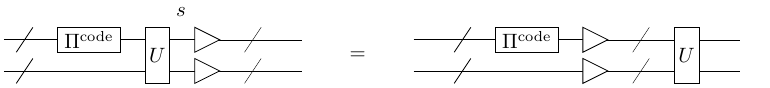},\nonumber\\
  &\includegraphics{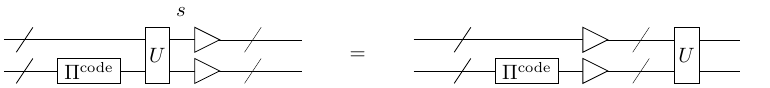},\nonumber\\
  &\quad\text{and when $s=0,1$}\nonumber\\&\includegraphics{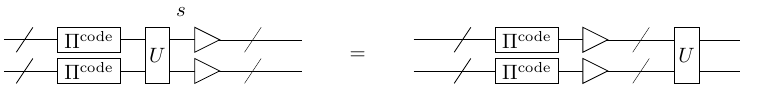},
\end{align}
where $U$ is the logical gate applied by the gadget.
An error-correction gadget is fault-tolerant if it satisfies
\begin{align}
\label{eq:ecA}
  &\text{\textbf{EC A\@:} when $s=0$}\nonumber\\&\includegraphics{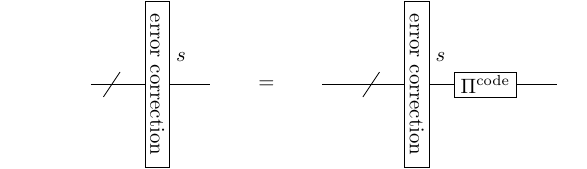},\\
\label{eq:ecB}
  &\text{\textbf{EC B\@:} when $s=0$}\nonumber\\&\includegraphics{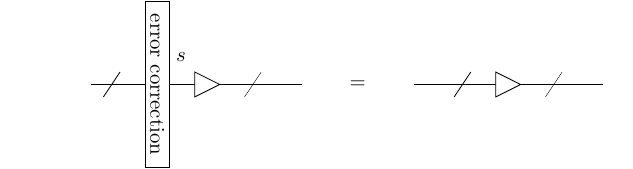},\nonumber\\
  &\quad\text{and when $s=0,1$}\nonumber\\&\includegraphics{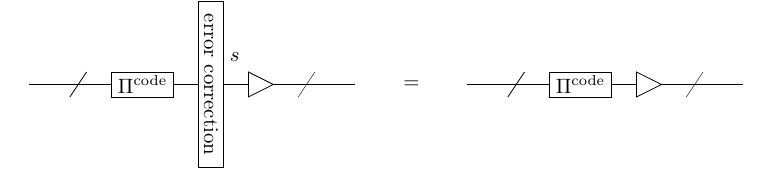}.
\end{align}
These conditions of fault tolerance are used in place of the conditions of fault tolerance in Ref.~\cite{G}, where the labels of the conditions here in the bold text show the corresponding conditions in Ref.~\cite{G} with the same labels.
\end{widetext}

Given that every gadget is fault-tolerant, the existence of a threshold can be proved for the local statistic error model~\cite{G}.
As will be explained in Sec.~\ref{sec:threshold_proof}, it is possible to apply the same argument as that in Ref.~\cite{G} to prove that the level-$l$ circuit at every level $l=0,1,\ldots, L$ follows the local stochastic error model with a bound $p_l>0$ on the logical error rate, i.e., the probability of having a fault at a given level-$l$ location.
To determine the scaling of $p_l$ in Sec.~\ref{sec:threshold_proof}, it is necessary to have a bound on $G(l)$ defined in~\eqref{eq:G_l_def}, i.e., the maximum depth of the level-$(l-1)$ circuit for a level-$l$ gadget, where the maximum is taken over all level-$l$ gadgets including elementary operations and error correction.

In the explanation of each gadget below,
we will check that
\begin{equation}
  \label{eq:G_l}
  G(l)=O(\mathrm{poly}(N_{r_l})).
\end{equation}
From~\eqref{eq:N_r_scaling}, it then follows that
\begin{equation}
  \label{eq:G_l_scaling}
  G(l)=\exp(O(l)).
\end{equation}
The depth of each gadget may include wait operations to wait for classical computation such as those for the decoder and the gate teleportation.
During our protocol, the number of parallel processes for classical computation is limited to
\begin{equation}
  \label{eq:classical_parallel}
  O(\poly(N^{(l)})),
\end{equation}
which will turn out to satisfy the condition~\eqref{eq:parallel_process_number} on the number of parallel processes, as will be shown in Sec.~\ref{sec:overhead}.
We will also show that the depth of the level-$l$ circuit for a level-$l$ abbreviation is upper bounded by
\begin{equation}
  O(\log(N^{(l)})),
\end{equation}
including the wait operations to wait for the nonzero-time classical computation in the gate teleportation.
To obtain these bounds, our analysis in the following uses
\begin{align}
  K^{(l)}&\approx N^{(l)},\\
  K_{r_l}&\approx N_{r_l},
\end{align}
due to~\eqref{eq:rate} and~\eqref{eq:rate_Q_r_l}, and
\begin{align}
  &r_l\approx \log(N_{r_l})=O(l)\\
  &\ll \log(N^{(l)})=O(l^2)\\
  &\ll \poly(N_{r_l})=\exp(O(l))\\
  &\ll \poly(N^{(l)})=\exp(O(l^2))
\end{align}
due to~\eqref{eq:N_r_scaling} and~\eqref{eq:n_L_scaling}.

To bound the space overhead, it is also necessary to have bounds on the number of registers in each level-$l$ gadget.
As designed in Sec.~\ref{sec:gadget}, in our protocol, the number of level-$(l-1)$ registers used in a level-$l$ gadget is to be bounded by
\begin{equation}
  \label{eq:qubit_requirement}
  N_{r_l}+8+2=N_{r_l}+O(1)
\end{equation}
per encoded level-$l$ register.
The number of bits used for each level-$l$ gadget is to be bounded by
\begin{equation}
  \label{eq:classical_bit_requirement}
  O(\mathrm{poly}(N^{(l)})).
\end{equation}

\textbf{Two-register Clifford-gate abbreviation}:
The level-$l$ two-register Clifford-gate abbreviation based on gate teleportation~\cite{gottesmanchuang1999,K5,K6} is given by Fig.~\ref{fig:inblock_clifford}, where the gate teleportation is also presented.
The implementation is assisted by auxiliary level-$l$ registers in a state $(\mathbbm{1}^{B_1B_2}\otimes U_\mathrm{C}^{B_3B_4})\ket{\Phi^{(l)}}^{B_1B_2B_3B_4}$,
which is prepared by the Clifford-state preparation operation in~\eqref{eq:clifford_state_preparation}.
Given the $4K^{(l)}$-bit measurement outcome,
correction in quantum teleportation~\cite{PhysRevLett.70.1895} is given by an operator in the form of
\begin{equation}
  \label{eq:quantum_teleportation_correction_clifford}
  U_\mathrm{C}\left(\bigotimes_{k^{(l)}=1}^{K^{(l)}}P_{k^{(l)}}^{B_3}\otimes\bigotimes_{k^{(l)}=1}^{K^{(l)}}P_{k^{(l)}}^{B_4}\right)U_\mathrm{C}^\dag,
\end{equation}
where $P_{k^{(l)}}^{B_j}\in\{X,Z,Y,\mathbbm{1}\}$ for each $j\in\{3,4\}$ and $k^{(l-1)}\in\{1,\ldots,K^{(l-1)}\}$ is a Pauli gate acting on the $k^{(l)}$th qubit of the level-$l$ register $B_j$ with labeling~\eqref{eq:k_Q_l}.
Since $U_\mathrm{C}$ is Clifford,
the operator in~\eqref{eq:quantum_teleportation_correction_clifford} is a tensor product of single-qubit Pauli operators.
Conditioned on the measurement outcome, the tensor product of Pauli operators~\eqref{eq:quantum_teleportation_correction_clifford} acting on the level-$l$ registers is calculated using an efficient decoder and applied as the correction of the quantum teleportation by the Pauli-gate operation in~\eqref{eq:pauli_gate}~\cite{K5,K6}.
The $4K^{(l)}$-bit measurement outcome yields a $4K^{(l)}$-dimensional binary row vector representing the $2K^{(l)}$-qubit Pauli operator $\bigotimes_{k^{(l)}=1}^{K^{(l)}}P_{k^{(l)}}^{B_3}\otimes\bigotimes_{k^{(l)}=1}^{K^{(l)}}P_{k^{(l)}}^{B_4}$ in~\eqref{eq:quantum_teleportation_correction_clifford}, and we can compute its conjugation~\eqref{eq:quantum_teleportation_correction_clifford} by the Clifford unitary $U_\mathrm{C}$ via multiplication of a $4K^{(l)}\times 4K^{(l)}$ binary matrix representing $U_\mathrm{C}$ (from the right of the row vector), as shown in Ref.~\cite{PhysRevA.70.052328}.
The resulting $4K^{(l)}$-dimensional binary row vector of~\eqref{eq:quantum_teleportation_correction_clifford} is used as the classical argument in invoking the Pauli-gate operation for the correction.

The level-$l$ circuit of the abbreviation has the depth bounded by
\begin{equation}
  \label{eq:runtime_clifford}
  O(\log(N^{(l)})).
\end{equation}
The depth is dominated by the wait operations to wait for classical computation to obtain the tensor product of Pauli operators~\eqref{eq:quantum_teleportation_correction_clifford} since the other part has a constant depth.
Regarding the runtime of classical computation, using $O(N^{(l)2})$ parallel processes, we can perform the multiplication of the $4K^{(l)}\times 4K^{(l)}$ binary matrix for the conjugation~\eqref{eq:quantum_teleportation_correction_clifford} within runtime $O(\log(K^{(l)}))=O(\log(N^{(l)}))$, which leads to~\eqref{eq:runtime_clifford}.
The required number of bits is $O(K^{(l)2})=O(N^{(l)2})$, which is dominated by those for storing the $4K^{(l)}\times 4K^{(l)}$ matrix representing $U_\mathrm{C}$.

As shown in~\eqref{eq:argument_clifford}, the level-$l$ two-register Clifford-gate abbreviation can be invoked with the argument of the classical description of $U_\mathrm{C}$, and we here describe its on-demand function.
In the on-demand case, the argument is passed to the level-$l$ Clifford-state preparation operation used in the abbreviation.
Since we represent $U_\mathrm{C}$ in~\eqref{eq:argument_clifford} as $K^{(l)}$ $4\times 4$ binary matrices,
we can compute the conjugation~\eqref{eq:quantum_teleportation_correction_clifford} within $O(1)$ runtime, using $O(N^{(l)})$ parallel processes of multiplying these $K^{(l)}=O(N^{(l)})$ constant-size binary matrices in parallel.
Therefore, the part of the level-$l$ circuit relevant to processing the argument has the depth $O(1)$,
using $O(K^{(l)})=O(N^{(l)})$ bits.
As a whole,~\eqref{eq:runtime_clifford} is an upper bound of the depth even in the on-demand case.

\textbf{$R_y(\pm\nicefrac{\pi}{4})$-gate abbreviation}:
The level-$l$ $R_y(\pm\nicefrac{\pi}{4})$-gate abbreviation based on gate teleportation is given by Fig.~\ref{fig:interblock_T}, where the gate teleportation is also presented.
The implementation is assisted by two types of auxiliary level-$l$ registers $A_1$ and $A_2$.
The register $A_1$ works like the auxiliary qubit in the gate teleportation of Fig.~\ref{fig:interblock_T} and is initially prepared in a magic state ${(R_y(\nicefrac{\pi}{4})\ket{0})}^{\otimes K^{(l)}}$ by the magic-state preparation operation in~\eqref{eq:magic_state_preparation}.
Note that, since the magic-state preparation operation prepares two registers in ${({(R_y(\nicefrac{\pi}{4})\ket{0})}^{\otimes K^{(l)}})}^{\otimes 2}$, the circuit in Fig.~\ref{fig:interblock_T} discards one of the prepared magic states for simplicity of presentation; however, we may use the magic state discarded in Fig.~\ref{fig:interblock_T} as that of $A_1$ for another $R_y(\pm\nicefrac{\pi}{4})$-gate abbreviation to improve the efficiency.
The other register $A_2$ is used for applying the $R_y(\nicefrac{\pi}{4})$ gates only to the selected qubits and is initially prepared in a state ${\ket{0}}^{\otimes K^{(l)}}$ by the initial-state preparation operation in~\eqref{eq:initial_state_preparation}.
For each qubit on which $U_{R_y(\pm\nicefrac{\pi}{4})}$ acts trivially as $\mathbbm{1}$,
a single-qubit \textsc{SWAP} gate is applied between $A_1$ and $A_2$, so as to avoid the subsequent application of the $R_y(\nicefrac{\pi}{4})$ gate.
These \textsc{SWAP} gates are implemented collectively by a single use of a two-register Clifford-gate abbreviation in Fig.~\ref{fig:inblock_clifford}.
For example, to implement $U_{R_y(\pm\nicefrac{\pi}{4})}={(R_y(\pm\nicefrac{\pi}{4}))}^{\otimes K^{(l)}}$, no \textsc{SWAP} gate is applied, and to implement $U_{R_y(\pm\nicefrac{\pi}{4})}={R_y(\pm\nicefrac{\pi}{4})}^{\otimes (K^{(l)}-1)}\otimes\mathbbm{1}$, a \textsc{SWAP} gate is applied between a pair of the qubits.
Then the gate teleportation to apply ${R_y(\nicefrac{\pi}{4})}$ is performed by the circuit at the top of Fig~\ref{fig:interblock_T}.
Conditioned on the measurement outcome, Clifford gates for the correction, i.e., $HZ$ gates, are applied.
At this point, to the qubits on which $U_{R_y(\pm\nicefrac{\pi}{4})}$ dictates to apply $R_y(-\nicefrac{\pi}{4})$, we have applied $R_y(\nicefrac{\pi}{4})$ gates.
Thus, an extra Clifford gate $R_y(-\nicefrac{\pi}{2})=ZH$ is applied to each of these qubits in the end.
Each of the corrections of $HZ$ gates and the application of extra $R_y(-\nicefrac{\pi}{2})$ gates are implemented by a two-register Clifford-gate abbreviation in Fig.~\ref{fig:inblock_clifford}.
These gates may act nontrivially only on a single register, but for each $k^{(l)}\in\{1,\ldots, K^{(l)}\}$, we here represent a $HZ$ and $R_y(-\nicefrac{\pi}{2})$ (or $\mathbbm{1}$) gate for the correction acting on the $k^{(l)}$th qubit in the level-$l$ register as a two-qubit Clifford unitary $HZ\otimes\mathbbm{1}$ and $R_y(-\nicefrac{\pi}{2})\otimes\mathbbm{1}$ (or $\mathbbm{1}\otimes\mathbbm{1}$) acting trivially on the $k^{(l)}$th qubit of another level-$l$ register as well (i.e., $A_2$ in Fig.~\ref{fig:interblock_T}).
The correction of $HZ$ gates is applied by the on-demand two-register Clifford gate abbreviation with~\eqref{eq:argument_clifford}.

The level-$l$ circuit for the implementation has the depth
\begin{equation}
  \label{eq:runtime_magic}
  O(\log(N^{(l)})).
\end{equation}
The depth is dominated by the two-register Clifford-gate abbreviations since the other part has a constant depth.
Using the $K^{(l)}$-bit outcome of the measurement, we perform the classical computation to decide the correction by $HZ$ gates as shown in Fig.~\ref{fig:interblock_T}.
Performing this classical computation for all $k^{(l)}$, we obtain the $K^{(l)}$ $4\times 4$ matrices as in~\eqref{eq:argument_clifford}, which we use as the argument to invoke the two-register Clifford-gate abbreviation for performing the correction.
Using $K^{(l)}=O(N^{(l)})$ parallel processes, the runtime of this computation is $O(1)$, where, from each of the $K^{(l)}$ bits of the measurement outcome, one of the $K^{(l)}$ processes computes each $4\times 4$ (i.e., constant-size) matrix.
Apart from the classical computation, the depth of each two-register Clifford-gate abbreviation is upper bounded by $O(\log(N^{(l)}))$ as shown in~\eqref{eq:runtime_clifford}, which is dominant.
The required number of bits is $O(K^{(l)})=O(N^{(l)})$, which is dominated by those for storing the $K^{(l)}$ $4\times 4$ matrices used for the two-register Clifford gates.

\textbf{Measurement gadget}:
The level-$l$ measurement gadget is constructed as in Fig.~\ref{fig:measurement}.
The gadget is implemented by transversally performing level-$(l-1)$ measurement operations, followed by an efficient decoder.
The fault tolerance, i.e.,~\eqref{eq:measA}, follows from the transversality.

We here provide the construction of the efficient decoder for the quantum Hamming codes to analyze its runtime.
In particular, we provide a protocol for calculating the measurement outcomes for $K^{(l)}$ qubits in a level-$l$ register from those in the $N_{r_l}$ level-$(l-1)$ registers, where the outcomes at the physical level ($l-1=0$) are those of the measurement in the $Z$ basis of the physical qubits.
Let $o_{n,k^{(l-1)}}\in\{0,1\}$ be the measurement outcome of the qubit $(n,k^{(l-1)})$ with~\eqref{eq:n_Q_r_l_k_Q_l_1} in the level-$(l-1)$ registers ($n\in\{1,\ldots,N_{r_l}\}$ and $k^{(l-1)}\in\{1,\ldots,K^{(l-1)}\}$), which is determined by the level-$(l-1)$ measurement operations.
For each $k^{(l-1)}$, we estimate the true outcome by the following procedure.
We calculate the syndrome $e_{i,k^{(l-1)}}\in\{0,1\}$ for each $i\in\{1,\ldots,r_l\}$ by
\begin{equation}
  \label{eq:error_syndrome}
  e_{i,k^{(l-1)}}=\sum_{n=1}^{N_{r_l}}o_{n,k^{(l-1)}}b_{i,n}\mod 2,
\end{equation}
where $b_{i,n}$ is the element of the parity-check matrix used in~\eqref{eq:Z_generator} for $\Q_{r_l}$.
This calculation provides the label of the presumably erroneous qubit written as
\begin{equation}
  \label{eq:error_qubit}
  \tilde{n}_{k^{(l-1)}}\coloneqq\sum_{i=1}^{r_l}e_{i,k^{(l-1)}}2^{i-1},
\end{equation}
which is the bit string stored in the memory after the calculation of~\eqref{eq:error_syndrome}.
If $\tilde{n}_{k^{(l-1)}}=0$, then we estimate that $o_{1,k^{(l-1)}},\ldots,o_{N_{r_l},k^{(l-1)}}$ have no error;
otherwise, we estimate that the qubit $(\tilde{n}_{k^{(l-1)}},k^{(l-1)})$ has an error.
To calculate the measurement outcome for the level-$l$ register,
we use the labeling~\eqref{eq:k_Q_l}, i.e., $k^{(l)}\mapsto(k,k^{(l-1)})$, for the $k^{(l)}$th qubit in the level-$l$ register, where $k^{(l)}\in\{1,\ldots,K^{(l)}\}$ and $k\in\{1,\ldots,K_{r_l}\}$.
\begin{widetext}
Under this labeling $k^{(l)}\mapsto(k,k^{(l-1)})$,
for each $k^{(l)}$, we calculate the measurement outcome $o_{k^{(l)}}$ of the $k^{(l)}$th qubit in the level-$l$ register from those of $(n,k^{(l-1)})$ in the level-$(l-1)$ registers as
\begin{align}
  \label{eq:outcome}
  o_{k^{(l)}}=\begin{cases}
    \sum_{n=1}^{N_{r_l}}o_{n,k^{(l-1)}}b_n^{(k)}&\mod 2,~\text{if $\tilde{n}_{k^{(l-1)}}=0$},\\
    \sum_{n=1}^{N_{r_l}}o_{n,k^{(l-1)}}b_n^{(k)}+b_{\tilde{n}_{k^{(l-1)}}}^{(k)}&\mod 2,~\text{otherwise},
             \end{cases}
\end{align}
where $b_n^{(k)}$ is the element of the bit string in~\eqref{eq:kth_logical} for representing the $k$th logical qubit of $\Q_{r_l}$.
Then, the decoder outputs the level-$l$ measurement outcomes $o_{k^{(l)}}$ while all the level-$(l-1)$ measurement outcomes $o_{n,k^{(l-1)}}$ are deallocated.
\end{widetext}

The gadget has a constant depth
\begin{equation}
  O(1),
\end{equation}
which is followed by classical computation for the decoder to calculate the measurement outcome for the encoded level-$l$ register.
We also bound the runtime of this classical computation in the following.
To calculate the measurement outcomes with the error correction,
we use $r_l K^{(l-1)}\times N_{r_l}$ processes of classical computation in parallel to obtain the syndromes $e_{i,k^{(l-1)}}$ in~\eqref{eq:error_syndrome} for all $i\in\{1,\ldots,r_l\}$ and $k^{(l-1)}\in\{1,\ldots,K^{(l-1)}\}$.
We divide these $r_l K^{(l-1)}\times N_{r_l}$ processes into $r_l K^{(l-1)}$ sets of $N_{r_l}$ processes.
Using each set, we calculate $e_{i,k^{(l-1)}}$ by parallel classical computation with the $N_{r_l}$ processes in the set.
The runtime of the sum in~\eqref{eq:error_syndrome} for this calculation is $O(\log(N_{r_l}))$.
Therefore, using at most $r_l K^{(l-1)}\times N_{r_l}=O(N^{(l)}\log(N_{r_l}))$ processes (where we use $K^{(l-1)}N_{r_l}=O(N^{(l-1)}N_{r_l}=N^{(l)})$), we can obtain $\tilde{n}_{k^{(l-1)}}$ in~\eqref{eq:error_qubit} in the memory within runtime
$O(\log(N_{r_l}))$.
In addition, we use $K^{(l)}\times N_{r_l}$ processes of classical computation in parallel to obtain the measurement outcomes $o_{k^{(l)}}$ in~\eqref{eq:outcome} for all $k^{(l)}$.
We divide these $K^{(l)}\times N_{r_l}$ processes into $K^{(l)}$ sets of $N_{r_l}$ processes.
Using each set, we calculate the sum over $n$ and the error correction of $b_{\tilde{n}_{k^{(l-1)}}}^{(k)}$ in~\eqref{eq:outcome} by parallel classical computation with the $N_{r_l}$ processes in the set, which can be performed within the runtime of $O(\log(N_{r_l}))$.
As a whole, using at most $K^{(l)}\times N_{r_l}=O(N^{(l)}N_{r_l})$ processes, we can obtain $o_{k^{(l)}}$ within runtime
\begin{equation}
\label{eq:decoding_runtime}
  O(\log(N_{r_l})).
\end{equation}
The classical computation uses $O(N^{(l)})$ bits, which are dominated by those for storing the measurement outcomes for all the $O(N^{(l)})$ qubits in the registers.

To use the measurement outcome in the other level-$l$ gadgets,
it is sufficient to guarantee that we finish calculating the measurement outcome always before starting the next level-$l$ gadget.
Indeed, our protocol performs level-$l$ gadgets in a synchronized way by inserting wait operations as shown in Fig.~\ref{fig:circuit_conversion}.
Since the runtime~\eqref{eq:decoding_runtime} of the decoder is smaller than the maximum depth $G(l)$ of the level-$l$ gadgets in~\eqref{eq:G_l_scaling},
the measurement outcome is always available for the next level-$l$ gadget.

\textbf{Gate gadgets}:
The level-$l$ $H$-, $\textsc{CNOT}$-, $CZ$- and Pauli-gate gadgets are constructed as in Fig.~\ref{fig:transversal_gate}.
The level-$l$ $H$-, $\textsc{CNOT}$-, and $CZ$-gate gadgets are implemented by transversal level-$(l-1)$ $H$-, $\textsc{CNOT}$-, and $CZ$-gate operations, respectively, on all the level-$(l-1)$ registers.
To implement the level-$l$ Pauli-gate gadget, each level-$l$ Pauli operator $P_{k^{(l)}}$ in~\eqref{eq:pauli_l} is represented as a logical operator in the form of a tensor product of Pauli operators acting on the $N_{r_l}$ level-$(l-1)$ registers, and the product of these logical operators for all $P_{k^{(l)}}$ becomes a tensor product of the Pauli operators on the $N_{r_l}$ level-$(l-1)$ registers, i.e.,
\begin{equation}
  \label{eq:pauli_l_1}
  \bigotimes_{n=1}^{N_{r_l}}\bigotimes_{k^{(l-1)}=1}^{K^{(l-1)}}P_{n,k^{(l-1)}},
\end{equation}
where $P_{n,k^{(l-1)}}\in\{X,Z,Y,\mathbbm{1}\}$ is a Pauli gate acting on the $k^{(l-1)}$th qubit in the $n$th level-$(l-1)$ register, and the global phase is ignored;
then, the operator in~\eqref{eq:pauli_l_1} is applied using the level-$(l-1)$ Pauli-gate operations transversally for all the $N_{r_l}$ level-$(l-1)$ registers.
The fault tolerance, i.e.,~\eqref{eq:gateA},~\eqref{eq:gateB},~\eqref{eq:gate2A}, and~\eqref{eq:gate2B}, follows from the transversality.

The level-$l$ $H$-, $\textsc{CNOT}$-, $CZ$-gate gadgets have a constant depth
\begin{equation}
  O(1).
\end{equation}
By contrast, for the level-$l$ Pauli-gate gadget, we need to obtain the Pauli operators~\eqref{eq:pauli_l_1} from the $K^{(l)}$ Pauli operators $P_{k^{(l)}}$ by classical computation.
If we can perform this classical computation during the compilation, the Pauli-gate gadget has the constant depth
\begin{equation}
  O(1).
\end{equation}
On the other hand, as stated in~\eqref{eq:argument_pauli}, there are cases where we need to compute~\eqref{eq:pauli_l_1} on the fly while inserting wait operations in the level-$l$ Pauli-gate gadget to wait for the runtime of this classical computation, which we bound in the following.
As in the two-register Clifford-gate abbreviation in Fig.~\ref{fig:inblock_clifford}, we represent the input $K^{(l)}$ Pauli operators as the $2K^{(l)}$-dimensional binary row vector.
Since $\mathcal{Q}^{(l)}$ has $K^{(l-1)}$ blocks of $\mathcal{Q}_{r_l}$, we can compute the corresponding logical operator~\eqref{eq:pauli_l_1} by multiplying $2K^{(l-1)}$ copies of the $K_{r_l}\times N_{r_l}$ matrices $G_{r_l}$ in~\eqref{eq:generator_matrix} (from the right of the corresponding $K_{r_l}$-bit segment of this vector).
Thus, using $2K^{(l-1)}\times (K_{r_l}\times N_{r_l})=O(K^{(l)}N_{r_l})=O(N^{(l)}N_{r_l})$ parallel processes, we can perform this matrix multiplication in time $O(\log(N_{r_l}))$.
Therefore, the depth of the Pauli-gate gadget including the wait operations is
\begin{equation}
  \label{eq:pauli_depth}
  O(\log(N_{r_l})).
\end{equation}
The required number of bits is $O(N^{(l)})$, dominated by those for storing the $2K^{(l)}$-dimensional vectors.

\textbf{Initial-state preparation gadget}:
The level-$l$ initial-state preparation gadget is constructed as in Fig.~\ref{fig:preparation_0}.
The initial part of the gadget is aimed at implementing an encoding unitary $U_\mathrm{encode}$ to transform the initialized $N_{r_l}$ level-$(l-1)$ registers into an encoded logical state of $\Ket{0}^{\otimes K^{(l)}}$.
Under the decomposition~\eqref{eq:H} with the $K^{(l-1)}$ code blocks of $\Q_{r_l}$,
the unitary is written in the form
\begin{equation}
  \label{eq:encode}
  U_\mathrm{encode} = V^{\otimes K^{(l-1)}},
\end{equation}
where $V$ is an encoding unitary for the code $\Q_{r_l}$ satisfying
\begin{equation}
  V (\ket{\psi}\otimes \ket{0}^{\otimes (N_{r_l}-K_{r_l})})= \ket{\psi_{\mathrm{logical}}}.
\end{equation}
Here, $\ket{\psi}$ on the left-hand side is any $K_{r_l}$-qubit state, and $\ket{\psi_{\mathrm{logical}}}$ is the state of $N_{r_l}$ qubits representing the logical state $\ket{\psi}$.
In the circuit of Fig.~\ref{fig:preparation_0}, $N_{r_l}$ level-$(l-1)$ registers are initially prepared in $\Ket{0}^{\otimes K^{(l-1)}}$ by the level-$(l-1)$ initial-state preparation operations in~\eqref{eq:initial_state_preparation}, which $U_\mathrm{encode}$ transforms into logical $\Ket{0}^{\otimes K^{(l)}}$.

For any stabilizer code, e.g., the code used here, the unitary $U_\mathrm{encode}$ for encoding is Clifford; in particular, to implement $U_\mathrm{encode}$, the analysis here uses the fact shown in Ref.~\cite{PhysRevA.56.76} that an encoding unitary $V$ for the stabilizer code with $N_{r_l}$ physical qubits and $K_{r_l}$ logical qubits can be implemented by an $N_{r_l}$-qubit stabilizer circuit using $O(N_{r_l}(N_{r_l}-K_{r_l}))=O(N_{r_l}\log(N_{r_l}))$ one- and two-qubit Clifford gates, which lead to the depth at most $O(N_{r_l}\log(N_{r_l}))$.
Then a circuit for $U_\mathrm{encode}$ is constructed by replacing each gate $G$ in the stabilizer circuit for $V$ with $G^{\otimes K^{(l-1)}}$ according to~\eqref{eq:encode}.
The stabilizer circuit for $V$ constructed by the technique in Ref.~\cite{PhysRevA.56.76} is composed only of $H$, \textsc{CNOT}, $CZ$, and $Z$ (i.e., Pauli) gates.
Thus, in the stabilizer circuit, we use the level-$(l-1)$ gate operations corresponding to these gates; that is, we do not need to use the two-register Clifford-gate abbreviation in Fig.~\ref{fig:inblock_clifford}.
As a result, $U_\mathrm{encode}$ is implemented by a level-$(l-1)$ circuit with depth $O(N_{r_l}\log(N_{r_l}))$.

The level-$(l-1)$ stabilizer circuit for the encoding by itself is non-fault-tolerant in the sense that one fault among the level-$(l-1)$ locations of the circuit may lead to an uncorrectable error at the end of the circuit;
to remedy this, verification is performed by measuring the logical $Z$ operators of all the logical qubits, and all the $Z$ generators of the stabilizer, for each of the $K^{(l-1)}$ code blocks of $\Q_{r_l}$.
To measure these operators without destroying the state itself, another set of $N_{r_l}$ level-$(l-1)$ registers are prepared in logical $\Ket{0}^{\otimes K^{(l)}}$ by the same non-fault-tolerant level-$(l-1)$ stabilizer circuit, and all the syndromes are extracted at once by the constant-depth \textsc{CNOT}-gate operations in the level-$(l-1)$ circuit of Fig.~\ref{fig:preparation_0}.
We obtain each syndrome from the measurement outcomes by classical computation using the description of the logical $Z$ operators and the $Z$ stabilizer generators.

If no error is detected from these syndromes, i.e., in the case of success in the verification, then
the gadget outputs the logical $\Ket{0}^{\otimes K^{(l)}}$ prepared in this first run.
Otherwise, i.e., if the verification in this first run detects an error,
then the logical $\Ket{0}^{\otimes K^{(l)}}$ prepared in this first run is discarded.
This ensures that a fault in a single location leading to incorrigible multiple $X$ errors is always detected and discarded.
Note that since $\Ket{0}$ is stabilized by $Z$, i.e., $Z\Ket{0}=\Ket{0}$, detection of $Z$ errors is unnecessary for verifying $\Ket{0}$; that is, multiple $Z$ errors before the ideal decoder in~\eqref{eq:prepB} may lead to a logical $Z$ error, but do not change the decoded state.
If the first run has discarded the state, the gadget performs the same non-fault-tolerant level-$(l-1)$ stabilizer circuit for the preparation and outputs the logical $\Ket{0}^{\otimes K^{(l)}}$ prepared in the second run without verification.
In this case, a single fault should be present in the part of the first run, and thus the part of the preparation in the second run has no fault as long as the gadget has at most one faulty location.
Therefore, in any of these two cases, the gadget satisfies the conditions~\eqref{eq:prepA} and~\eqref{eq:prepB} of fault tolerance.

The depth of the level-$(l-1)$ circuit for implementing the initial-state preparation gadget is dominated by the encoding part for implementing $U_\mathrm{encode}$.
The depth of the gadget including both the first and second runs is at most twice as long as the first run, and thus in the following, we analyze the depth of the first run.
The required number of $H$-, \textsc{CNOT}-, $CZ$-, and Pauli-gate operations in the encoding part is $O\left(N_{r_l}\log(N_{r_l})\right)$,
which is an upper bound of the depth of this part.
As for the other parts, the $O(N_{r_l})$ preparations of $\Ket{0}^{\otimes K^{(l-1)}}$ in the gadget have the depth $O(N_{r_l})$.
Since $\mathcal{Q}^{(l)}$ has $K^{(l-1)}$ blocks of $\mathcal{Q}_{r_l}$,
the classical computation for the post-selection computes the inner product between the $(N_{r_l}\times K^{(l-1)})$-bit measurement outcome and each row of $K^{(l-1)}$ copies of $H_{r_l}$ and $G_{r_l}$ in~\eqref{eq:parity_check_matrix} and~\eqref{eq:generator_matrix}, respectively.
Here, $H_{r_l}$ has $r_l$ rows, and $G_{r_l}$ has $K_{r_l}$ rows.
Thus, using $(r_l+K_{r_l})\times (N_{r_l}\times K^{(l-1)})=O(N_{r_l}N^{(l)})$ parallel processes (where we use $K_{r_l}K^{(l-1)}=K^{(l)}=O(N^{(l)})$), we can perform all the inner products in runtime $O(\log(N_{r_l}))$.
At this point, we have the $(r_l+K_{r_l})\times K^{(l-1)}$-bit string representing these inner products.
By taking bitwise \textsc{OR} over this bit string by parallel processes, we check whether the bit string includes $1$ within the runtime of $O(\log((r_l+K_{r_l})\times K^{(l-1)}))=O(\log(N^{(l)}))$.
Consequently, the depth of the initial-state preparation gadget is
\begin{equation}
  \label{eq:depth_encoding}
  O\left(N_{r_l}\log(N_{r_l})\right).
\end{equation}
The required number of bits is $O(N_{r_l}\times K^{(l-1)})=O(N^{(l)})$, dominated by those for storing and processing the $(N_{r_l}\times K^{(l-1)})$-bit measurement outcome.

Note that in place of the stabilizer circuit for the encoding unitary $V$ in Ref.~\cite{PhysRevA.56.76}, Refs.~\cite{10.5555/2481569.2481579,Steane_2002} provide more optimized stabilizer circuits for preparing logical $\Ket{0}$, which could also be used here.
Moreover, the depth of a stabilizer circuit could be shortened by parallelizing the circuit using auxiliary qubits~\cite{10.1137/S0097539799355053,10.5555/3381089.3381102}, but such further optimization is not considered here.
Our analysis of the existence of a threshold and the space and time overheads holds without such an optimization.

\textbf{Error-correction gadget}:
The level-$l$ error-correction gadget is constructed as in Fig.~\ref{fig:error_correction} based on Knill's error correction~\cite{K5,K6}.
The gadget is assisted by $2(N_{r_l}+1)$ level-$(l-1)$ registers.
Using the same circuit as the initial-state preparation gadget in Fig.~\ref{fig:preparation_0} on these $2(N_{r_l}+1)$ level-$(l-1)$ registers,
we prepare $N_{r_l}$ out of these $2(N_{r_l}+1)$ in logical $\ket{0}^{\otimes K^{(l)}}$.
Using this logical $\ket{0}^{\otimes K^{(l)}}$,
the circuit in Fig.~\ref{fig:error_correction} conducts quantum teleportation of the state of the encoded logical level-$l$ register.
The $2K^{(l)}$-bit outcome from the measurement gadgets is fed to an efficient decoder to calculate a logical operator used as correction in quantum teleportation~\cite{PhysRevLett.70.1895}
\begin{equation}
  \label{eq:quantum_teleportation_correction}
 \bigotimes_{k^{(l)}=1}^{K^{(l)}}P_{k^{(l)}},
\end{equation}
where $P_{k^{(l)}}\in\{X,Z,Y,\mathbbm{1}\}$ is a Pauli gate acting on the $k^{(l)}$th qubit in the level-$l$ register with labeling~\eqref{eq:k_Q_l}.
As in the level-$l$ measurement gadget, the measurement in this quantum teleportation needs wait operations to wait for the $O(\log(N_{r_l}))$ runtime of classical computation.
This logical operator is implemented by invoking the level-$l$ Pauli-gate gadget in Fig.~\ref{fig:transversal_gate} with the argument.
With no fault, this gadget carries out error correction whereas the fault tolerance follows from the transversality, satisfying~\eqref{eq:ecA} and~\eqref{eq:ecB}.

The depth of the level-$(l-1)$ circuit is
\begin{equation}
  \label{eq:depth_error_correction}
  O(N_{r_l}\log(N_{r_l})),
\end{equation}
dominated by the depth~\eqref{eq:depth_encoding} of the level-$l$ initial-state preparation gadgets.
As for the other parts, the runtime of classical computation for decoders from the measurement outcomes is $O(\log(N_{r_l}))$ as discussed above, and that for the level-$l$ Pauli-gate gadget is also $O(\log(N_{r_l}))$.
All the other parts have a constant depth.
The required number of bits is dominated by $O(N^{(l)})$ to store and process the measurement outcome.

\textbf{Clifford-state preparation gadget}:
The level-$l$ Clifford-state preparation gadget is constructed as in Figs.~\ref{fig:preparation_clifford} and~\ref{fig:verification_clifford}.
The gadget starts with a level-$(l-1)$ stabilizer circuit for preparing logical $(\mathbbm{1}^{B_1B_2}\otimes U_\mathrm{C}^{B_3B_4})\ket{\Phi^{(l)}}^{B_1B_2B_3B_4}$, where $\ket{\Phi^{(l)}}^{B_1B_2B_3B_4}$ is the maximally entangled state between $B_1B_2$ and $B_3B_4$ given by~\eqref{eq:phi}.
Each of the level-$l$ registers $B_1,B_2,B_3,B_4$ is encoded in $N_{r_l}$ level-$(l-1)$ registers in the level-$(l-1)$ circuit, and these $N_{r_l}$ level-$(l-1)$ registers will also be collectively called $B_1,B_2,B_3,B_4$, respectively.
After preparing logical $\ket{\Phi^{(l)}}^{B_1B_2B_3B_4}$, application of logical $U_\mathrm{C}^{B_3B_4}$ is implemented by using the level-$(l-1)$ elementary operations in the following way.
Using the Clifford unitary $U_\mathrm{encode}$ in~\eqref{eq:encode}, logical $U_\mathrm{C}$ can be implemented by a Clifford unitary
\begin{align}
  \label{eq:Clifford}
  &(U_\mathrm{encode}^{B_3}\otimes U_\mathrm{encode}^{B_4})\times\nonumber\\
  &(U_\mathrm{C}\otimes\mathbbm{1}^{\otimes 2(N_{r_l}K^{(l-1)}-K^{(l)})})\times\nonumber\\
  &{(U_\mathrm{encode}^{B_3}\otimes U_\mathrm{encode}^{B_4})}^\dag,
\end{align}
where $U_\mathrm{C}$ acts on the $2K^{(l)}$ qubits that hold the decoded state of $B_3$ and $B_4$, and $\mathbbm{1}^{\otimes 2(N_{r_l}K^{(l-1)}-K^{(l)})}$ is the identity operator acting on the remaining $2(N_{r_l}-K_{r_l})K^{(l-1)}=2(N_{r_l}K^{(l-1)}-K^{(l)})$ qubits in the $2N_{r_l}$ level-$(l-1)$ registers for $B_3$ and $B_4$.
We decompose $U_\mathrm{encode}$ into elementary operations as discussed for the initial-state preparation gadget in Fig.~\ref{fig:preparation_0} and also decompose $U_\mathrm{C}$ into elementary operations, which we will discuss later when we evaluate the depth of the gadget.

Similar to the encoding in the initial-state preparation gadget of Fig.~\ref{fig:preparation_0}, the level-$(l-1)$ stabilizer circuit for preparing logical $(\mathbbm{1}^{B_1B_2}\otimes U_\mathrm{C}^{B_3B_4})\ket{\Phi^{(l)}}^{B_1B_2B_3B_4}$ by itself is non-fault-tolerant in the sense that one fault among level-$(l-1)$ locations in the circuit may lead to an uncorrectable error at the end of the circuit, which necessitates a part for performing verification.
The verification part starts with applying the level-$l$ error-correction gadgets in Fig.~\ref{fig:error_correction} to $B_1$, $B_2$, $B_3$, and $B_4$, so as to ensure that the state is in the code space of $\Q_{r_l}$ regardless of faults in the preparation part.
The circuit is followed by measurements of the logical stabilizer operators of logical $(\mathbbm{1}\otimes U_\mathrm{C})\ket{\Phi^{(l)}}$ to detect logical errors with post-selection.
There are $4K^{(l)}$ stabilizer operators of $(\mathbbm{1}\otimes U_\mathrm{C})\ket{\Phi^{(l)}}$, i.e.,
\begin{align}
  \label{eq:xx}
  &{(X_{j,k^{(l)}}\otimes\mathbbm{1}^{\otimes (2K^{(l)}-1)})}^{B_1 B_2}\otimes\nonumber\\
  &\quad{(U_\mathrm{C}(X_{j+2,k^{(l)}}\otimes\mathbbm{1}^{\otimes (2K^{(l)}-1)})U_\mathrm{C}^\dag)}^{B_3 B_4},\\
  \label{eq:zz}
  &{(Z_{j,k^{(l)}}\otimes\mathbbm{1}^{\otimes (2K^{(l)}-1)})}^{B_1B_2}\otimes\nonumber\\
  &\quad{(U_\mathrm{C}(Z_{j+2,k^{(l)}}\otimes\mathbbm{1}^{\otimes (2K^{(l)}-1)})U_\mathrm{C}^\dag)}^{B_3 B_4},
\end{align}
where $j\in\{1,2\}$, and $k^{(l)}\in\{1,\ldots,K^{(l)}\}$.
In~\eqref{eq:xx} and~\eqref{eq:zz}, $X_{j,k^{(l)}}$ and $Z_{j,k^{(l)}}$ are the logical $X$ and $Z$ operators, respectively, of the $k^{(l)}$th logical qubits in $B_j$ for $j\in\{1,2,3,4\}$, and $\mathbbm{1}^{\otimes (2K^{(l)}-1)}$ is the identity operator acting on the rest of logical qubits.

To measure the $2K^{(l)}$ logical operators in~\eqref{eq:xx} without destroying the state itself,
we extract the syndromes with the circuit shown in Fig.~\ref{fig:verification_clifford},
using $2K_{r_l}$ auxiliary level-$(l-1)$ registers twice.
Note that the measurement of the other $2K^{(l)}$ logical operators~\eqref{eq:zz} is performed in the same way as~\eqref{eq:xx} using $Z$ in place of $X$, and hence the following explanation will focus on~\eqref{eq:xx}.
The unitary gates $U_X^{A_1B_1}$, $U_X^{A_1B_2}$, $U_{X,\mathrm{C},3}^{A_1B_3B_4}$, and $U_{X,\mathrm{C},4}^{A_1B_3B_4}$ used in the circuit of Fig.~\ref{fig:verification_clifford} will be defined later in~\eqref{eq:U_X} and~\eqref{eq:U_XC}.
Among the $2K_{r_l}$ auxiliary level-$(l-1)$ registers, the first $K_{r_l}$ level-$(l-1)$ registers are collectively denoted by $A_1$, and the second by $A_2$.
Each of these $2K_{r_l}$ auxiliary level-$(l-1)$ registers is prepared in the state $\Ket{0}^{\otimes K^{(l-1)}}$ by the level-$(l-1)$ initial-state preparation operation in~\eqref{eq:initial_state_preparation}.
Each of $A_1$ and $A_2$ has $K_{r_l}\times K^{(l-1)}=K^{(l)}$ qubits in its $K_{r_l}$ level-$(l-1)$ registers.
In each of the two uses of the $K_{r_l}$ auxiliary level-$(l-1)$ registers of $A_1$, we measure $K^{(l)}$ out of the $2K^{(l)}$ logical operators in~\eqref{eq:xx}, where a logical error can be detected as a bit flip of the measurement outcomes in $A_1$.
The other $K_{r_l}$ auxiliary level-$(l-1)$ registers of $A_2$ are used as flag qubits to make these syndrome measurements fault-tolerant~\cite{PhysRevLett.121.050502}, where an error on $A_1$ that may propagate and cause a logical error is detected as a bit flip of the measurement outcomes in $A_2$.
The flip of each of the $2K^{(l)}$-bit measurement outcomes can be checked using $O(N^{(l)})$ parallel processes immediately, i.e., in runtime $O(1)$.

If no error is detected either from the $2K^{(l)}$ logical stabilizer operators in~\eqref{eq:xx} and the flag qubits in this verification of $X\otimes X$, and also either from the $2K^{(l)}$ logical stabilizer operators in~\eqref{eq:zz} and the flag qubits in the subsequent verification of $Z\otimes Z$, then the gadget outputs the logical $(\mathbbm{1}^{B_1B_2}\otimes U_\mathrm{C}^{B_3B_4})\ket{\Phi^{(l)}}^{B_1B_2B_3B_4}$ prepared in this first run.
Otherwise, the gadget discards the state prepared in the first run and performs the same non-fault-tolerant level-$(l - 1)$ stabilizer circuit for the preparation and outputs the logical $(\mathbbm{1}^{B_1B_2}\otimes U_\mathrm{C}^{B_3B_4})\ket{\Phi^{(l)}}^{B_1B_2B_3B_4}$ prepared in the second run without verification.

The Clifford-state preparation gadget becomes fault-tolerant, that is, satisfies the conditions~\eqref{eq:prepCA} and~\eqref{eq:prepCB}.
Since the case with no fault ($s=0$) is obvious, here we explain the reason why~\eqref{eq:prepCB} with $s=1$ holds true.
The gadget can have only the first run (with the success in the verification) or both the first and second runs.
Whenever the gadget needs to perform the second run, the single fault should be present in the part of the first run; then, the output state in the second run has no error.
Thus, in the following, we consider the cases where the gadget has only the first run without detecting any error in the verification.
The cases with $s=1$ belong to either of the following two cases: (i) a case where the single fault occurs on one of the level-$(l-1)$ locations in the part of the level-$(l-1)$ circuit in Fig.~\ref{fig:preparation_clifford} for preparing logical $(\mathbbm{1}^{B_1B_2}\otimes U_\mathrm{C}^{B_3B_4})\ket{\Phi^{(l)}}^{B_1B_2B_3B_4}$, and (ii) a case where the single fault occurs on one of the level-$(l-1)$ locations in the part of the level-$(l-1)$ circuit in Fig.~\ref{fig:preparation_clifford} for the verification.
In case (i), since the error-correction gadgets have no fault, these gadgets leave $B_1$ through $B_4$ in a state in the code space of $4K^{(l)}$ logical qubits; in addition, the measurements of the $4K^{(l)}$ logical stabilizer operators in~\eqref{eq:xx} and~\eqref{eq:zz} are also faultless, and thus finding no logical error specifies the logical state of all the $4K^{(l)}$ logical qubits as $(\mathbbm{1}\otimes U_\mathrm{C})\ket{\Phi^{(l)}}$ in the code space.
As for case (ii), the prepared state before the verification is the logical state $(\mathbbm{1}^{B_1B_2}\otimes U_\mathrm{C}^{B_3B_4})\ket{\Phi^{(l)}}^{B_1B_2B_3B_4}$ without any error, and the fault in the verification does not cause any uncorrectable error at the end of the level-$(l-1)$ circuit.
In particular, a fault in the error-correction gadgets does not cause any uncorrectable error owing to transversality.
On the other hand, in the measurement of the logical operators, an error on the registers in $A_1$ may propagate and spread, leading to uncorrectable errors.
Nonetheless, this fault causes no problem because such an error is always detected by the measurements on the registers $A_2$, thanks to the technique of flag qubits~\cite{PhysRevLett.121.050502}.
Finally, there may be a question of whether a correctable error on $B_1,B_2,B_3,B_4$ would be aggravated by the subsequent faultless measurement of the logical operators, but since the bit-flip or phase-flip error propagating from $B_1,B_2,B_3,B_4$ to $A_1,A_2$ never propagates back from $A_1,A_2$ to $B_1,B_2,B_3,B_4$ by construction, the error remains correctable.
These two cases show fault tolerance.

The unitary gates $U_X^{A_1B_1}$, $U_X^{A_1B_2}$, $U_{X,\mathrm{C},3}^{A_1B_3B_4}$, and $U_{X,\mathrm{C},4}^{A_1B_3B_4}$ used for the measurements of the logical operators in~\eqref{eq:xx} are defined as follows, where we may omit the identity operator $\mathbbm{1}$ for simplicity of presentation.
In the following, we may use the mapping
\begin{equation}
  \label{eq:map_k}
  k^{(l)}\mapsto(k,k^{(l-1)})
\end{equation}
defined as~\eqref{eq:k_Q_l} for simplicity of notation.
For each $k^{(l)}\in\{1,\ldots,K^{(l)}\}$,
let $U_{X,k^{(l)}}^{B_1}$ denote an operator representing the logical operator $X_{1,k^{(l)}}$ appearing in~\eqref{eq:xx} for the $k^{(l)}$th logical qubit, in the form of a tensor product of $X$ and $\mathbbm{1}$ acting on the $N_{r_l}K^{(l-1)}$ qubits of the $N_{r_l}$ level-$(l-1)$ registers that form $B_1$;
under the mapping~\eqref{eq:map_k}, we can write $U_{X,k^{(l)}}^{B_1}$ as
\begin{align}
  \label{eq:U_X_k}
  U_{X,k^{(l)}}^{B_1}=\prod_{n=1}^{N_{r_l}}P_{X,k,n}^{B_1,n,k^{(l-1)}}
\end{align}
where
\begin{equation}
  \label{eq:P_X}
  P_{X,k,n}\in\{X,\mathbbm{1}\}
\end{equation}
is determined in such a way that $\bigotimes_{n=1}^{N_{r_l}}P_{X,k,n}$ should be the logical $X$ operator of the $k$th logical qubit of $\Q_{r_l}$, and the superscript of $P_{X,k,n}^{B_1,n,k^{(l-1)}}$ shows that $P_{X,k,n}^{B_1,n,k^{(l-1)}}$ acts on the $k^{(l-1)}$th qubit in the $n$th level-$(l-1)$ register of $B_1$.
The unitaries $U_{X,k^{(l)}}$ for all $k^{(l)}\in\{1,\ldots,K^{(l)}\}$ commute with each other since each $U_{X,k^{(l)}}$ is a tensor product only of $X$ and $\mathbbm{1}$.
Similarly, let $U_{X,\mathrm{C},3,k^{(l)}}^{B_3B_4}$ denote an operator representing the logical operator ${(U_\mathrm{C}(X_{3,k^{(l)}}\otimes\mathbbm{1}^{\otimes (2K^{(l)}-1)})U_\mathrm{C}^\dag)}^{B_3 B_4}$ in~\eqref{eq:xx}.
Since $U_\mathrm{C}$ is Clifford, $U_{X,\mathrm{C},3,k^{(l)}}^{B_3B_4}$ can be written under the mapping~\eqref{eq:map_k} as a tensor product of Pauli operators
\begin{align}
  \label{eq:U_XC_k}
  &U_{X,\mathrm{C},3,k^{(l)}}^{B_3B_4}=\nonumber\\
  &\quad\prod_{n=1}^{N_{r_l}}\prod_{{k^{(l-1)}}^\prime=1}^{K^{(l-1)}}\Big(P_{X,3,3,k^{(l-1)},k,n,{k^{(l-1)}}^\prime}^{B_3,n,{k^{(l-1)}}^\prime}\otimes\nonumber\\
  &\qquad P_{X,3,4,k^{(l-1)},k,n,{k^{(l-1)}}^\prime}^{B_4,n,{k^{(l-1)}}^\prime}\Big),
\end{align}
where
\begin{align}
  \label{eq:P_X3}
  P_{X,3,j,k^{(l-1)},k,n,{k^{(l-1)}}^\prime}\in\{\pm X,\pm Z,\pm Y,\pm \mathbbm{1}\}
\end{align}
for $j\in\{3,4\}$, and $P_{X,3,j,k^{(l-1)},k,n,{k^{(l-1)}}^\prime}^{B_j,n,{k^{(l-1)}}^\prime}$ acts on the ${k^{(l-1)}}^\prime$th qubit in the $n$th level-$(l-1)$ register of $B_j$.
The unitaries $U_{X,\mathrm{C},3,k^{(l)}}$ for all $k^{(l)}\in\{1,\ldots,K^{(l)}\}$ commute with each other due to the commutativity of the logical operators ${U_\mathrm{C}(X_{3,k^{(l)}}\otimes\mathbbm{1}^{\otimes (2K^{(l)}-1)})U_\mathrm{C}^\dag}$.

Using the operators~\eqref{eq:U_X_k} and~\eqref{eq:U_XC_k} given by the tensor product of Pauli operators,
controlled Pauli gates used for measuring~\eqref{eq:xx} are defined as follows.
For a Pauli operator $P^{B_j,n,{k^{(l-1)}}^\prime}$ used in~\eqref{eq:U_X_k} and~\eqref{eq:U_XC_k} (where ${k^{(l-1)}}^\prime=k^{(l-1)}$ for~\eqref{eq:U_X_k}), let
\begin{equation}
  C^{A_{1},k,k^{(l-1)}}P^{B_j,n,{k^{(l-1)}}^\prime}
\end{equation}
denote a controlled Pauli operator that applies $P^{B_j,n,{k^{(l-1)}}^\prime}$ to the target qubit, i.e., the ${k^{(l-1)}}^\prime$th qubit in the $n$th level-$(l-1)$ register of $B_j$, controlled by the $k^{(l-1)}$th qubit in the $k$th level-$(l-1)$ register of $A_{1}$.
Define $CU_{X,k^{(l)}}^{A_1B_1}$ as a controlled unitary operator given under the mapping~\eqref{eq:map_k} by
\begin{equation}
  \label{eq:CU_Xk}
  CU_{X,k^{(l)}}^{A_1B_1}=\prod_{n=1}^{N_{r_l}}C^{A_1,k,k^{(l-1)}}P_{X,k,n}^{B_1,n,k^{(l-1)}}.
\end{equation}
The controlled unitaries $CU_{X,k^{(l)}}$ for all $k^{(l)}\in\{1,\ldots,K^{(l)}\}$ commute with each other due to the commutativity of $U_{X,k^{(l)}}$.
Define $CU_{X,\mathrm{C},3,k^{(l)}}^{A_1B_3B_4}$ in the same way as $CU_{X,k^{(l)}}^{A_1B_1}$ using $U_{X,\mathrm{C},3,k^{(l)}}^{B_3B_4}$ in place of $U_{X,k^{(l)}}^{B_1}$; that is, $CU_{X,\mathrm{C},3,k^{(l)}}^{A_1B_3B_4}$ is given under the mapping~\eqref{eq:map_k} by
\begin{align}
  \label{eq:CU_XCk}
  &CU_{X,\mathrm{C},3,k^{(l)}}^{A_1B_3B_4}=\nonumber\\
  &\quad\prod_{n=1}^{N_{r_l}}\prod_{{k^{(l-1)}}^\prime=1}^{K^{(l-1)}}\Big(C^{A_1,k,k^{(l-1)}}P_{X,3,3,k^{(l-1)},k,n,{k^{(l-1)}}^\prime}^{B_3,n,{k^{(l-1)}}^\prime}\otimes\nonumber\\
  &\qquad C^{A_1,k,k^{(l-1)}}P_{X,3,4,k^{(l-1)},k,n,{k^{(l-1)}}^\prime}^{B_4,n,{k^{(l-1)}}^\prime}\Big).
\end{align}
The controlled unitaries $CU_{X,\mathrm{C},3,k^{(l)}}$ for all $k^{(l)}\in\{1,\ldots,K^{(l)}\}$ commute with each other due to the commutativity of $U_{X,\mathrm{C},3,k^{(l)}}$.
Define $CU_{X,\mathrm{C},4,k^{(l)}}^{A_1B_3B_4}$ in the same way as $CU_{X,\mathrm{C},3,k^{(l)}}^{A_1B_3B_4}$ using $X_{4,k^{(l)}}$ in place of $X_{3,k^{(l)}}$.
Then, define $U_X^{A_1B_1}$ in Fig.~\ref{fig:verification_clifford} as the product of commuting operators $CU_{X,k^{(l)}}^{A_1B_1}$ for all $k^{(l)}\in\{1,\ldots,K^{(l)}\}$, i.e., under the mapping~\eqref{eq:map_k},
\begin{align}
  \label{eq:U_X}
  U_X^{A_1B_1}&\coloneqq\prod_{k^{(l)}=1}^{K^{(l)}}CU_{X,k^{(l)}}^{A_1B_1}\nonumber\\
              &=\prod_{k=1}^{K_{r_l}}\prod_{n=1}^{N_{r_l}}\left(\prod_{k^{(l-1)}=1}^{K^{(l-1)}}C^{A_1,k,k^{(l-1)}}P_{X,k,n}^{B_1,n,k^{(l-1)}}\right),
\end{align}
and $U_X^{A_1B_2}$ in the same way as $U_X^{A_1B_1}$ using $B_2$ in place of $B_1$.
Due to~\eqref{eq:P_X},
$U_X^{A_1B_1}$ in~\eqref{eq:U_X} can be implemented by at most $N_{r_l}K_{r_l}$ level-$(l-1)$ \textsc{CNOT}-gate operations, and so can $U_X^{A_1B_2}$.
Similarly,
define $U_{X,\mathrm{C},3}^{A_1B_3B_4}$ in Fig.~\ref{fig:verification_clifford} as the product of commuting operators $CU_{X,\mathrm{C},3,k^{(l)}}^{A_1B_3B_4}$ for all $k^{(l)}\in\{1,\ldots,K^{(l)}\}$, i.e., under the mapping~\eqref{eq:map_k},
\begin{widetext}
\begin{align}
  \label{eq:U_XC}
  &U_{X,\mathrm{C},3}^{A_1B_3B_4}\nonumber\\
  &\coloneqq\prod_{k^{(l)}=1}^{K^{(l)}}CU_{X,\mathrm{C},3,k^{(l)}}^{A_1B_3B_4}\nonumber\\
  &=\prod_{k=1}^{K_{r_l}}\prod_{n=1}^{N_{r_l}}\left(\prod_{k^{(l-1)}=1}^{K^{(l-1)}}\prod_{{k^{(l-1)}}^\prime=1}^{K^{(l-1)}}\left(C^{A_1,k,k^{(l-1)}}P_{X,3,3,k^{(l-1)},k,n,{k^{(l-1)}}^\prime}^{B_3,n,{k^{(l-1)}}^\prime}\otimes C^{A_1,k,k^{(l-1)}}P_{X,3,4,k^{(l-1)},k,n,{k^{(l-1)}}^\prime}^{B_4,n,{k^{(l-1)}}^\prime}\right)\right),
\end{align}
\end{widetext}
and $U_{X,\mathrm{C},4}^{A_1B_3B_4}$ as the product of $CU_{X,\mathrm{C},4,k^{(l)}}^{A_1B_3B_4}$ for all $k^{(l)}\in\{1,\ldots,K^{(l)}\}$.
Note that in the last equality of~\eqref{eq:U_XC}, we can reorder $\prod_{k^{(l-1)}=1}^{K^{(l-1)}}\prod_{n=1}^{N_{r_l}}$ into $\prod_{n=1}^{N_{r_l}}\prod_{k^{(l-1)}=1}^{K^{(l-1)}}$,
since operators $P_{X,3,3,k^{(l-1)},k,n,{k^{(l-1)}}^\prime}^{B_3,n,{k^{(l-1)}}^\prime}\otimes P_{X,3,4,k^{(l-1)},k,n,{k^{(l-1)}}^\prime}^{B_4,n,{k^{(l-1)}}^\prime}$ with different $n$ act on different level-$(l-1)$ registers of $B_1$ and $B_2$, and thus their controlled versions $C^{A_1,k,k^{(l-1)}}P_{X,3,3,k^{(l-1)},k,n,{k^{(l-1)}}^\prime}^{B_3,n,{k^{(l-1)}}^\prime}\otimes C^{A_1,k,k^{(l-1)}}P_{X,3,4,k^{(l-1)},k,n,{k^{(l-1)}}^\prime}^{B_4,n,{k^{(l-1)}}^\prime}$ for all $k^{(l-1)}$ and $n$ commute with each other.
Due to~\eqref{eq:P_X3},
$U_{X,\mathrm{C},3}^{A_1B_3B_4}$ in~\eqref{eq:U_XC} can be implemented by at most $2N_{r_l}K_{r_l}$ level-$(l-1)$ two-register Clifford-gate abbreviations, and so can $U_{X,\mathrm{C},4}^{A_1B_3B_4}$.

In the following,
we will count the number of one- and two-register gates in the part for implementing the Clifford unitary~\eqref{eq:Clifford},
which has been deferred in the above explanations.
We here prove that any Clifford unitary acting on the $2N_{r_l}K^{(l-1)}$ qubits in $2N_{r_l}$ level-$(l-1)$ registers, including~\eqref{eq:Clifford}, can be implemented by a level-$(l-1)$ circuit composed of two-register Clifford gates with their number bounded by
\begin{equation}
  \label{eq:num_gates}
  O\left(N_{r_l}^2\right).
\end{equation}
Note that in a special case where each level-$(l-1)$ register reduces to only one qubit, Ref.~\cite{PhysRevA.70.052328} has shown an improved bound $O\left(\nicefrac{N_{r_l}^2}{\log(N_{r_l})}\right)$ that may be better than~\eqref{eq:num_gates} by a logarithmic factor, but we leave as an open problem whether such an improvement is still possible in our more general case where each level-$(l-1)$ register may have more than one qubit.
The depth of a stabilizer circuit could be shortened by parallelizing the circuit using auxiliary qubits~\cite{10.1137/S0097539799355053,10.5555/3381089.3381102}, but this further optimization is not considered here.

Our proof of~\eqref{eq:num_gates} uses the fact that any stabilizer circuit for implementing a Clifford unitary can be rewritten into an equivalent stabilizer circuit consisting of $11$ rounds in a sequence, where each round is given by a circuit composed only of one type of Clifford gate, i.e., $H$, $\textsc{CNOT}$, $S$, $\textsc{CNOT}$, $S$, $\textsc{CNOT}$, $H$, $S$, $\textsc{CNOT}$, $S$, and $\textsc{CNOT}$, in this order~\cite{PhysRevA.70.052328}.
This decomposition of the stabilizer circuit into one- and two-qubit gates can be performed feasibly using Gaussian elimination as shown in Ref.~\cite{PhysRevA.70.052328}.
Note that Ref.~\cite{8335339} also provides a similar equivalence of stabilizer circuits, which may also be used here.
Each round of $H$ or $S$ gates is performed with the level-$(l-1)$ two-register Clifford-gate abbreviation in Fig.~\ref{fig:inblock_clifford} acting on each of the $2N_{r_l}$ level-$(l-1)$ registers at most once.
As for a round of $\textsc{CNOT}$ gates, we here show that we can implement an arbitrary $2N_{r_l}K^{(l-1)}$-qubit $\textsc{CNOT}$ circuit by a circuit composed only of $O(N_{r_l}^2)$ two-register Clifford gates on $2N_{r_l}$ level-$(l-1)$ registers.

To analyze the round of $\textsc{CNOT}$ gates, as in Ref.~\cite{10.5555/2011763.2011767}, we represent the $2N_{r_l}K^{(l-1)}$-qubit $\textsc{CNOT}$ circuit as a linear reversible transformation over the finite field $\mathbb{F}_2$, i.e., a $2N_{r_l}K^{(l-1)}\times 2N_{r_l}K^{(l-1)}$ binary matrix $A$, in such a way that the $\textsc{CNOT}$ circuit linearly transforms the standard basis $\{\Ket{x}\}$ of these $2N_{r_l}K^{(l-1)}$ qubits as
\begin{equation}
  \Ket{x}\mapsto\Ket{xA},
\end{equation}
where $x=(x_1,\ldots,x_{2N_{r_l}K^{(l-1)}})\in\mathbb{F}_2^{2N_{r_l}K^{(l-1)}}$ is a row binary vector that is ordered in such a way that its $i$th element $x_i$ with
\begin{equation}
i=(j-1)N_{r_l}K^{(l-1)}+(n-1)K^{(l-1)}+k^{(l-1)}
\end{equation}
represents the standard basis $\{\Ket{x_i}:x_i\in\mathbb{F}_2=\{0,1\}\}$ of the $k^{(l-1)}$th qubit in the $n$th level-$(l-1)$ register of $B_{j+2}$,
for $k^{(l-1)}\in \{1,\ldots,K^{(l-1)}\}$, $n \in  \{1,\ldots,N_{r_l}\}$, and $j\in \{1,2\}$.
We will decompose $A$ into linear reversible transformations implementable by two-register Clifford gates, using Gaussian elimination.

\begin{widetext}
Our decomposition of $A$ representing the round of $\textsc{CNOT}$ gates is as follows.
We consider $A$ to be a $2N_{r_l}\times 2N_{r_l}$ block matrix, where each $K^{(l-1)}\times K^{(l-1)}$ block represents the transformation involving the qubits in one or two level-$(l-1)$ registers.
Any linear reversible transformation is implementable by $\textsc{CNOT}$ gates as shown in Ref.~\cite{10.5555/2011763.2011767},
but our decomposition will be different from that in Ref.~\cite{10.5555/2011763.2011767} in that we here use the two-register Clifford gates rather than single-qubit $\textsc{CNOT}$ gates.
Each two-register Clifford gate can apply an arbitrarily large number of $\textsc{CNOT}$ gates between two level-$(l-1)$ registers at once, implementing an arbitrary linear reversible transformation of the basis of the qubits in the two level-$(l-1)$ registers.
To show the decomposition, we use the lower-upper (LU) decomposition of $A$, i.e.,
\begin{equation}
  A=PLU,
\end{equation}
where $P$ is a permutation matrix, and $L$ and $U$ are lower and upper triangular matrices, respectively.
The permutation matrix $P$ has $2N_{r_l}\times 2N_{r_l}$ blocks and is implementable with $\binom{2N_{r_l}}{2}$ two-register Clifford gates, where each two-register Clifford gate performs \textsc{SWAP} gates between qubits in the two level-$(l-1)$ registers.
The lower triangular matrix $L$ can be written in the form of the block matrix as
\begin{equation}
  L=\left(\begin{matrix}
      L_{1,1} & 0       & 0       & \cdots & 0\\
      L_{2,1} & L_{2,2} & 0       & \cdots & 0\\
      L_{3,1} & L_{3,2} & L_{3,3} & \ddots & \vdots\\
      \vdots  & \vdots  & \ddots  & \ddots & 0\\
      L_{2N_{r_l},1} & L_{2N_{r_l},2} & \cdots & L_{2N_{r_l},2N_{r_l}-1} & L_{2N_{r_l},2N_{r_l}}
  \end{matrix}\right),
\end{equation}
where each block $L_{n,n^\prime}$ is a $K^{(l-1)}\times K^{(l-1)}$ matrix, and $L_{n,n}$ on the diagonal is a nonsingular lower triangular matrix since $A$ is reversible.
Then we decompose
\begin{equation}
  L=DL^\prime,
\end{equation}
where $D$ is a block diagonal matrix
\begin{equation}
  D=\left(\begin{matrix}
      L_{1,1}     & 0       & 0       & \cdots & 0\\
      0 & L_{2,2} & 0       & \cdots & 0\\
      0 & 0 & L_{3,3} & \ddots & \vdots\\
      \vdots  & \vdots  & \ddots  & \ddots & 0\\
      0 & 0 & \cdots & 0 & L_{2N_{r_l},2N_{r_l}}
  \end{matrix}\right),
\end{equation}
and $L^\prime$ is a block matrix with identity matrices on the diagonal
\begin{equation}
  L^\prime=\left(\begin{matrix}
      \mathbbm{1} & 0       & 0       & \cdots & 0\\
      L^\prime_{2,1} & \mathbbm{1} & 0       & \cdots & 0\\
      L^\prime_{3,1} & L^\prime_{3,2} & \mathbbm{1} & \ddots & \vdots\\
      \vdots  & \vdots  & \ddots  & \ddots & 0\\
      L^\prime_{2N_{r_l},1} & L^\prime_{2N_{r_l},2} & \cdots & L^\prime_{2N_{r_l},2N_{r_l}-1} & \mathbbm{1}
  \end{matrix}\right).
\end{equation}
To implement $D$, we use $N_{r_l}$ two-register Clifford gates, where each two-register Clifford gate implements a linear reversible transformation acting nontrivially on two blocks on the diagonal.
As for $L^\prime$, each off-diagonal block is implementable with one two-register Clifford gate; for example, we can decompose
\begin{align}
  \label{eq:gaussian_elimination}
  &\left(\begin{matrix}
      \mathbbm{1} & 0       & 0       & \cdots & 0\\
      L^\prime_{2,1} & \mathbbm{1} & 0       & \cdots & 0\\
      L^\prime_{3,1} & L^\prime_{3,2} & \mathbbm{1} & \ddots & \vdots\\
      \vdots  & \vdots  & \ddots  & \ddots & 0\\
      L^\prime_{2N_{r_l},1} & L^\prime_{2N_{r_l},2} & \cdots & L^\prime_{2N_{r_l},2N_{r_l}-1} & \mathbbm{1}
  \end{matrix}\right)\\
  &=
  \left(\begin{matrix}
      \mathbbm{1} & 0       & 0       & \cdots & 0\\
      L^\prime_{2,1} & \mathbbm{1} & 0       & \cdots & 0\\
      0 & 0 & \mathbbm{1} & \ddots & \vdots\\
      \vdots  & \vdots  & \ddots  & \ddots & 0\\
      0 & 0 & \cdots & 0 & \mathbbm{1}
  \end{matrix}\right)
  \left(\begin{matrix}
      \mathbbm{1} & 0       & 0       & \cdots & 0\\
      0 & \mathbbm{1} & 0       & \cdots & 0\\
      L^\prime_{3,1} & L^\prime_{3,2} & \mathbbm{1} & \ddots & \vdots\\
      \vdots  & \vdots  & \ddots  & \ddots & 0\\
      L^\prime_{2N_{r_l},1} & L^\prime_{2N_{r_l},2} & \cdots & L^\prime_{2N_{r_l},2N_{r_l}-1} & \mathbbm{1}
  \end{matrix}\right)\\
  &=\left(\begin{matrix}
      \mathbbm{1} & 0       & 0       & \cdots & 0\\
      L^\prime_{2,1} & \mathbbm{1} & 0       & \cdots & 0\\
      0 & 0 & \mathbbm{1} & \ddots & \vdots\\
      \vdots  & \vdots  & \ddots  & \ddots & 0\\
      0 & 0 & \cdots & 0 & \mathbbm{1}
  \end{matrix}\right)
\left(\begin{matrix}
      \mathbbm{1} & 0       & 0       & \cdots & 0\\
      0 & \mathbbm{1} & 0       & \cdots & 0\\
      L^\prime_{3,1} & 0 & \mathbbm{1} & \ddots & \vdots\\
      \vdots  & \vdots  & \ddots  & \ddots & 0\\
      0 & 0 & \cdots & 0 & \mathbbm{1}
  \end{matrix}\right)
  \left(\begin{matrix}
      \mathbbm{1} & 0       & 0       & \cdots & 0\\
      0 & \mathbbm{1} & 0       & \cdots & 0\\
      0 & L^\prime_{3,2} & \mathbbm{1} & \ddots & \vdots\\
      \vdots  & \vdots  & \ddots  & \ddots & 0\\
      L^\prime_{2N_{r_l},1} & L^\prime_{2N_{r_l},2} & \cdots & L^\prime_{2N_{r_l},2N_{r_l}-1} & \mathbbm{1}
  \end{matrix}\right)=\cdots,
\end{align}
where the linear reversible transformation represented by each matrix with only one off-diagonal block, such as $L^\prime_{2,1}$ and $L^\prime_{3,1}$ in the last line, is implementable with one two-register Clifford gate.
By repeating this decomposition from left to right and from up to bottom as in the Gaussian elimination~\cite{10.5555/2011763.2011767}, we can implement all the $\nicefrac{2N_{r_l}(2N_{r_l}-1)}{2}$ off-diagonal blocks of $L^\prime$ with $\nicefrac{2N_{r_l}(2N_{r_l}-1)}{2}$ two-register Clifford gates.
We can decompose $U$ in the same way as $L$ using the upper triangular matrices in place of the lower triangular matrices.
As a result, the required number of two-register Clifford gates for implementing $A$ is bounded by
\begin{equation}
  \binom{2N_{r_l}}{2}+2\times\left(N_{r_l}+\frac{2N_{r_l}(2N_{r_l}-1)}{2}\right)=O(N_{r_l}^2).
\end{equation}
That is, the required number of two-register Clifford gates for implementing each round of $\textsc{CNOT}$ gates is bounded by $O(N_{r_l}^2)$.
\end{widetext}

Therefore, the number of one- and two-register gates in the part for implementing the Clifford unitary~\eqref{eq:Clifford} is bounded by $O(N_{r_l}^2)$ as claimed in~\eqref{eq:num_gates}, dominated by those for the rounds of $\textsc{CNOT}$ gates.
Since each two-register Clifford-gate abbreviation may have runtime at most $O(\log(N^{(l-1)}))=O(\log(N^{(l)}))$ as shown in~\eqref{eq:runtime_clifford},
the depth of the part for implementing the Clifford unitary~\eqref{eq:Clifford} is
\begin{equation}
  \label{eq:runtime_clifford_preparation}
  O(N_{r_l}^2\log(N^{(l)})).
\end{equation}

The part of $U_X^{A_1B_1}$, $U_X^{A_1B_2}$, $U_{X,\mathrm{C}}^{(3)A_1B_3B_4}$, and $U_{X,\mathrm{C}}^{(4)A_1B_3B_4}$ in the verification have $O(N_{r_l}K_{r_l})=O(N_{r_l}^2)$ two-register Clifford gates as can be seen from~\eqref{eq:U_X} and~\eqref{eq:U_XC}.
Due to the $O(\log(N^{(l)}))$ runtime of a two-register Clifford-gate abbreviation,
the required depth for implementing all these gates is also
\begin{equation}
  \label{eq:runtime_clifford_preparation_verification}
  O(N_{r_l}^2\log(N^{(l)})).
\end{equation}

As for other non-dominant parts,
due to~\eqref{eq:depth_encoding}, each level-$l$ initial-state preparation gadget has the $O(N_{r_l}\log(N_{r_l}))$ depth.
Due to~\eqref{eq:depth_error_correction}, each level-$l$ error-correction gadget has the $O(N_{r_l}\log(N_{r_l}))$ depth.
The other parts are implementable within a constant depth.
Finally, the depth of the second run of the gadget is smaller than that of the first run; thus, the depth of the gadget including the first and second runs is at most twice as long as the first run.

Consequently, the depth of the level-$(l-1)$ circuit in Fig.~\ref{fig:preparation_clifford} is
\begin{equation}
  \label{eq:depth_Clifford_all}
  O(N_{r_l}^2\log(N^{(l)})),
\end{equation}
dominated by~\eqref{eq:runtime_clifford_preparation} and~\eqref{eq:runtime_clifford_preparation_verification}.
The required number of bits for classical computation is $O(N^{(l)2})$, dominated by those for storing the $O(N^{(l)})\times O(N^{(l)})$ matrix.

As in~\eqref{eq:argument_clifford}, the Clifford-state preparation operation can be invoked with an argument for representing $U_\mathrm{C}$, and below we discuss the on-demand function of the corresponding gadget.
Note that, as a special case, $U_\mathrm{C}$ in the form of~\eqref{eq:argument_clifford} covers the implementation of the single-qubit $HZ$ gates on a level-$l$ register (while acting trivially on the other) used for the correction in the level-$l$ $R(\pm\nicefrac{\pi}{4})$-gate abbreviation.
The required function for the on-demand gadget is to implement a logical two-register Clifford unitary $U_\mathrm{C}$ acting on the two encoded level-$l$ registers in the tensor-product form of~\eqref{eq:argument_clifford}, i.e.,
$U_\mathrm{C}=\bigotimes_{k^{(l)}=1}^{K^{(l)}}U_{\mathrm{C},l}^{(k^{(l)})}$,
where $U_{\mathrm{C},l}^{(k^{(l)})}$ is a two-qubit Clifford unitary acting on the $k^{(l)}$th qubit in each of the two level-$l$ registers.
In the on-demand case,
several parts of the level-$(l-1)$ circuit of the gadget in Fig.~\ref{fig:preparation_clifford} change depending on $U_\mathrm{C}$.
The parts that may change are the Clifford unitary~\eqref{eq:Clifford} for implementing logical $U_\mathrm{C}$ and the control unitaries $U_X^{A_1B_1}$, $U_X^{A_1B_2}$, $U_{X,\mathrm{C},3}^{A_1B_3B_4}$, and $U_{X,\mathrm{C},4}^{A_1B_3B_4}$ in~\eqref{eq:U_X} and~\eqref{eq:U_XC} in the verification for measuring the logical $X$ operators in~\eqref{eq:xx} (as well as those of $Z$ in~\eqref{eq:zz}).
In this case, all the level-$(l-1)$ two-register Clifford-gate abbreviations that depend on $U_\mathrm{C}$ in Fig.~\ref{fig:preparation_clifford} must be on-demand ones.
In particular, the level-$(l-1)$ on-demand two-register Clifford-gate abbreviations are used for the Clifford unitary~\eqref{eq:Clifford} and the control unitaries $U_{X,\mathrm{C},3}^{A_1B_3B_4}$ and $U_{X,\mathrm{C},4}^{A_1B_3B_4}$ of~\eqref{eq:U_XC} (as well as those of $Z$), as shown in Fig.~\ref{fig:preparation_clifford}.
In the following, we show that these level-$(l-1)$ on-demand abbreviations are available.

To show the availability of the level-$(l-1)$ on-demand abbreviations in implementing logical $U_\mathrm{C}$ in the tensor-product form~\eqref{eq:argument_clifford},
we here show that the unitaries of these level-$(l-1)$ on-demand abbreviations are always written in a tensor-product form
\begin{equation}
  \label{eq:decomposition_correction_clifford}
  \bigotimes_{k^{(l-1)}=1}^{K^{(l-1)}}U_{\mathrm{C},l-1}^{(k^{(l-1)})},
\end{equation}
as required at level $(l-1)$.
To show this,
under the mapping $k^{(l)}\mapsto (k,k^{(l-1)})$ of~\eqref{eq:k_Q_l},
we write $U_\mathrm{C}=\bigotimes_{k^{(l)}=1}^{K^{(l)}}U_{\mathrm{C},l}^{(k^{(l)})}$ as
\begin{equation}
  \label{eq:U_Cl_level_l-1}
  U_\mathrm{C}=\bigotimes_{k^{(l-1)}=1}^{K^{(l-1)}}\left(\bigotimes_{k=1}^{K_{r_l}}U_{\mathrm{C},l}^{(k,k^{(l-1)})}\right),
\end{equation}
where $U_{\mathrm{C},l}^{(k,k^{(l-1)})}\coloneqq U_{\mathrm{C},l}^{(k^{(l)})}$ acts on the $k$th logical qubits of the $k^{(l-1)}$th code blocks of $\Q_{r_l}$ for the two level-$l$ registers.
Moreover, the Clifford unitary $U_\mathrm{encode}=V^{\otimes K^{(l-1)}}$ for the encoding in~\eqref{eq:encode} is also a tensor product of the Clifford unitary $V$ for the encoding of $\Q_{r_l}$.
Thus, the Clifford unitary~\eqref{eq:Clifford} to implement logical $U_\mathrm{C}$ in the tensor-product form~\eqref{eq:argument_clifford} has a tensor-product form
\begin{align}
  \label{eq:Clifford_tensor_product}
  &\bigotimes_{k^{(l-1)}=1}^{K^{(l-1)}}\Big[(V\otimes V)\nonumber\\
  &\quad\left(\left(\bigotimes_{k=1}^{K_{r_l}}U_{\mathrm{C},l}^{(k,k^{(l-1)})}\right)\otimes \mathbbm{1}^{\otimes 2(N_{r_l}-K_{r_l})}\right){(V\otimes V)}^\dag\Big].
\end{align}
Then, by performing gate decomposition of each Clifford unitary
\begin{equation}
  (V\otimes V)\left(\left(\bigotimes_{k=1}^{K_{r_l}}U_{\mathrm{C},l}^{(k,k^{(l-1)})}\right)\otimes \mathbbm{1}^{\otimes 2(N_{r_l}-K_{r_l})}\right){(V\otimes V)}^\dag
\end{equation}
acting on the $k^{(l-1)}$th code blocks of $\Q_{r_l}$,
we see that each two-register Clifford gate in the level-$(l-1)$ circuit also has the tensor-product form~\eqref{eq:decomposition_correction_clifford} on level-$(l-1)$ registers, as required.

As for the availability of the level-$(l-1)$ on-demand abbreviations in implementing the control unitary $U_{X,\mathrm{C},3}^{A_1B_3B_4}$ of~\eqref{eq:U_XC} for the verification,
we here show that the unitaries of these level-$(l-1)$ on-demand abbreviations are also written in a tensor-product form~\eqref{eq:decomposition_correction_clifford}.
In the on-demand case,
due to $U_\mathrm{C}=\bigotimes_{k^{(l)}=1}^{K^{(l)}}U_{\mathrm{C},l}^{(k^{(l)})}$ in~\eqref{eq:argument_clifford} as in the above analysis,
the stabilizer operator ${(U_\mathrm{C}(X_{3,k^{(l)}}\otimes\mathbbm{1}^{\otimes (2K^{(l)}-1)})U_\mathrm{C}^\dag)}^{B_3 B_4}$ in~\eqref{eq:xx} to be measured in the verification has the form of
\begin{equation}
  \label{eq:xx_on_demand}
  (P_{3,k^{(l)}}\otimes P_{4,k^{(l)}})\otimes\mathbbm{1}^{\otimes (2K^{(l)}-2)},
\end{equation}
where $P_{3,k^{(l)}}$ and $P_{4,k^{(l)}}$ are Pauli operators acting on the $k^{(l)}$th qubit in the level-$l$ register of $B_3$ and $B_4$, respectively, and $\mathbbm{1}^{\otimes (2K^{(l)}-2)}$ acts on the rest of the qubits.
Then, $U_{X,\mathrm{C},3,k^{(l)}}^{B_3B_4}$ in~\eqref{eq:U_XC_k} for representing the logical operator in the form of~\eqref{eq:xx_on_demand} reduces to
\begin{align}
  \label{eq:U_XC_k_simplified}
  &U_{X,\mathrm{C},3,k^{(l)}}^{B_3B_4}=\nonumber\\
  &\quad\prod_{n=1}^{N_{r_l}}P_{X,3,3,k^{(l-1)},k,n,{k^{(l-1)}}}^{B_3,n,{k^{(l-1)}}}\otimes P_{X,3,4,k^{(l-1)},k,n,{k^{(l-1)}}}^{B_4,n,{k^{(l-1)}}}.
\end{align}
Substituting $U_{X,\mathrm{C},3,k^{(l)}}^{B_3B_4}$ in~\eqref{eq:U_XC_k} with~\eqref{eq:U_XC_k_simplified} in the construction of  $U_{X,\mathrm{C},3}^{A_1B_3B_4}$ in~\eqref{eq:U_XC},
we see that each of the $2N_{r_l}K_{r_l}$ level-$(l-1)$ two-register Clifford gates for implementing $U_{X,\mathrm{C},3}^{A_1B_3B_4}$ is in the form of
\begin{align}
  \label{eq:U_XC_simplified_3}
  &\prod_{k^{(l-1)}=1}^{K^{(l-1)}}C^{A_1,k,k^{(l-1)}}P_{X,3,3,k^{(l-1)},k,n,{k^{(l-1)}}}^{B_3,n,{k^{(l-1)}}}\nonumber\\
  &=\bigotimes_{k^{(l-1)}=1}^{K^{(l-1)}}C^{A_1,k,k^{(l-1)}}P_{X,3,3,k^{(l-1)},k,n,{k^{(l-1)}}}^{B_3,n,{k^{(l-1)}}}
\end{align}
or
\begin{align}
  \label{eq:U_XC_simplified_4}
  &\prod_{k^{(l-1)}=1}^{K^{(l-1)}}C^{A_1,k,k^{(l-1)}}P_{X,3,4,k^{(l-1)},k,n,{k^{(l-1)}}}^{B_4,n,{k^{(l-1)}}}\nonumber\\
  &=\bigotimes_{k^{(l-1)}=1}^{K^{(l-1)}}C^{A_1,k,k^{(l-1)}}P_{X,3,4,k^{(l-1)},k,n,{k^{(l-1)}}}^{B_4,n,{k^{(l-1)}}},
\end{align}
where $\prod$ becomes $\bigotimes$ since the controlled Pauli gates act on different qubits.
Therefore, $U_{X,\mathrm{C},3}^{A_1B_3B_4}$ for the verification is composed of the two-register Clifford gates in the tensor-product form~\eqref{eq:decomposition_correction_clifford} on level-$(l-1)$ registers, as required.

For this on-demand function, the runtime of classical computation for the gate decomposition is bounded as follows.
To decompose the Clifford unitary into a stabilizer circuit in the rounds composed of one of the $H$, $S$, and \textsc{CNOT} gates,
we use Gaussian elimination to convert the binary matrices into a canonical form as shown in Ref.~\cite{PhysRevA.70.052328}.
As discussed with~\eqref{eq:argument_clifford}, $U_\mathrm{C}$ in the tensor-product form is represented by $K^{(l)}$ $4\times 4$ binary matrices.
To obtain each $U_{\mathrm{C},l-1}^{(k^{(l-1)})}$ in~\eqref{eq:decomposition_correction_clifford} from $U_{\mathrm{C},l}^{(k^{(l)})}$ for all $k^{(l)}$, we use the encoding unitary $V$ for each code block $\Q_{r_l}$ rather than $U_\mathrm{encode}$ in~\eqref{eq:encode}.
By representing the Clifford unitary $V$ as the $O(N_{r_l})\times O(N_{r_l})$ binary matrices, we perform the Gaussian elimination of the $O(N_{r_l})\times O(N_{r_l})$ matrices.
As a whole, the Gaussian elimination of the $O(N_{r_l})\times O(N_{r_l})$ matrix requires $O(N_{r_l}^3)$ arithmetic operations, and using $O(N_{r_l}^2)$ parallel processes, we can run the Gaussian elimination within runtime $O(N_{r_l})$, which is dominant here.
Therefore, for the part of preparing logical $(\mathbbm{1}\otimes U_\mathrm{C})\ket{\Phi^{(l)}}$,
we obtain the level-$(l-1)$ circuit composed of the rounds of the $H$, $S$, or \textsc{CNOT} gates within runtime $O(N_{r_l})$.
As for the runtime of classical computation for the gate decomposition of the part for the verification,
we obtain~\eqref{eq:xx_on_demand} by calculating the conjugation of the single-qubit Pauli operator (and the single-qubit identity operator) by the two-qubit Clifford gate.
The logical operator of~\eqref{eq:xx_on_demand} is given by the tensor product of logical operators for at most one logical qubit in each of a pair of code blocks of $\Q_{r_l}$.
Thus in the on-demand case, the controlled Pauli gates of~\eqref{eq:U_XC_simplified_3} and~\eqref{eq:U_XC_simplified_4} (as well as~\eqref{eq:U_X}) for the verification can be obtained from logical operators of a pair of logical qubits of $\Q_{r_l}$ that appear in these controlled Pauli gates.
The logical operator for each logical qubit of $\Q_{r_l}$ has been calculated during compilation,
and in the same way,
each pair of logical operators that may appear in these controlled Pauli gates in the on-demand case can be calculated and stored during compilation, which we here read from the classical memory during execution.
As a whole, the runtime for the classical computation during the on-demand Clifford-state preparation gadget is
\begin{equation}
  \label{eq:runtime_gate_decomposition_clifford}
  O(N_{r_l}),
\end{equation}
which is included in the time overhead in our setting via the insertion of the wait operations in Fig.~\ref{fig:preparation_clifford}.
Nevertheless, the runtime of this classical computation is smaller than the depth of the other dominant parts, i.e.,~\eqref{eq:runtime_clifford_preparation} and~\eqref{eq:runtime_clifford_preparation_verification}.
Thus, even for the on-demand Clifford-state preparation operation, the bound of the depth of the corresponding Clifford-state preparation gadget remains the same as~\eqref{eq:depth_Clifford_all}.

\textbf{Magic-state preparation gadget}:
The level-$l$ magic-state preparation gadget is constructed as in Fig.~\ref{fig:preparation_magic}.
The gadget starts with a level-$(l-1)$ stabilizer circuit for encoding two sets of $N_{r_l}$ level-$(l-1)$ registers into logical magic states.
For each set, we use $U_\mathrm{encode}$ in~\eqref{eq:encode} to transform $K_{r_l}$ level-$(l-1)$ registers prepared in ${(R_y(\nicefrac{\pi}{4})\ket{0})}^{\otimes K^{(l-1)}}$ and $(N_{r_l}-K_{r_l})$ level-$(l-1)$ registers prepared in $\ket{0}^{\otimes K^{(l-1)}}$ into a logical state ${(R_y(\nicefrac{\pi}{4})\ket{0})}^{\otimes K^{(l)}}$ as a whole, in the same way as the initial-state preparation gadget in Fig.~\ref{fig:preparation_0}.
As a result, we have logical ${(R_y(\nicefrac{\pi}{4})\ket{0})}^{\otimes K^{(l)}}$ in each of the two sets.
We call these two sets of $N_{r_l}$ level-$(l-1)$ registers $A$ and $B$, respectively, as shown in Fig.~\ref{fig:preparation_magic}.

Similar to the initial- and Clifford-state preparation gadgets, the encoding part is non-fault-tolerant, and hence, we need verification.
In the verification,
after ensuring that the state is in the code space of $\Q_{r_l}$,
we measure all the logical stabilizer operators of logical ${(R_y(\nicefrac{\pi}{4})\ket{0})}^{\otimes K^{(l)}}$ to detect logical errors with post-selection.
Each state $R_y(\nicefrac{\pi}{4})\ket{0}$ is stabilized by
\begin{equation}
  H(R_y(\nicefrac{\pi}{4})\ket{0})=R_y(\nicefrac{\pi}{4})\ket{0}.
\end{equation}
To specify the logical state of these two encoded registers to be ${(R_y(\nicefrac{\pi}{4})\ket{0})}^{\otimes K_{r_l}}\otimes {(R_y(\nicefrac{\pi}{4})\ket{0})}^{\otimes K_{r_l}}$, we use a set of the $2K_{r_l}$ stabilizer operators of ${(R_y(\nicefrac{\pi}{4})\ket{0})}^{\otimes K_{r_l}}\otimes {(R_y(\nicefrac{\pi}{4})\ket{0})}^{\otimes K_{r_l}}$ given by
\begin{align}
  \label{eq:H_stabiliziers}
  &\Big\{(H^{\otimes k}\otimes\mathbbm{1}^{\otimes(K_{r_l}-k)})\otimes(\mathbbm{1}^{\otimes k}\otimes H^{\otimes(K_{r_l}-k)}),\nonumber\\
  &{\quad(\mathbbm{1}^{\otimes k}\otimes H^{\otimes(K_{r_l}-k)})\otimes(H^{\otimes k}\otimes \mathbbm{1}^{\otimes(K_{r_l}-k)})\Big\}}_{k\in\{1,\ldots,K_{r_l}\}},
\end{align}
where $\mathbbm{1}^{\otimes 0}=1$ and $H^{\otimes 0}=1$.
To measure the logical operators of these stabilizer operators in a transversal manner,
we here use the fact that the logical operator $H^{\otimes K_{r_l}}$ acting on all the logical qubits of each code block $\Q_{r_l}$ is given by $H^{\otimes N_{r_l}}$ acting on all the qubits in the level-$(l-1)$ registers.
To measure different operators in~\eqref{eq:H_stabiliziers},
for $k=1,\ldots,K_{r_l}$,
we iteratively perform a logical \textsc{SWAP} gate between the $k$th logical qubit of each code block $\Q_{r_l}$ in $A$ and $B$, and then measure logical $H^{\otimes K_{r_l}}$ of each code block $\Q_{r_l}$.
After finishing all the $K_{r_l}$ iterations, it is to be verified that all the logical qubits in $A$ and $B$ should be in a logical state stabilized by~\eqref{eq:H_stabiliziers}, i.e., in logical ${(R_y(\nicefrac{\pi}{4})\ket{0})}^{\otimes K^{(l)}}\otimes{(R_y(\nicefrac{\pi}{4})\ket{0})}^{\otimes K^{(l)}}$.

More specifically, for each $k\in\{1,\ldots,K_{r_l}\}$, the circuit in Fig.~\ref{fig:preparation_magic} iteratively performs a logical \textsc{SWAP} gate between a pair of the $k$th logical qubits in each code block $\Q_{r_l}$ of $A$ and $B$, so that the first $k$ logical qubits of $\Q_{r_l}$ that were initially in $B$ are brought to $A$ after the $k$th logical \textsc{SWAP} gate.
Each of the logical \textsc{SWAP} gates is implemented by the level-$l$ two-register Clifford-gate abbreviation in Fig.~\ref{fig:inblock_clifford}.
The first \textsc{SWAP} gate also serves the purpose of ensuring that the state is in the code space of $\Q_{r_l}$.
In each iteration, the circuit in Fig.~\ref{fig:preparation_magic} performs a measurement of logical $H^{\otimes K_{r_l}}$ of each code block $\Q_{r_l}$ in $A$ and $B$.
This measurement is assisted by two auxiliary level-$(l-1)$ registers prepared in $\Ket{0}^{\otimes K^{(l-1)}}$, where one is used for extracting the measurement outcomes, and the other is used as the flag qubits~\cite{PhysRevLett.121.050502}.
This measurement can be performed via controlled $H$ gates on the target in $A$ and $B$ controlled by the auxiliary level-$(l-1)$ register for the extraction, as in Refs.~\cite{yamasaki2020polylogoverhead,G6,Chamberland2019faulttolerantmagic}.
By decomposing $H=R_y(\nicefrac{\pi}{4})Z{R_y(-\nicefrac{\pi}{4})}$,
we implement these controlled $H$ gates using $R_y(\pm\nicefrac{\pi}{4})$ gates in Fig.~\ref{fig:interblock_T} and also using the $X$ basis as the basis of the control qubits, which leads to the $R_y(\pm\nicefrac{\pi}{4})$ and \textsc{CNOT} part of the circuit in Fig.~\ref{fig:preparation_magic}.
For each $k^{(l-1)}\in\{1,\ldots,K^{(l-1)}\}$,
the outcome of measuring the operator $H^{\otimes N_{r_l}}$ on the $N_{r_l}$ qubits forming the $k^{(l-1)}$th code block $\Q_{r_l}$ is extracted as that on the $k^{(l-1)}$th qubit in the first auxiliary level-$(l-1)$ register.
Given that the measured state is in the code space, this outcome coincides with that of measuring logical $H^{\otimes K_{r_l}}$ in each code block $\Q_{r_l}$.
In addition, an error on this auxiliary level-$(l-1)$ register that may propagate and cause an uncorrectable error is detected as a bit flip of the measurement outcomes on the flag qubits.
Consequently, after the $K_{r_l}$ iterations without detecting any error, it is to be verified that $A$ and $B$ should be in logical ${(R_y(\nicefrac{\pi}{4})\ket{0})}^{\otimes K^{(l)}}\otimes {(R_y(\nicefrac{\pi}{4})\ket{0})}^{\otimes K^{(l)}}$ stabilized by~\eqref{eq:H_stabiliziers}, as long as the state of $A$ and $B$ is in the code space.
As in the Clifford-state preparation gadget, the flip of each of these $O(N^{(l)})$-bit measurement outcomes can be checked using $O(N^{(l)})$ parallel processes immediately, i.e., in runtime $O(1)$.

If no error is detected either from the logical stabilizer operators~\eqref{eq:H_stabiliziers} and the flag qubits in these iterations,
then the gadget outputs the logical ${(R_y(\nicefrac{\pi}{4})\ket{0})}^{\otimes K^{(l)}}\otimes {(R_y(\nicefrac{\pi}{4})\ket{0})}^{\otimes K^{(l)}}$ prepared in this first run.
Otherwise, the gadget discards the state prepared in the first run and performs the same non-fault-tolerant level-$(l - 1)$ stabilizer circuit for the preparation and outputs the logical ${(R_y(\nicefrac{\pi}{4})\ket{0})}^{\otimes K^{(l)}}\otimes {(R_y(\nicefrac{\pi}{4})\ket{0})}^{\otimes K^{(l)}}$ prepared in the second run without verification.

The magic-state preparation gadget is fault-tolerant, that is, satisfies the conditions~\eqref{eq:prepmA} and~\eqref{eq:prepmB}.
The gadget can have only the first run (with the success in the verification) or both the first and second runs.
but in the same way as the analysis of the fault tolerance of the Clifford-state preparation gadgets,
we here explain the cases where the gadget has only the first run with $s=1$.
In particular, the analysis reduces to the following three cases:
(i) a case where a fault occurs on one of the level-$(l-1)$ locations on $A$ in the part of the level-$(l-1)$ circuit in Fig.~\ref{fig:preparation_magic} for preparing logical ${(R_y(\nicefrac{\pi}{4})\ket{0})}^{\otimes K^{(l)}}$, (ii) a case where a fault occurs on one of the level-$(l-1)$ locations on $B$ in the part for preparing logical ${(R_y(\nicefrac{\pi}{4})\ket{0})}^{\otimes K^{(l)}}$, and (iii) a case where a fault occurs on one of the level-$(l-1)$ locations in the part of the level-$(l-1)$ circuit in Fig.~\ref{fig:preparation_magic} for the verification.
In case (i), since the single fault occurs at a location on $A$, the register $B$ is prepared in logical ${(R_y(\nicefrac{\pi}{4})\ket{0})}^{\otimes K^{(l)}}$ with no error after the preparation part.
In addition, the level-$l$ two-register Clifford-gate abbreviation in Fig.~\ref{fig:inblock_clifford} for the logical \textsc{SWAP} gate ensures that the state in $A$ and $B$ is in the code space of $\Q_{r_l}$ since this Clifford-gate part has no fault.
Then the measurements of the $2K_{r_l}$ logical stabilizer operators~\eqref{eq:H_stabiliziers} in each of the $K^{(l-1)}$ code space $\Q_{r_l}$ are faultlessly carried out.
Conditioned on the success of post-selection,
it is guaranteed that the logical state of all the $2K_{r_l}\times K^{(l-1)}=K^{(l)}$ logical qubits should be ${(R_y(\nicefrac{\pi}{4})\ket{0})}^{\otimes K^{(l)}}\otimes {(R_y(\nicefrac{\pi}{4})\ket{0})}^{\otimes K^{(l)}}$.
In case (ii), the fault tolerance can be shown in the same way as case (i) with reversal of $A$ and $B$.
As for case (iii), each of $A$ and $B$ before the verification is in the logical state ${(R_y(\nicefrac{\pi}{4})\ket{0})}^{\otimes K^{(l)}}$ without any error, and the level-$(l-1)$ error in the verification does not cause uncorrectable errors at the end of the level-$(l-1)$ circuit.
In particular, a fault in implementing the level-$l$ two-register Clifford-gate abbreviation in Fig.~\ref{fig:inblock_clifford} does not cause uncorrectable errors.
To see this, if we replace level-$l$ elementary operations in the abbreviation of Fig.~\ref{fig:inblock_clifford}
with the corresponding level-$l$ gadgets and assume that these gadgets satisfy the fault-tolerant properties from~\eqref{eq:measA} to~\eqref{eq:gate2B}, then we can confirm that the resulting level-$(l-1)$ circuit for the abbreviation implementing the logical two-register gate satisfies~\eqref{eq:gate2A} and~\eqref{eq:gate2B}.
Moreover, conditioned on the success of the post-selection, a fault among level-$(l-1)$ locations acting on auxiliary level-$(l-1)$ registers in the verification does not propagate and cause logical errors since such error propagation can be detected due to the technique of flag qubits~\cite{PhysRevLett.121.050502}, as in protocols in Refs.~\cite{yamasaki2020polylogoverhead,G6,Chamberland2019faulttolerantmagic} for magic state preparation with this technique.
A correctable error before the faultless two-register Clifford-gate abbreviation is corrected in the same way as level-$l$ error-correction gadgets in Fig.~\ref{fig:error_correction} correcting such an error.
Finally, a correctable error before the faultless measurement of the logical stabilizer operators is kept correctable in the same way as the Clifford-state preparation gadget.
These cases show the fault tolerance.

The depth of the level-$(l-1)$ circuit for the level-$l$ magic-state preparation gadget is dominated by the level-$l$ two-register Clifford-gate abbreviations for the logical \textsc{SWAP} gates in the verification.
The verification part iterates the level-$l$ two-register Clifford-gate abbreviation and the measurements of the logical stabilizer operators $K_{r_l}$ times in total.
The part corresponding to the level-$l$ two-register Clifford-gate abbreviation in Fig.~\ref{fig:inblock_clifford} has the depth of $O(N_{r_l}^{2}\log(N^{(l)}))$, dominated by the level-$l$ Clifford-state preparation gadget as shown in~\eqref{eq:depth_Clifford_all}.
As for the other parts in the verification, each level-$(l-1)$ $R_y(\pm\nicefrac{\pi}{4})$-gate abbreviation has the depth of $O(\log(N^{(l)}))$ as shown in~\eqref{eq:runtime_magic}, and in each iteration, the level-$(l-1)$ $R_y(\pm\nicefrac{\pi}{4})$-gate abbreviations are used sequentially $O(N_{r_l})$ times; that is, the depth of the level-$(l-1)$ $R_y(\pm\nicefrac{\pi}{4})$-gate abbreviations in each iteration is $O(N_{r_l}\log(N^{(l)}))$.
The depth of performing the \textsc{CNOT}-gate operations for measuring the logical stabilizer operators is $O(N_{r_l})$ in each iteration.
The other parts in each iteration have a constant depth including the $O(1)$ runtime of classical computation for the post-selection.
Repeating $K_{r_l}$ times, the verification part as a whole has the depth of $K_{r_l}\times O(N_{r_l}^{2}\log(N^{(l)}))$, which dominate the depth of the gadget.
As for other non-dominant parts, the depth of the part for implementing $U_\mathrm{encode}$ is $O(N_{r_l}\log(N_{r_l}))$ as shown in~\eqref{eq:depth_encoding}.
The preparation operations at the beginning of the gadget are performed sequentially in the $O(N_{r_l})$ depth.
Finally, the depth of the second run of the gadget is smaller than that of the first run; thus, the depth of the gadget including the first and second runs is at most twice as long as the first run.
Consequently, the depth of the gadget is
\begin{equation}
  \label{eq:depth_magic}
K_{r_l}\times O(N_{r_l}^{2}\log(N^{(l)}))=O(N_{r_l}^{3}\log(N^{(l)})).
\end{equation}
The required number of bits for the classical computation is dominated by $O(N^{(l)2})$ bits used for the level-$l$ Clifford-state preparation gadgets included in this gadget.

\section{\label{sec:threshold_proof}Proof of existence of threshold for doubly exponential error suppression}

In this section, we present the proof of the existence of a threshold for doubly exponential error suppression in the concatenation level $L$ achieved by the fault-tolerant protocol constructed in Secs.~\ref{sec:gadget} and~\ref{sec:fault_tolerant_gadget}.
To prove the existence of a threshold,
the existing fault-tolerant protocol with constant space overhead in Refs.~\cite{gottesman2014faulttolerant,PhysRevA.87.020304,8555154} assumes that classical computation for the decoder for the quantum LDPC codes runs instantaneously in zero time, and it has been unknown whether constant-space-overhead FTQC is possible if the classical computation has nonzero runtime that grows on the large scales, as in our setting.
The problem of requiring the zero (i.e., non-growing) runtime of classical computation is rooted even in conventional protocols based on vanishing-rate quantum LDPC codes such as the $2$D surface code,
where decoders must process the stream of syndrome data at the rate it is received even if the code distance grows~\cite{10.1109/ISCA45697.2020.00053,https://doi.org/10.48550/arxiv.2208.01178,https://doi.org/10.48550/arxiv.2209.08552,https://doi.org/10.48550/arxiv.2209.09219,bombin2023modular}.
After all, if we need to wait for a growing runtime of the decoder for the large-size quantum LDPC codes for sufficient error suppression, physical qubits suffer from more errors during waiting; in this case, the assumption of having a constant physical error rate between performing the error corrections may be violated, and hence, the proof technique for the threshold theorem based on the quantum LDPC codes is no longer applicable in a straightforward way.
For example, in the local stochastic error model as considered in the analysis of the existing constant-space-overhead protocol~\cite{gottesman2014faulttolerant} and our work, time-correlated errors at physical error rate $p$ can occur in such a way that each time period of time length $\nicefrac{1}{p}$ on each physical qubit always includes exactly one faulty location, uniformly at random over the $\nicefrac{1}{p}$ locations.
Then, when the runtime $T_\mathrm{decoder}$ of the decoder on a large scale exceeds a constant time
\begin{equation}
  \label{eq:growing_runtime}
  T_\mathrm{decoder}>\nicefrac{1}{p},
\end{equation}
physical qubits simultaneously suffer from errors while waiting for the decoder with unit probability, which are uncorrectable.
To correct a general class of errors in such a situation, a novel technique should be needed, such as another nontrivial time scheduling of quantum error correction and gate implementation, progressing beyond those for the existing constant-space-overhead protocol based on quantum LDPC codes~\cite{gottesman2014faulttolerant,PhysRevA.87.020304,8555154}.
By contrast, the crucial contribution here is to develop the technique of using the concatenated code with the non-vanishing rate, rather than the quantum LDPC codes.
Owing to this technique, the assumption that the runtime of classical computation should never grow on large scales can be avoided.
Our analysis proves the existence of a threshold in the local stochastic error model even if we take into account the growth of the nonzero runtime of classical computation.

Note that, for the quantum expander code used for the existing constant-space-overhead protocol~\cite{PhysRevA.87.020304,gottesman2014faulttolerant,8555154}, Ref.~\cite{Grospellier} claims to construct a decoder that has $O(1)$ runtime on arbitrary large scales, which would avoid the above growth of the runtime of the decoder if implemented without any architectural time overhead.
However, the $O(1)$ runtime of this decoder depends on an assumption that the decoder can receive an arbitrarily long bit string as the input (e.g., by reading it from a classical memory at the hardware level) and can return an arbitrarily long bit string as the output (e.g., by writing it into the memory) within only a constant time on the arbitrarily large scales.
In particular, the decoder receives measurement syndromes for the large-size code block of the quantum LDPC code as the input, runs many parallel processes in such a way that each process runs in $O(1)$ time, and then, from the results of these processes, returns estimation of the overall errors in the large-size code block as the output.
If such an input/output model of the decoder were implemented naively, to meet the requirement of the constant time in Ref.~\cite{Grospellier},
the input/output speed between the decoder and the memory would diverge to infinity on large scales.
In practice, hardware implementations of decoders in physical experiments are challenged by architectural overheads~\cite{9586326,Bourassa2021blueprintscalable}, and at the hardware level, classical input/output interface may incur a growing architectural time overhead on large scales.
It is unknown how one can implement the input/output model of the above decoder within a constant time at the hardware level.
Problematically, it has been difficult for the existing constant-space-overhead protocol~\cite{PhysRevA.87.020304,gottesman2014faulttolerant,8555154} to tolerate such growing architectural time overheads in implementing the decoder due to~\eqref{eq:growing_runtime}.
By contrast, the following analysis of the existence of a threshold shows that our fault-tolerant protocol can tolerate polynomially growing architectural time overheads, as discussed later in~\eqref{eq:poly_overhead_threshold}.

\begin{figure*}[t]
  \centering
  \includegraphics[width=7.0in]{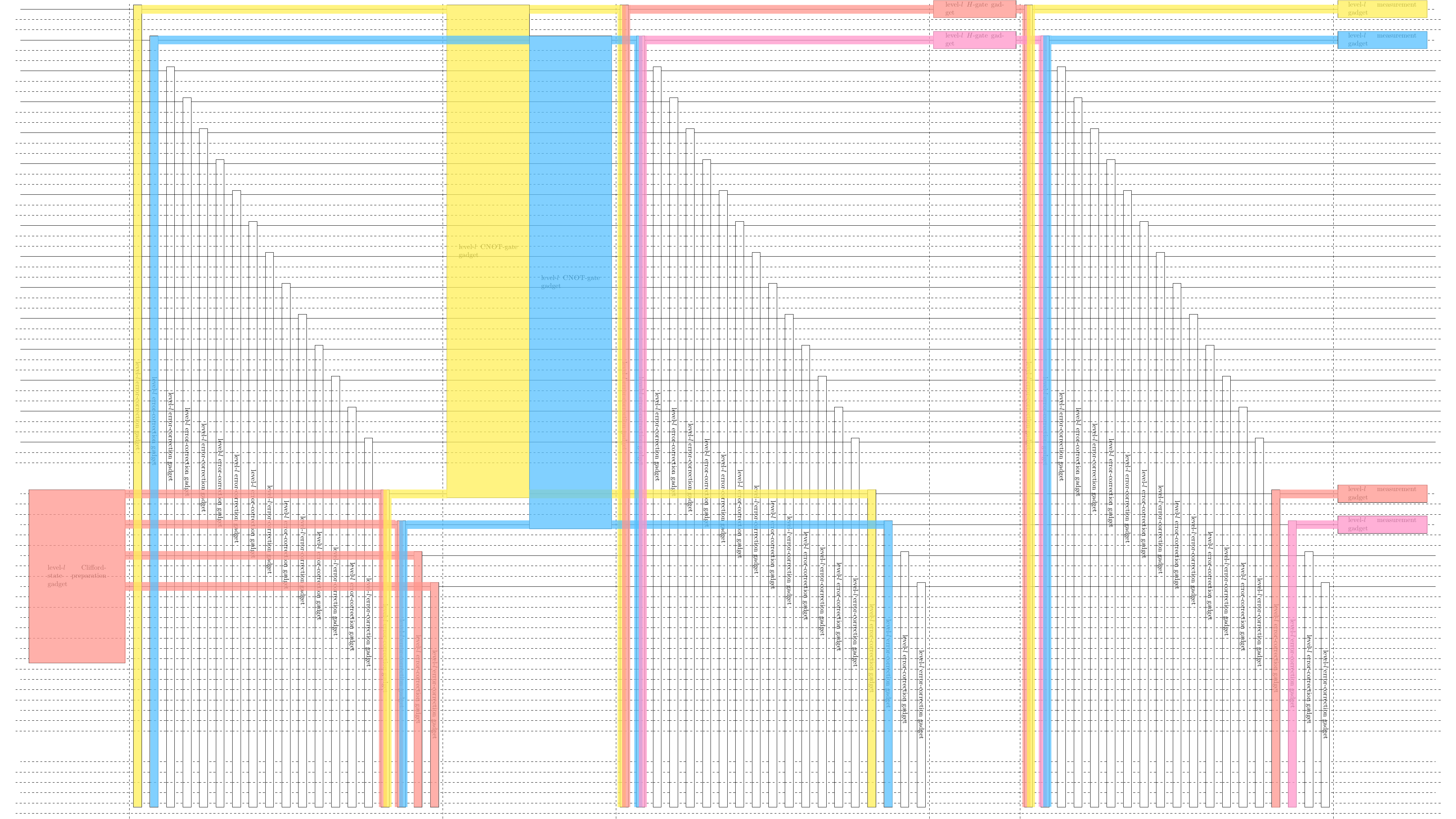}
  \caption{\label{fig:exrec}Extended rectangles (ExRecs) in the level-$(l-1)$ circuit on the right of Fig.~\ref{fig:circuit_conversion} for implementing a logical level-$l$ circuit. Each level-$l$ ExRec corresponding to a level-$l$ location is a part of the level-$(l-1)$ circuit (colored in the figure) consisting of the level-$l$ gadget corresponding to a level-$l$ location, all the level-$l$ error-correction gadgets placed between the location and the adjacent locations, and the level-$(l-1)$ wait operations inserted between these gadgets. Note that the figure does not color the level-$l$ ExRecs corresponding to level-$l$ wait operations for simplicity.}
\end{figure*}

In particular, by suitably modifying the proof of the threshold theorem for the concatenated code~\cite{G},
our proof is done by counting the number of locations in extended rectangles (ExRecs), so as to bound a probability of faults at a concatenation level in terms of that at a lower level.
Given a level-$l$ circuit and the corresponding level-$(l-1)$ circuit, for each level-$l$ location of the level-$l$ circuit, the level-$l$ ExRec corresponding to the level-$l$ location is defined as a part of the level-$(l-1)$ circuit consisting of the level-$l$ gadget corresponding to the level-$l$ location, all the level-$l$ error-correction gadgets placed between the location and the adjacent locations, and the level-$(l-1)$ wait operations inserted between these gadgets.
A difference from Ref.~\cite{G} arises from the fact that the error correction on each set of $O(N_{r_l})$ encoded level-$l$ registers in our fault-tolerant protocol are performed in a synchronized way as shown in Fig.~\ref{fig:circuit_conversion},
and for the synchronization, our protocol inserts level-$(l-1)$ wait operations between the gadgets.
Due to this difference, a level-$l$ ExRec for a level-$l$ location in our protocol includes the level-$(l-1)$ wait operations between the level-$l$ gadget for the location and each adjacent level-$l$ error-correction gadget.
See Fig.~\ref{fig:exrec} for an illustration of ExRecs in our protocol.

Let $A(l)$ be the maximum number of pairs of locations in the level-$(l-1)$ circuit of a level-$l$ ExRec, where the maximum is taken over all the possible choices of the level-$l$ ExRecs\@.
In the following, we discuss an upper bound of $A(l)$ in the cases of $l<L$ while the case of $l=L$ will also be discussed subsequently.
To bound $A(l)$, it is crucial to use the fact that the maximum depth $G(l)$ of the level-$(l-1)$ circuit for any level-$l$ gadget used in our fault-tolerant protocol is bounded polynomially in $N_{r_l}$ by~\eqref{eq:G_l} and~\eqref{eq:G_l_scaling}, i.e.,
\begin{equation}
  G(l)=O(\poly(N_{r_l}))=\exp(O(l)),
\end{equation}
which has been checked in Sec.~\ref{sec:fault_tolerant_gadget} and is dominated by~\eqref{eq:depth_magic} of the magic-state preparation gadgets in the worst case.
Note that $G(l)$ is an upper bound of the sum of the depth of the gadget and that of wait operations inserted for synchronizing error correction as shown in Fig.~\ref{fig:circuit_conversion}.
We stress that $G(l)$ in our analysis of Secs.~\ref{sec:gadget} and~\ref{sec:fault_tolerant_gadget} includes the wait operations to wait for nonzero-time classical computation such as those in the decoder and the gate teleportation.
A level-$l$ elementary operation acts on at most $8$ level-$l$ registers, where $8$ are used for the level-$l$ magic-state preparation in~\eqref{eq:magic_state_preparation}.
A level-$l$ gadget corresponding to a level-$l$ elementary operation uses $(N_{r_l}+8)$ level-$(l-1)$ registers per encoded level-$l$ register.
Thus,
the number of level-$(l-1)$ locations of the level-$l$ gadget and the wait operations inserted for synchronizing error correction is at most
(see also Fig.~\ref{fig:circuit_conversion})
\begin{equation}
  8(N_{r_l}+8)G(l).
\end{equation}
As for error correction, a level-$l$ error-correction gadget uses at most $(3N_{r_l}+2)$ level-$(l-1)$ registers as shown in~\eqref{eq:error_correction}.
Thus, the number of level-$(l-1)$ locations used for each level-$l$ error-correction gadget is
\begin{equation}
  (3N_{r_l}+2)G(l).
\end{equation}
As shown in Fig.~\ref{fig:circuit_conversion},
each time period for error correction performs at most $(N_{r_{l+1}}+8)$ error-correction gadgets sequentially.
Each level-$l$ elementary operation has at most $4$ adjacent level-$l$ elementary operations, as in the case of~\eqref{eq:cnot_gate},~\eqref{eq:cz_gate}, and~\eqref{eq:clifford_state_preparation}; that is, each level-$l$ ExRec has at most $4$ level-$l$ error-correction gadgets in addition to one level-$l$ gadget to implement a level-$l$ elementary operation.
As a whole, the number of level-$(l-1)$ locations in a level-$l$ ExRec is upper bounded by (see also Fig.~\ref{fig:circuit_conversion})
\begin{align}
  \label{eq:exrec_bound}
  &8(N_{r_l}+8)G(l)+4(N_{r_{l+1}}+8)(3N_{r_l}+2)G(l)\nonumber\\
  &=[8(N_{r_l}+8)+4(N_{r_{l+1}}+8)(3N_{r_l}+2)]G(l).
\end{align}
Therefore, in the cases of $l<L$, it holds that
\begin{align}
  A(l)&\leqq \binom{[8(N_{r_l}+8)+4(N_{r_{l+1}}+8)(3N_{r_l}+2)]G(l)}{2}\\
      &=O(\poly(N_{r_l}))\\
      &=\exp({O(l)}).
\end{align}
As for the case of $l=L$,
since the number of error-correction gadgets sequentially used in each time period of error correction is at most $9$ ($\leqq N_{r_{L+1}}+8$) as shown in Fig.~\ref{fig:circuit_conversion},
the above upper bound of $A(l)$ holds for $l=L$.
Thus, to simplify the presentation, our analysis will use a constant factor $\alpha>0$ such that
\begin{equation}
  \label{eq:A_n}
  A(l)\leqq 2^{\alpha l}
\end{equation}
for all $l\in\{1,\ldots,L\}$.
Note that these upper bounds may not be tight but are sufficient for the following analysis.

To bound the error rate of a level-$l$ circuit in terms of those in the corresponding level-$(l-1)$ circuit,
a conventional argument in Ref.~\cite{G} under the assumption of the local stochastic error model is applied here.
In the same way as the local stochastic error model at the physical level as in~\eqref{eq:local_stochastic_error_model_physical},
for a level-$l$ circuit at each level $l\in\{1,\ldots, L\}$, we say that the circuit undergoes a local statistical error model if the faults occurring in the circuit satisfy the following: (i) a set $S$ of faulty level-$l$ locations in the level-$l$  circuit is chosen with probability $p^{(l)}(S)$, and errors occur at the level-$l$ locations in $S$ in such a way that the level-$l$ elementary operations at the level-$l$ locations in $S$ are replaced with an arbitrary quantum channel $\mathcal{E}$ that is consistent with the causal order of the locations in $S$; (ii) each level-$l$ location $i$ in the circuit has a parameter $p_{l,i}$ such that for any set $R$ of level-$l$ locations, the probability $\Pr\{S\supseteqq R\}$ of having faults at every level-$l$ location in $R$ is at most
\begin{equation}
  \prod_{i\in R} p_{l,i}.
\end{equation}
We call the parameter $p_{l,i}$ a logical error rate for a level-$l$ location $i$ in the level-$l$ circuit.
For simplicity of presentation, we consider a single parameter $p_{l}$ such that $p_{l,i}\leqq p_{l}$ holds for all $i$.
If obvious from the context, we may simply call $p_0$ the physical error rate, and $p_l$ the logical error rate at level $l$.

For each $l$, suppose that level-$l$ gadgets are fault-tolerant in the sense of the definitions of the equivalence relations shown in Sec.~\ref{sec:gadget}, and that the level-$(l-1)$ circuit undergoes a local stochastic error model.
Then, the argument in Ref.~\cite{G} proves that the level-$l$ circuit also undergoes a local stochastic error model.
As discussed in Sec.~\ref{sec:setting} with~\eqref{eq:local_stochastic_error_model_physical}, the level-$0$ circuit is assumed to undergo the local stochastic error model in our setting.
Thus, using such an argument recursively, the level-$l$ circuits at every level $l\in\{1,\ldots,L\}$ are shown to undergo a local stochastic error model.
Moreover, since no single fault at a level-$(l-1)$ location in a level-$l$ ExRec leads to a fault of the corresponding level-$l$ location, the argument in Ref.~\cite{G} shows that the logical error rate $p_{l}$ at level $l$ is bounded in terms of $p_{l-1}$ at level $(l-1)$ by
\begin{equation}
  p_{l}\leqq A(l){(p_{l-1})}^2.
\end{equation}

Therefore, it follows from~\eqref{eq:A_n} that
\begin{align}
  \label{eq:bound}
  p_{l}&\leqq 2^{\alpha l}{(p_{l-1})}^2;
\end{align}
that is, $p_l$ satisfies
\begin{equation}
  \label{eq:bound_g_l}
  2^{g(l)}p_l \leqq {(2^{g(l-1)}p_{l-1})}^2,
\end{equation}
where
\begin{equation}
  \label{eq:g}
  g(l)\coloneqq \alpha(l+2).
\end{equation}
Using~\eqref{eq:bound_g_l} recursively, it holds that
\begin{align}
  p_L&\leqq 2^{-g(L)}{(2^{g(0)}p_0)}^{2^L}\\
     &=\frac{{(2^{2\alpha}p_0)}^{2^L}}{2^{\alpha(L+2)}}.
\end{align}
Consequently, there exists a threshold
\begin{equation}
  \label{eq:p_th}
  p_\mathrm{th}\geqq 2^{-2\alpha}(>0)
\end{equation}
such that for any nonzero physical error rate $p_0\in(0,p_\mathrm{th})$ below the threshold,
the logical error rate $p_L$ decreases doubly exponentially as $L$ increases, i.e.,
\begin{equation}
  \label{eq:p_L}
  p_L\leqq{\left(\frac{p_0}{p_\mathrm{th}}\right)}^{2^L}p_\mathrm{th},
\end{equation}
achieving the same scaling as the doubly exponential error suppression in $L$ with the conventional concatenated codes~\cite{G}.
This conclusion is summarized as Proposition~1 in the main text.

From the argument here, $2\alpha$ on the right-hand side of~\eqref{eq:p_th} is crucial for obtaining a concrete value of a lower bound of $p_\mathrm{th}$ while the exact evaluation of $p_\mathrm{th}$ may require numerical simulation.
To make this lower bound of $p_\mathrm{th}$ larger, $\alpha$ should be small, and hence, it is essential to make the size of ExRecs (or $A(l)$ in~\eqref{eq:A_n}) as small as possible.
For further optimization of the ExRecs, it would be crucial to reduce the size of Clifford- and magic-state preparation gadgets and also optimize the scheduling of error correction rather than just performing it sequentially between level-$l$ operations (as in Fig.~\ref{fig:circuit_conversion}).

Note that the above proof of the doubly exponential error suppression in $L$ holds even in more general cases of
\begin{equation}
  \label{eq:poly_overhead_threshold}
  A(l)=\exp({O(\poly(l))}).
\end{equation}
For example, suppose that classical computation for the decoder or the gate teleportation in a gadget had a polynomial runtime $O(N^{(l)})=\exp({O(l^2)})$ in the code size $N^{(l)}$ of $\Q^{(l)}$, which would be much slower than our protocol but may be the case where the number of parallel processes in classical computation per level-$l$ register is limited.
Even in such a case, we would have $A(l)=\exp({O(l^2)})\leqq 2^{\alpha l^2}$ for some $\alpha$, and with a choice of $g(l)=\alpha(l^2+4l+6)$ in place of $g$ in~\eqref{eq:g}, we can prove $p_L\leqq\nicefrac{{(2^{6\alpha}p_0)}^{2^L}}{2^{\alpha(L^2+4L+6)}}$ from the same argument as the above.
Then, a threshold $p_\mathrm{th}\geqq 2^{-6\alpha} (>0)$ in place of~\eqref{eq:p_th} still exists.
In summary, the above argument leads to the following theorem.
\begin{theorem}[\label{thm:general_logical_error_rate}Threshold theorem and logical error rate]
  Suppose that we have a fault-tolerant protocol using concatenation of $[[n_l,k_l,2t+1]]$ quantum codes with $n_l=\exp(O(\poly(l)))$ at each concatenation level $l\in\{1,2,\ldots\}$, where $t\geqq 1$,
  the level-$l$ gadgets are fault-tolerant in terms of the equivalence relations translated from Ref.~\cite{G} as those in Sec.~\ref{sec:fault_tolerant_gadget} in the case of $t=1$, and each level-$l$ ExRec is composed of $\exp(O(\poly(l)))$ level-$(l-1)$ elementary operations.
  Then, under the local stochastic error model at physical error rate $p_0>0$,
  there exists a nonzero threshold $p_\mathrm{th}>0$ such that the level-$l$ circuit for each $l$ undergoes the local stochastic error model, and the logical error rate $p_l$ at concatenation level $l$ is bounded by
  \begin{equation}
    p_l\leqq{\left(\frac{p_0}{p_\mathrm{th}}\right)}^{{(t+1)}^l}p_\mathrm{th}.
  \end{equation}
\end{theorem}

Remarkably, a practical threshold better than $p_\mathrm{th}$ in~\eqref{eq:p_th} is achievable with minor modifications of our fault-tolerant protocol.
Our results make it possible to achieve such a practical threshold with constant space overhead and parallelizability.
Below we show techniques for such improvement of the threshold.

Firstly, instead of starting concatenation as $\Q_{r_1},\Q_{r_2},\ldots$, the $7$-qubit code $\Q_3$ can be used in the first \textit{constant} times of the concatenation; in this case, the sequence of codes in the concatenation becomes $\Q_3,\ldots,\Q_3,\Q_{r_1},\Q_{r_2},\ldots$.
That is, we can use the logical qubit of the constant-size concatenated $7$-qubit code in place of each physical qubit of $\Q^{(L)}$.
With this modification, even if the physical error rate $p_0$ is larger than the threshold $p_\mathrm{th}$ for $\Q^{(L)}$ itself,
the concatenated $7$-qubit code can reduce the logical error rate from $p_0$ to a rate below $p_\mathrm{th}$, so that the rest of the concatenation $\Q_{r_1},\Q_{r_2},\ldots$ can arbitrarily suppress the logical error rate within the constant space overhead.
In this way, the threshold of concatenating $\Q_3,\ldots,\Q_3,\Q_{r_1},\Q_{r_2},\ldots$ reduces to that of the concatenated $7$-qubit code, with only increasing a constant factor of the space overhead for concatenating $\Q_3$ constant times.

Secondly, we can use $\Q_3$ in this concatenation as the error-detecting code to detect two errors, rather than the error-correcting code to correct one error.
To use $\Q_3$ as the error-detecting code,  we discard the states whenever we detect any error in the decoder.
As long as this post-selection for error detection is performed only during the first constant times of the concatenation, the overhead of the post-selections is still constant.
The post-selection techniques combined with the concatenated code can conventionally achieve \textit{state-of-the-art high thresholds}~\cite{K5, K6,yamasaki2020polylogoverhead}, which can be even better than other leading candidates such as the topological codes.

Finally, we can also concatenate an \textit{arbitrary} quantum code and $\Q^{(L)}$ as long as logical operations that we use in place of physical operations for our fault-tolerant protocol at concatenation level $0$ are implementable for such a quantum code.
For example, the surface code has well-established procedures for implementing logical operations for universal quantum computation~\cite{Horsman_2012,Litinski2019gameofsurfacecodes}; thus, we can use the logical qubit of a constant-size surface code in place of each physical qubit of $\Q^{(L)}$ to achieve the same threshold as that of the surface code, and at the same time attain the constant overhead asymptotically.
Indeed, any quantum code with one logical qubit can be concatenated with $\Q^{(L)}$ by using the logical qubit of the code in place of each physical qubit of $\Q^{(L)}$, as long as the code can implement a set of logical operations that our fault-tolerant protocol uses on physical qubits, namely, a measurement in the $Z$ basis, the $H$, $S$, $\textsc{CNOT}$, $CZ$, Pauli, and $R_y(\pm\nicefrac{\pi}{4})$ gates, and preparation of $\Ket{0}$.

Given the fact that current quantum technology is insufficient for realizing physical error rate below a threshold in a scalable way, the significance of these modifications of our protocol is to provide flexibility in the protocol design from the theoretical side to reduce the technological demands in terms of the threshold while maintaining the constant overhead.
However, these modifications may increase the space and time overheads of the fault-tolerant protocol up to a constant factor.
Therefore, depending on the physical error rate achievable in experiments, the protocol with these modifications should be optimized to balance the trade-off between the overhead and the threshold.
The optimization of the protocol may require further assumptions on technological advances, which we leave for future work.
The techniques developed here constitute a fundamental step for further optimization of the fault-tolerant protocols.

\section{\label{sec:overhead}Overhead}

In this section, we prove that our fault-tolerant protocol achieves the constant space overhead.
Furthermore, we also clarify the time overhead of our protocol.
In particular, recalling that the task of FTQC formulated in Sec.~\ref{sec:setting} is to simulate the given $W(M)$-qubit $D(M)$-depth original circuit within an error $\epsilon$,
we will show that the time overhead of our protocol is quasi-polylogarithmic, i.e.,
\begin{equation}
  \exp\left(O\left(\log^2\left(\log\left(\frac{M}{\epsilon}\right)\right)\right)\right).
\end{equation}
Subsequently, we will also discuss how we can modify our protocol to improve the scaling of the time overhead to
\begin{equation}
  \exp\left(O\left(\log\log\left(\frac{M}{\epsilon}\right)\log\log\log\left(\frac{M}{\epsilon}\right)\right)\right).
\end{equation}

Starting from the $W(M)$-qubit $D(M)$-depth original circuit, our protocol in Fig.~\ref{fig:circuit_compilation} uses $\lceil\nicefrac{W(M)}{K^{(L)}}\rceil$ level-$L$ registers to represent the $W(M)$ qubits and further adds $8+2=10$ auxiliary level-$L$ registers per register; in total, the level-$L$ circuit has
\begin{equation}
  \label{eq:level_L_space}
  \text{$11\lceil\nicefrac{W(M)}{K^{(L)}}\rceil$ level-$L$ registers}.
\end{equation}
The depth of the level-$L$ circuit is bounded as follows.
The $W(M)$-qubit $D(M)$-depth original circuit consists of $D(M)$ one-depth parts.
For each of the one-depth parts, we perform the Clifford gates in this part using the level-$L$ Clifford-gate abbreviations~\eqref{eq:clifford_abbreviation} and perform $R_y(\pm\nicefrac{\pi}{4})$ gates in this part by the level-$L$ $R_y(\pm\nicefrac{\pi}{4})$-gate abbreviations~\eqref{eq:r_y_abbreviation}.
A single use of level-$L$ Clifford-gate abbreviation can apply two-qubit Clifford gates only between qubits in a pair of level-$L$ registers on which the abbreviation acts.
To perform two-qubit Clifford gates to qubits in different pairs of the level-$L$ registers, we use different level-$L$ Clifford-gate abbreviations.
To bound the required depth of the level-$L$ circuit for applying all the combinations of Clifford gates in a one-depth part of the original circuit,
we consider, for each one-depth part, a graph with $\lceil\nicefrac{W(M)}{K^{(L)}}\rceil$ vertices, where each vertex represents one of the level-$L$ registers, and an edge connects a pair of vertices if and only if a two-qubit Clifford gate in this one-depth part connects the qubits in the corresponding pair of level-$L$ registers.
Since each level-$L$ register has $K^{(L)}$ qubits, the qubits in a level-$L$ register may be connected with those in at most $K^{(L)}$ other level-$L$ registers by two-qubit Clifford gates in each one-depth part.
Thus, this graph has at most degree $K^{(L)}$; that is, the graph is $(K^{(L)}+1)$-edge colorable~\cite{B22}.
For all the pairs of the vertices connected by the edges in the same color, we can in parallel perform the level-$L$ Clifford-gate abbreviations on the corresponding pairs of level-$L$ registers.
Thus, we can perform any possible combination of Clifford gates in each one-depth part of the original circuit by at most $(K^{(L)}+1)$ parallel uses of the level-$L$ two-register Clifford-gate abbreviations in the level-$L$ circuit.
Furthermore, we can perform the $R_y(\pm\nicefrac{\pi}{4})$ gates in each one-depth part by one parallel use of the level-$L$ $R_y(\pm\nicefrac{\pi}{4})$-gate abbreviations.
As a result, from the $W(M)$-qubit $D(M)$-depth original circuit, we obtain the compiled level-$L$ circuit on $11\lceil\nicefrac{W(M)}{K^{(L)}}\rceil$ level-$L$ registers composed of $O(K^{(L)}D(M))$ parallel uses of level-$L$ abbreviations.
Since each level-$L$ abbreviation has at most $O(\log(N^{(L)}))$ depth due to~\eqref{eq:runtime_clifford} and~\eqref{eq:runtime_magic},
the depth of the level-$L$ circuit is bounded by
\begin{equation}
  \label{eq:level_L_time}
  O(K^{(L)}\log(N^{(L)})\times D(M)).
\end{equation}
The number of locations in the level-$L$ circuit is
\begin{align}
  \label{eq:locations_level_L}
  &11\lceil\nicefrac{W(M)}{K^{(L)}}\rceil\times O(K^{(L)}\log(N^{(L)})\times D(M))\nonumber\\
  &=O(W(M) D(M)\log(N^{(L)}))\nonumber\\
  &\leqq C W(M) D(M)L^2,\quad\text{for large $M$},
\end{align}
where the last line follows from $N^{(L)}=\exp(O(L^2))$ in~\eqref{eq:n_L_scaling}, and $C$ is a constant factor.

For the fixed target error $\epsilon$ in simulating the original circuit,
we then derive the required concatenation level $L_M$.
On one hand, to achieve the target error $\epsilon$ with~\eqref{eq:locations_level_L},
the required logical error rate is
\begin{equation}
  \label{eq:required_p_L}
  p_L\leqq \frac{\epsilon}{C W(M) D(M)L^2}.
\end{equation}
On the other hand, the result~\eqref{eq:p_L} of our threshold analysis shows that,
to achieve~\eqref{eq:required_p_L}, it suffices to choose $L$ such that
\begin{align}
  2^L\geqq\log_{\nicefrac{p_\mathrm{th}}{p_0}}\left(\frac{C W(M) D(M)}{\nicefrac{\epsilon}{p_\mathrm{th}}}\right)+2\log_{\nicefrac{p_\mathrm{th}}{p_0}}\left(L\right),
\end{align}
i.e.,
\begin{align}
  L&=\Omega\left(\log\left(\log\left(\frac{W(M) D(M)}{\epsilon}\right)\right)\right)\\
   &=\Omega\left(\log\left(\log\left(\frac{M}{\epsilon}\right)\right)\right),
\end{align}
where the last line follows from $\log(W(M)D(M))=\log(\poly(M))\approx\log(M)$.
To keep the code $\mathcal{Q}^{(L)}$ used for our fault-tolerant protocol as small as possible, we here choose, up to a constant factor,
\begin{align}
  \label{eq:L}
  L_M&=\Theta\left(\log\left(\log\left(\frac{M}{\epsilon}\right)\right)\right).
\end{align}
For later convenience,
the number of physical and logical qubits of $\Q_{r_{L_M}}$ grows, due to $N_{r_L}=\exp(O(L))$ in~\eqref{eq:N_r_scaling} and $K_{r_L}=\exp(O(L))$ in~\eqref{eq:K_r_scaling}, as
\begin{align}
  N_{r_{L_M}}&=\exp\left(O\left(\log\left(\log\left(\frac{M}{\epsilon}\right)\right)\right)\right)\\
             &=O\left(\polylog\left(\frac{M}{\epsilon}\right)\right),\\
  K_{r_{L_M}}&=\exp\left(O\left(\log\left(\log\left(\frac{M}{\epsilon}\right)\right)\right)\right)\\
             &=O\left(\polylog\left(\frac{M}{\epsilon}\right)\right).
\end{align}
On the other hand, due to $N^{(L)}=\exp(O(L^2))$ in~\eqref{eq:n_L_scaling} and $K^{(L)}=\exp(O(L^2))$ in~\eqref{eq:k_L_scaling},
each level-$L_M$ register has a quasi-polylogarithmic size in $M$, i.e.,
\begin{align}
\label{eq:N_L_overhead}
N^{(L_M)}&=\exp\left(O\left(\log^2\left(\log\left(\frac{M}{\epsilon}\right)\right)\right)\right),\\
\label{eq:K_L_overhead}
K^{(L_M)}&=\exp\left(O\left(\log^2\left(\log\left(\frac{M}{\epsilon}\right)\right)\right)\right).
\end{align}

With the above choice of $L_M$,
our protocol achieves the constant space overhead as shown in the following.
Starting from the $11\lceil\nicefrac{W(M)}{K^{(L_M)}}\rceil$ level-$L_M$ registers in~\eqref{eq:level_L_space}, we recursively replace each level-$l$ register with $(N_{r_l}+8+2)=(N_{r_l}+10)$ level-$(l-1)$ registers as in~\eqref{eq:qubit_requirement},
except that we replace a level-$1$ register with $N_{r_1}$ physical qubits without auxiliary ones.
Thus, the total required number of physical qubits is given by
\begin{align}
  W_\mathrm{FT}(M)&=11\lceil\nicefrac{W(M)}{K^{(L_M)}}\rceil\times \left(N_{r_1}\prod_{l=2}^{L_M}(N_{r_l}+10)\right);
\end{align}
i.e., we have
\begin{align}
  &\frac{W_\mathrm{FT}(M)}{W(M)}\\
  &=\frac{11\lceil\nicefrac{W(M)}{K^{(L_M)}}\rceil\times (N_{r_1}\prod_{l=2}^{L_M}(N_{r_l}+10))}{W(M)}\\
  &\leqq\frac{11(\nicefrac{W(M)}{K^{(L_M)}}+1)\times (N_{r_1}\prod_{l=2}^{L_M}(N_{r_l}+10))}{W(M)}\\
  &= 11\left(\frac{N_{r_1}}{K_{r_1}}\prod_{l=2}^{L_M}\frac{N_{r_l}+10}{K_{r_l}}\right)\left(1+\frac{K^{(L_M)}}{W(M)}\right)\\
  \label{eq:space_overhead_result}
  &\leqq 11\left(\frac{N_{r_1}}{K_{r_1}}\prod_{l=2}^{L_M}\frac{N_{r_l}+10}{K_{r_l}}\right)\left(1+\frac{K^{(L_M)}}{M}\right),
\end{align}
where the last line follows from $W(M)\geqq M$ in~\eqref{eq:W}.
The factor $\left(1+\nicefrac{K^{(L_M)}}{M}\right)$ in~\eqref{eq:space_overhead_result} converges to $1$ as $M\to\infty$ since $K^{(L_M)}$ is quasi-polylogarithmic in $M$ as shown in~\eqref{eq:K_L_overhead}.
To see that the dominant factor in~\eqref{eq:space_overhead_result}, i.e.,
\begin{equation}
  \label{eq:dominant_space_overhead}
  11\left(\frac{N_{r_1}}{K_{r_1}}\prod_{l=2}^{L_M}\frac{N_{r_l}+10}{K_{r_l}}\right)
\end{equation}
is $O(1)$ as $M\to\infty$,
recall that the crucial condition in deriving the non-vanishing rate~\eqref{eq:rate} of $\mathcal{Q}^{(L)}$ has been the convergence of the infinite sum~\eqref{eq:infinite_sum}.
Since this condition also holds in the case here, i.e.,
\begin{equation}
\label{eq:infinite_sum_plus_ten}
  \sum_{l=2}^{\infty}\left|\frac{(N_{r_l}+10)-K_{r_l}}{K_{r_l}}\right|<\infty,
\end{equation}
in the same way as showing~\eqref{eq:rate},
it holds that
\begin{equation}
  \label{eq:space_formula}
  \prod_{l=2}^{L_M}\frac{N_{r_l}+10}{K_{r_l}}=O(1),\quad\text{as $M\to\infty$}.
\end{equation}
Therefore, the space overhead~\eqref{eq:space_overhead} is
\begin{equation}
\label{eq:space_overhead_conclusion}
  \frac{W_\mathrm{FT}(M)}{W(M)}=O(1),\quad\text{as $M\to\infty$},
\end{equation}
achieving the constant space overhead.

\begin{figure}[t]
  \centering
  \includegraphics[width=3.4in]{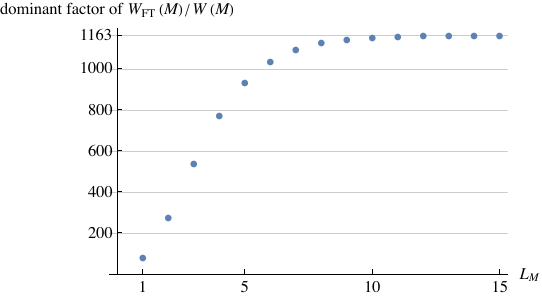}
  \caption{\label{fig:space_overhead}The dominant factor of the space overhead (i.e.,~\eqref{eq:dominant_space_overhead}) in our fault-tolerant protocol including all the auxiliary qubits used at a time. The space overhead at each time step is upper bounded by the factor of $1163$ for an arbitrarily large concatenation level $L_M$ while the doubly exponential suppression of the logical error rate is achieved as $L_M$ increases.}
\end{figure}

For concreteness, in Fig.~\ref{fig:space_overhead}, we numerically evaluate the dominant factor~\eqref{eq:dominant_space_overhead} of the space overhead in~\eqref{eq:space_overhead_result}.
The figure shows that even if we include all the auxiliary qubits,
the space overhead of our fault-tolerant protocol at each time step is upper bounded by the factor of $1163$ as a whole on arbitrarily large scales.
Note that the significance of our work is to develop the fundamental techniques for our protocol to achieve the constant space overhead, and we leave further optimization of the constant factor for future research.

Regarding the number of bits,
our protocol is designed to use $O(\poly(N^{(l)}))$ bits in a level-$l$ gadget for any $l$, as in~\eqref{eq:classical_bit_requirement}.
The level-$l$ circuits for $l\in\{L_M,L_M-1,\ldots,0\}$ in our protocol are composed of $O(\poly(M))$ elementary operations.
Therefore, the number of bits to store all data used in the gadgets in these circuits at all the concatenation levels is upper bounded by
\begin{equation}
  \sum_{l=1}^{L_M}O(\poly(N^{(l)}))\times O(\poly(M))=O(\poly(M)),
\end{equation}
satisfying~\eqref{eq:classical_memory}.

As for the number of parallel processes of classical computation,
a level-$l$ gadget uses at most $O(\poly(N^{(l)}))$ processes at a time, as in~\eqref{eq:classical_parallel}.
Moreover, the level-$l$ gadget is implemented using at most $O(N_{r_l})$ level-$(l-1)$ elementary operations at a time,
where each level-$(l-1)$ gadget for a level-$(l-1)$ elementary operation uses at most $O(\poly(N^{(l-1)}))$ processes at a time.
By repeating this counting recursively,
the number of processes for implementing each level-$L_M$ operation is
\begin{align}
  &O(\poly(N^{(L_M)}))+\nonumber\\
  &\quad O(N_{r_{L_M}})\times O(\poly(N^{(L_M-1)}))+\nonumber\\
  &\quad O(N_{r_{L_M}}N_{r_{L_M-1}})\times O(\poly(N^{(L_M-2)}))+\cdots\nonumber\\
  &\quad O(N_{r_{L_M}}N_{r_{L_M-1}}\cdots N_{r_{2}})\times O(\poly(N^{(1)}))\\
  &=O(\poly(N^{(L_M)})).
\end{align}
Since we have $O(\nicefrac{W(M)}{K^{(L_M)}})$ level-$L_M$ registers at a time,
the number of parallel processes used at a time is bounded by
\begin{align}
  &O\left(\frac{W(M)}{K^{(L_M)}}\times\poly(N^{(L_M)})\right)\\
  &=W(M)\times \exp(O(L_M^2))\\
  &=W(M)\times \exp\left(O\left(\log^2\left(\log\left(\frac{M}{\epsilon}\right)\right)\right)\right),
\end{align}
satisfying~\eqref{eq:parallel_process_number}.

As for the time overhead,
for each $l=L_M,\ldots,1$,
each level-$l$ operation is replaced with an $O(\poly(N_{r_l}))$-depth level-$(l-1)$ circuit, as in~\eqref{eq:G_l} (see also the counting in deriving~\eqref{eq:exrec_bound}).
Thus, in terms of the depth, the level-$(l-1)$ circuit is at most $O(\poly(N_{r_l}))$ times as deep as the level-$l$ circuit.
Therefore, starting from the $O(K^{(L_M)}\log(N^{(L_M)}) D(M))$-depth level-$L_M$ circuit in~\eqref{eq:level_L_time},
we multiply $O(\poly(N_{r_l}))$ recursively to obtain the depth of the resulting level-$0$ circuit as
\begin{align}
  \label{eq:D_FT}
  &D_\mathrm{FT}(M)\nonumber\\
  &=O\left(K^{(L_M)}\log(N^{(L_M)})D(M)\times \prod_{l=1}^{L_M}\poly(N_{r_l})\right)\\
  &=D(M)\times\left(\exp(O(L_M^2))\times \prod_{l=1}^{L_M}\exp(O(l))\right)\\
  &=D(M)\times\exp(O(L_M^2))\\
  &=D(M)\times\exp\left(O\left(\log^2\left(\log\left(\frac{M}{\epsilon}\right)\right)\right)\right).
\end{align}
As a result, the time overhead~\eqref{eq:time_overhead} is
\begin{align}
  \label{eq:time_overhead_conclusion}
  &\frac{D_\mathrm{FT}(M)}{D(M)}=\exp\left(O\left(\log^2\left(\log\left(\frac{M}{\epsilon}\right)\right)\right)\right),
\end{align}
achieving the quasi-polylogarithmic time overhead in $M$.
The conclusion in~\eqref{eq:L},~\eqref{eq:space_overhead_conclusion}, and~\eqref{eq:time_overhead_conclusion} is summarized as Proposition~2 in the main text.

We remark that the above argument on the space and time overhead works fairly in general.
In particular, the same argument leads to the following theorem.
\begin{theorem}[\label{thm:general_overhead}Space and time overhead]
  Under the assumptions of Theorems~\ref{thm:general_rate} and~\ref{thm:general_logical_error_rate},
  if the fault-tolerant protocol of Theorem~\ref{thm:general_logical_error_rate} uses at most $n_l+O(1)$ level-$(l-1)$ registers per encoded level-$l$ register,
  then for any $W(M)=O(\poly(M))$ and $D(M)=O(\poly(M))$,
  there exists a concatenation level
  \begin{equation}
    L_M=O\left(\log\left(\log\left(\frac{M}{\epsilon}\right)\right)\right)
  \end{equation}
  such that the fault-tolerant protocol can simulate any $W(M)$-qubit $D(M)$-depth original circuit within error $\epsilon>0$ in total variation distance with constant space overhead
  \begin{equation}
    \frac{W_\mathrm{FT}(M)}{W(M)}=O(1)
  \end{equation}
  and quasi-polylogarithmic time overhead
  \begin{equation}
    \frac{D_\mathrm{FT}(M)}{D(M)}=\exp\left(O\left(\polylog\left(\log\left(\frac{M}{\epsilon}\right)\right)\right)\right).
  \end{equation}
\end{theorem}

In particular, instead of taking the parameter $r_l$ of the code $\mathcal{Q}_{r_l}$ linearly in $l$, i.e., $r_l=l+2$ as we have done in~\eqref{seq:r_l}, we could grow $r_l$ more slowly to make the scaling of the time overhead smaller.
For example, for any constant $\beta>1$, choose
\begin{equation}
  \label{eq:r_l_log}
  r_l= \lceil\beta\log_2(l)\rceil.
\end{equation}
With the choice~\eqref{eq:r_l_log},
the number of physical qubits of $\mathcal{Q}_{r_l}$ in~\eqref{eq:N_r} and that of logical qubits in~\eqref{eq:K_r} are given respectively by
\begin{align}
  N_{r_l}&= 2^{\lceil\beta\log_2(l)\rceil}-1,\\
  K_{r_l}&= 2^{\lceil\beta\log_2(l)\rceil}-2\lceil\beta\log_2(l)\rceil-1.
\end{align}
Thus,
even if we replace $r_l=l+2$ with $r_l= \lceil\beta\log_2(l)\rceil$, we can still see that the infinite sum~\eqref{eq:infinite_sum_plus_ten} converges as
\begin{equation}
\sum_{l=2}^\infty\left|\frac{(N_{r_l}+10)-K_{r_l}}{N_{r_l}}\right|=\sum_{l=2}^\infty O\left(\frac{\log(l)}{l^\beta}\right)<\infty,
\end{equation}
which leads to~\eqref{eq:space_formula}.
Therefore,
the choice~\eqref{eq:r_l_log} still achieves the constant-space-overhead FTQC, i.e.,
\begin{equation}
  \frac{W_\mathrm{FT}(M)}{W(M)}=O(1).
\end{equation}
As for the time overhead, due to~\eqref{eq:n_L} and~\eqref{eq:k_L}, the number of physical qubits of $\mathcal{Q}^{(l)}$ and that of logical qubits scale respectively as
\begin{align}
  N^{(l)}&=\prod_{l^\prime=1}^{l}N_{r_{l^\prime}}=\exp(O(l\log(l))),\\
  K^{(l)}&=\prod_{l^\prime=1}^{l}K_{r_{l^\prime}}=\exp(O(l\log(l))).
\end{align}
Thus,
the level-$l$ gadgets with $r_l= \lceil\beta\log_2(l)\rceil$ are smaller than those with $r_l=l+2$ in terms of the scaling of the sizes; therefore, the scaling~\eqref{eq:p_L} of the logical error rate derived in our analysis is still valid, and we can choose the overall concatenation level $L_M$ in the same way as~\eqref{eq:L}.
Applying these scalings to~\eqref{eq:D_FT},
we achieve the time overhead scaling as
\begin{align}
  \label{eq:improved_time_overhead}
  &\frac{D_\mathrm{FT}(M)}{D(M)}\nonumber\\
  &=\exp\left(O\left(\log\log\left(\frac{M}{\epsilon}\right)\log\log\log\left(\frac{M}{\epsilon}\right)\right)\right).
\end{align}
As a result, our protocol with the modification of~\eqref{eq:r_l_log} can achieve a time overhead that is polylogarithmic up to the $\log\log\log\left(\nicefrac{M}{\epsilon}\right)$ factor in the exponent.
Note that, while improving the asymptotic scaling of the time overhead, this modification of our protocol may increase the constant factor of the space overhead.
The overall optimization of the protocol to balance the constant factor and the asymptotic scaling is left for future work.

For comparison, the existing constant-space-overhead fault-tolerant protocol~\cite{PhysRevA.87.020304,gottesman2014faulttolerant,8555154} incurs a polynomially large time overhead even though the existing protocol assumes that classical computation during the protocol can be performed instantaneously within zero time.
In particular, the existing protocol uses a concatenated code to prepare encoded auxiliary states used for gate teleportation, but due to the polylogarithmic space overhead in using the conventional concatenated code, the encoded auxiliary states cannot be prepared for applying the gates on all the code blocks in parallel at a time~\cite{gottesman2014faulttolerant}.
Since the size of each code block in the existing protocol is bounded polynomially by $O(M^\alpha)$ for fixed $\alpha>0$, the gates are applicable only to a polynomially small fraction of the code blocks within the constant space overhead; hence, the existing protocol incurs the polynomially large time overhead unless one improves the analysis in Ref.~\cite{gottesman2014faulttolerant} to reduce the required size of code blocks for a desired error suppression.
In addition, for the fair comparison with our protocol, the time overhead of the constant-space-overhead protocol using quantum LDPC codes also needs to be analyzed further by taking into account the wait operations to wait for the nonzero runtime of classical computation since our protocol is formulated and analyzed in this more demanding but practical setting.
The existing analysis of the constant-space-overhead protocol using quantum LDPC codes~\cite{PhysRevA.87.020304,gottesman2014faulttolerant,8555154} does not take into account such nonzero runtime of classical computation.
As discussed in Sec.~\ref{sec:threshold_proof}, to prove the existence of a threshold in the same setting as our protocol, one also needs to develop further new techniques beyond Ref.~\cite{gottesman2014faulttolerant} to deal with nonzero runtime of classical computation in the decoder of the quantum LDPC code.
Despite this more demanding setting, our analysis proves that our fault-tolerant protocol achieves the constant space overhead and the quasi-polylogarithmic time overhead at the same time.
Owing to the doubly exponential error suppression as the concatenation level increases,
we can keep the required code size small,
and hence, our protocol significantly improves the time overhead, from polynomial to quasi-polylogarithmic.
It would be an interesting open question whether a polylogarithmic time overhead or even a sub-polylogarithmic time overhead is achievable within the constant space overhead under reasonable assumptions including the nonzero runtime of classical computation in conducting FTQC\@.

\bibliography{citation_bibtex}

\end{document}